\documentclass[a4paper,fleqn,usenatbib]{mnras}
\usepackage{amsmath,amssymb,natbib,latexsym,times}
\usepackage[T1]{fontenc}
\usepackage{aecompl}

\usepackage{verbatim}
\usepackage[usenames,dvipsnames]{xcolor}
\usepackage{tabulary}
\usepackage{tabularx}
\usepackage{todonotes}
\usepackage{hyperref}
\numberwithin{equation}{section}

\newcommand{\assign}[1]{}
\newcommand{\contrib}[1]{}

\newcommand{\ngalngmix}{3.44}
\newcommand{\ngalimshape}{2.12}
\newcommand{\neffngmix}{5.7}
\newcommand{\neffimshape}{3.7}

\newcommand{\mrequirement}{0.03}
\newcommand{\crequirement}{\ensuremath{2 \times 10^{-3}}}

\newcommand{\greatdes}{GREAT-DES}
\newcommand{\healpix}{HEALPIX}
\newcommand{\fits}{FITS}
\newcommand{\meds}{MEDS}
\newcommand{\medsfull}{Multi-Epoch Data Structures}
\newcommand{\SE}{single-epoch}
\newcommand{\ME}{multi-epoch}

\newcommand{\uberseg}{{\"u}berseg}

\newcommand{\rband}{$r$-band}
\newcommand{\iband}{$i$-band}

\newcommand{\grizY}{$g$, $r$, $i$, $z$, $Y$}

\newcommand{\snr}{\ensuremath{S/N}}
\newcommand{\snrw}{\ensuremath{(S/N)_w}}
\newcommand{\snrr}{\ensuremath{(S/N)_r}}
\newcommand{\rgp}{\ensuremath{R_{gp}/R_p}}
\newcommand{\epsf}{\ensuremath{e_\textsc{psf}}}
\newcommand{\Tpsf}{\ensuremath{T_\textsc{psf}}}
\newcommand{\Tgal}{\ensuremath{T_\mathrm{gal}}}
\newcommand{\bfx}{\ensuremath{\mathbf{x}}}
\newcommand{\bfxpt}{\ensuremath{\mathbf{x} + \boldsymbol{\theta}}}
\newcommand{\dximax}{\ensuremath{\delta \xi^\mathrm{max}}}
\newcommand{\SN}{\ensuremath{\sigma_\textsc{sn}}}

\newcommand*\justify{%
  \fontdimen2\font=0.4em% interword space
  \fontdimen3\font=0.2em% interword stretch
  \fontdimen4\font=0.1em% interword shrink
  \fontdimen7\font=0.1em% extra space
  \hyphenchar\font=`\-% allowing hyphenation
}

\newcommand\code[1]{\texttt{\small\justify #1}}

\newcommand{\sex}{\textsc{SEx\-tractor}}
\newcommand{\psfex}{\textsc{PSF\-Ex}}
\newcommand{\aworld}{\code{A\_WORLD}}
\newcommand{\bworld}{\code{B\_WORLD}}
\newcommand{\frad}{\code{FLUX\_RADIUS}}
\newcommand{\magauto}{\code{MAG\_AUTO}}

\newcommand{\classstar}{\code{CLASS\_STAR}}
\newcommand{\spreadmodel}{\code{SPREAD\_MODEL}}

\newcommand{\imshape}{{\textsc{im3shape}}}
\newcommand{\ngmix}{\textsc{ngmix}}
\newcommand{\galsim}{\textsc{GalSim}}
\newcommand{\LevMar}{\textsc{LevMar}}
\newcommand{\scamp}{\textsc{SCamp}}
\newcommand{\swarp}{\textsc{SWarp}}
\newcommand{\lensfit}{\textsc{lensfit}}

\newcommand{\ngmixSN}{0.22}
\newcommand{\ngmixbands}{$r,i,z$}
\newcommand{\vece}{\mbox{\boldmath $e$}}
\newcommand{\vecg}{\mbox{\boldmath $g$}}

\newcommand{\coadd}{coadd}

\newcommand{\disk}{disc}

\newcommand{\nomepochs}{10}
\newcommand{\nomdepth}{24.1}

\newcommand{\spte}{SPT-E}
\newcommand{\sptearea}{139}

\newcommand{\devauc}{de Vaucouleurs}
\newcommand{\sersic}{S{\'e}rsic}

\newcommand{\levmar}{Levenberg-Marquardt}

\newcommand\eqn[1]{equation~\ref{#1}}
\newcommand\eqnb[2]{equations~\ref{#1}~\& \ref{#2}}
\newcommand\eqnc[2]{equations~\ref{#1}~--~\ref{#2}}
\newcommand\fig[1]{Figure~\ref{#1}}
\newcommand\figb[2]{Figures~\ref{#1}~\& \ref{#2}}
\newcommand\tab[1]{Table~\ref{#1}}

\newcommand\app[1]{Appendix~\ref{#1}}
\newcommand\edit[1]{#1}

\allowdisplaybreaks

\title{The DES Science Verification Weak Lensing Shear Catalogues }

\author[M. Jarvis, E. Sheldon, J. Zuntz, T. Kacprzak, S. Bridle, et al.]{
\parbox{\textwidth}{
\Large
M.~Jarvis$^{1}$,
E.~Sheldon$^{2}$,
J.~Zuntz$^{3}$,
T.~Kacprzak$^{4}$,
S.~L.~Bridle$^{3}$,
A.~Amara$^{4}$,
R.~Armstrong$^{5}$,
M.~R.~Becker$^{6,7}$,
G.~M.~Bernstein$^{1}$,
C.~Bonnett$^{8}$,
C.~Chang$^{4}$,
R.~Das$^{9}$,
J.~P.~Dietrich$^{10,11}$,
A.~Drlica-Wagner$^{12}$,
T.~F.~Eifler$^{1,13}$,
C.~Gangkofner$^{10,11}$,
D.~Gruen$^{14,15}$,
M.~Hirsch$^{16}$,
E.~M.~Huff$^{17,18}$,
B.~Jain$^{1}$,
S.~Kent$^{12}$,
D.~Kirk$^{16}$,
N.~MacCrann$^{3}$,
P.~Melchior$^{17,18}$,
A.~A.~Plazas$^{13}$,
A.~Refregier$^{4}$,
B.~Rowe$^{16}$,
E.~S.~Rykoff$^{7,19}$,
S.~Samuroff$^{3}$,
C.~S{\'a}nchez$^{8}$,
E.~Suchyta$^{17,18}$,
M.~A.~Troxel$^{3}$,
V.~Vikram$^{20}$,
T.~Abbott$^{21}$,
F.~B.~Abdalla$^{16,22}$,
S.~Allam$^{12}$,
J.~Annis$^{12}$,
A.~Benoit-L{\'e}vy$^{16}$,
E.~Bertin$^{23,24}$,
D.~Brooks$^{16}$,
E.~Buckley-Geer$^{12}$,
D.~L.~Burke$^{7,19}$,
D.~Capozzi$^{25}$,
A.~Carnero~Rosell$^{26,27}$,
M.~Carrasco~Kind$^{28,29}$,
J.~Carretero$^{30,8}$,
F.~J.~Castander$^{30}$,
J.~Clampitt$^{1}$,
M.~Crocce$^{30}$,
C.~E.~Cunha$^{7}$,
C.~B.~D'Andrea$^{25}$,
L.~N.~da Costa$^{26,27}$,
D.~L.~DePoy$^{31}$,
S.~Desai$^{10,11}$,
H.~T.~Diehl$^{12}$,
P.~Doel$^{16}$,
A.~Fausti Neto$^{26}$,
B.~Flaugher$^{12}$,
P.~Fosalba$^{30}$,
J.~Frieman$^{12,32}$,
E.~Gaztanaga$^{30}$,
D.~W.~Gerdes$^{9}$,
R.~A.~Gruendl$^{28,29}$,
G.~Gutierrez$^{12}$,
K.~Honscheid$^{17,18}$,
D.~J.~James$^{21}$,
K.~Kuehn$^{33}$,
N.~Kuropatkin$^{12}$,
O.~Lahav$^{16}$,
T.~S.~Li$^{31}$,
M.~Lima$^{34,26}$,
M.~March$^{1}$,
P.~Martini$^{17,35}$,
R.~Miquel$^{36,8}$,
J.~J.~Mohr$^{10,11,14}$,
E.~Neilsen$^{12}$,
B.~Nord$^{12}$,
R.~Ogando$^{26,27}$,
K.~Reil$^{19}$,
A.~K.~Romer$^{37}$,
A.~Roodman$^{7,19}$,
M.~Sako$^{1}$,
E.~Sanchez$^{38}$,
V.~Scarpine$^{12}$,
M.~Schubnell$^{9}$,
I.~Sevilla-Noarbe$^{38,28}$,
R.~C.~Smith$^{21}$,
M.~Soares-Santos$^{12}$,
F.~Sobreira$^{12,26}$,
M.~E.~C.~Swanson$^{29}$,
G.~Tarle$^{9}$,
J.~Thaler$^{39}$,
D.~Thomas$^{25}$,
A.~R.~Walker$^{21}$,
R.~H.~Wechsler$^{6,7,19}$
}
\vspace{0.4cm}
\\
\parbox{\textwidth}{
$^{1}$ Department of Physics and Astronomy, University of Pennsylvania, Philadelphia, PA 19104, USA\\
$^{2}$ Brookhaven National Laboratory, Bldg 510, Upton, NY 11973, USA\\
$^{3}$ Jodrell Bank Center for Astrophysics, School of Physics and Astronomy, University of Manchester, Oxford Road, Manchester, M13 9PL, UK\\
$^{4}$ Department of Physics, ETH Zurich, Wolfgang-Pauli-Strasse 16, CH-8093 Zurich, Switzerland\\
$^{5}$ Department of Astrophysical Sciences, Princeton University, Peyton Hall, Princeton, NJ 08544, USA\\
$^{6}$ Department of Physics, Stanford University, 382 Via Pueblo Mall, Stanford, CA 94305, USA\\
$^{7}$ Kavli Institute for Particle Astrophysics \& Cosmology, P. O. Box 2450, Stanford University, Stanford, CA 94305, USA\\
$^{8}$ Institut de F\'{\i}sica d'Altes Energies, Universitat Aut\`onoma de Barcelona, E-08193 Bellaterra, Barcelona, Spain\\
$^{9}$ Department of Physics, University of Michigan, Ann Arbor, MI 48109, USA\\
$^{10}$ Excellence Cluster Universe, Boltzmannstr.\ 2, 85748 Garching, Germany\\
$^{11}$ Faculty of Physics, Ludwig-Maximilians University, Scheinerstr. 1, 81679 Munich, Germany\\
$^{12}$ Fermi National Accelerator Laboratory, P. O. Box 500, Batavia, IL 60510, USA\\
$^{13}$ Jet Propulsion Laboratory, California Institute of Technology, 4800 Oak Grove Dr., Pasadena, CA 91109, USA\\
$^{14}$ Max Planck Institute for Extraterrestrial Physics, Giessenbachstrasse, 85748 Garching, Germany\\
$^{15}$ Universit\"ats-Sternwarte, Fakult\"at f\"ur Physik, Ludwig-Maximilians Universit\"at M\"unchen, Scheinerstr. 1, 81679 M\"unchen, Germany\\
$^{16}$ Department of Physics \& Astronomy, University College London, Gower Street, London, WC1E 6BT, UK\\
$^{17}$ Center for Cosmology and Astro-Particle Physics, The Ohio State University, Columbus, OH 43210, USA\\
$^{18}$ Department of Physics, The Ohio State University, Columbus, OH 43210, USA\\
$^{19}$ SLAC National Accelerator Laboratory, Menlo Park, CA 94025, USA\\
$^{20}$ Argonne National Laboratory, 9700 South Cass Avenue, Lemont, IL 60439, USA\\
$^{21}$ Cerro Tololo Inter-American Observatory, National Optical Astronomy Observatory, Casilla 603, La Serena, Chile\\
$^{22}$ Department of Physics and Electronics, Rhodes University, PO Box 94, Grahamstown, 6140, South Africa\\
$^{23}$ CNRS, UMR 7095, Institut d'Astrophysique de Paris, F-75014, Paris, France\\
$^{24}$ Sorbonne Universit\'es, UPMC Univ Paris 06, UMR 7095, Institut d'Astrophysique de Paris, F-75014, Paris, France\\
$^{25}$ Institute of Cosmology \& Gravitation, University of Portsmouth, Portsmouth, PO1 3FX, UK\\
$^{26}$ Laborat\'orio Interinstitucional de e-Astronomia - LIneA, Rua Gal. Jos\'e Cristino 77, Rio de Janeiro, RJ - 20921-400, Brazil\\
$^{27}$ Observat\'orio Nacional, Rua Gal. Jos\'e Cristino 77, Rio de Janeiro, RJ - 20921-400, Brazil\\
$^{28}$ Department of Astronomy, University of Illinois, 1002 W. Green Street, Urbana, IL 61801, USA\\
$^{29}$ National Center for Supercomputing Applications, 1205 West Clark St., Urbana, IL 61801, USA\\
$^{30}$ Institut de Ci\`encies de l'Espai, IEEC-CSIC, Campus UAB, Carrer de Can Magrans, s/n,  08193 Bellaterra, Barcelona, Spain\\
$^{31}$ George P. and Cynthia Woods Mitchell Institute for Fundamental Physics and Astronomy, and Department of Physics and Astronomy, Texas A\&M University, College Station, TX 77843,  USA\\
$^{32}$ Kavli Institute for Cosmological Physics, University of Chicago, Chicago, IL 60637, USA\\
$^{33}$ Australian Astronomical Observatory, North Ryde, NSW 2113, Australia\\
$^{34}$ Departamento de F\'{\i}sica Matem\'atica,  Instituto de F\'{\i}sica, Universidade de S\~ao Paulo,  CP 66318, CEP 05314-970, S\~ao Paulo, SP, Brazil\\
$^{35}$ Department of Astronomy, The Ohio State University, Columbus, OH 43210, USA\\
$^{36}$ Instituci\'o Catalana de Recerca i Estudis Avan\c{c}ats, E-08010 Barcelona, Spain\\
$^{37}$ Department of Physics and Astronomy, Pevensey Building, University of Sussex, Brighton, BN1 9QH, UK\\
$^{38}$ Centro de Investigaciones Energ\'eticas, Medioambientales y Tecnol\'ogicas (CIEMAT), Madrid, Spain\\
$^{39}$ Department of Physics, University of Illinois, 1110 W. Green St., Urbana, IL 61801, USA\\
}
}

\date{Accepted XXX. Received YYY; in original form ZZZ}

\pubyear{2015}

\begin{document}

\label{firstpage}
\pagerange{\pageref{firstpage}--\pageref{lastpage}}
\maketitle

\begin{abstract}
We present weak lensing shear catalogues for 
\sptearea\ square degrees of data taken during the Science Verification (SV) time
for the new Dark Energy Camera (DECam) being used for the Dark Energy Survey (DES).
We describe our object selection, point spread function estimation and shear measurement procedures using
two independent shear pipelines, \imshape\ and \ngmix, 
which produce catalogues
of \ngalimshape\ million 
and \ngalngmix\ million galaxies respectively.
We detail a set of null tests for the shear measurements and find that they pass the requirements for systematic 
errors at the level necessary for weak lensing science applications using the SV data.
We also discuss some of the planned algorithmic improvements that will be necessary to
produce sufficiently accurate shear catalogues
for the full 5-year DES, which is expected to cover 5000 square degrees.
\vspace{25pt}
\end{abstract}


% with so many sections, it is nice to see a table of contents
% comment this line for submission
%\tableofcontents

\section{Introduction}
\assign{Sarah}
\contrib{Mike}

Weak gravitational lensing provides a powerful statistical tool for studying the
distribution of mass in the Universe.  
Light traveling from distant galaxies to Earth is deflected
by the gravitational field of mass concentrations along the
path.  This deflection distorts the observed light distribution of galaxies,
and when this distortion is very small, 
stretching the surface brightness profile
by of order a few percent or less,
it is
referred to as ``weak lensing''.

The weak lensing distortion includes both a stretching component called
``shear'' and a dilation component called ``convergence''. 
Here we focus on the shear.
The observed shear field can be used to
make maps of the matter in the universe,
uncover the mass profiles of galaxies and
clusters of galaxies, and even 
test theoretical models of dark energy. 

In order to reach its full potential as a probe of dark matter and dark energy,
shear measurement must be extremely accurate.  Each galaxy is typically
stretched by about 
$2\%$, 
whereas the intrinsic unknown ellipticity of the
galaxy before being lensed is an order of magnitude larger.  
This ``shape
noise'' constitutes the primary statistical uncertainty for weak lensing
measurements.  
Nevertheless, by measuring the shapes of millions of galaxies,
the Dark Energy Survey (DES) and other current surveys can expect to 
make precise measurements of the mean shear with fractional statistical
uncertainties as low as 1\%. Future surveys may reach
0.1\%.
This implies that systematic errors
(i.e. biases)
 in the shape measurements need to
be controlled at a level approximately $3$ orders of magnitude smaller than the
shape noise on each measurement.

There are many potential sources of systematic error that can bias the shape
measurements used for estimating shears.  The galaxy images are blurred and
smeared
when the photons pass through
the atmosphere, the telescope optics, and the
detector, leading to a spatially and temporally variable point-spread function
(PSF).  The images are stretched by distortion from the telescope and sometimes by
features of the detector.  The images are pixellated and have various sources
of noise.  Detector defects, 
cosmic rays, satellite trails,
and other artefacts in the data can lead to some
pixels not being used, and measurement algorithms must deal properly with this
``missing data''.  Flux from nearby galaxies or stars can obfuscate the
determination of the observed intensity profile.  All of these phenomena must
be included in the analysis at very high accuracy if systematic uncertainties are to
be sub-dominant to statistical uncertainties.

Previous studies have taken a range of approaches to measuring galaxy shapes,
typically falling into one of two categories.  Moments-based methods
\citep[e.g.][]{KSB,Rhodes00, Melchior11} involve measuring second and
higher-order moments of the galaxy and the PSF.  Model-fitting methods
\citep[e.g.][]{Massey05, Nakajima07, Miller2013} involve fitting a
PSF-convolved galaxy model to the data. 
A number of blind challenges of shear measurements have been carried out to assess progress in a 
uniform way across the international shear measurement community: the Shear TEsting Programme 
\citep[STEP][]{step1,step2} and the 
GRavitational lEnsing Accuracy Testing (GREAT) Challenges 
\citep{great08handbook,great08results,great10handbook,great10results,great3handbook, great3results}.
The wide variety of shear measurement methods and their performance on these benchmarks are summarized there.
The two shear algorithms presented in
this work, \imshape\ \citep{im3shape}, and \ngmix, are both of the model-fitting variety
(cf.~\S\ref{sec:shear}).

Most shear measurement methods are biased in the low signal-to-noise (\snr) regime,
where the impact of pixel noise on the shape measurement of each galaxy becomes significant.
This ``noise bias'' effect was first discussed in \citet{BJ02} and \citet{Hirata04},
and was found to be the most significant of the effects studied in %SLB
the GREAT08 Challenge
\citep{great08handbook,great08results}.
It was derived analytically for maximum-likelihood methods in \citet{refregier12}, in the context of direct estimation in 
\citet{Melchior12}, and quantified in the context of future surveys in \citet{kacprzak12}.

Complex galaxy morphologies can also bias shear measurements
\citep{step2, Lewis09, Voigt10, Bernstein10, Melchior10, ZhangKomatsu11}.  This
``model bias'' can arise 
even for simple galaxy profiles if the model being used does not match reality.
Model bias was found to be around 1\% for bulge+\disk\ model fitting methods, and the interplay with
noise bias was found to be small \citep{Kacprzak14}.
The GREAT3 challenge
\citep{great3handbook, great3results} included realistic galaxy morphologies,
and those authors found that the mean model
bias was $\sim\!1$\% for a wide range of methods. 
The Fourier Domain Nulling approach \citep{Bernstein10} 
provides a potential solution to this problem, which 
may be able to avoid model bias altogether.

One strategy to account for these biases is to apply a multiplicative
correction factor calibrated from image simulations.  This can take the form of
a single constant bias correction applied to all galaxies \citep[e.g.][]{Schrabback07}, or it can 
vary according to galaxy properties such as the signal-to-noise ratio
\citep{Schrabback10,Gruen13} and size \citep{vonderLinden14}. For the
\imshape\ shear measurements, we calibrate biases as a function of both of these
parameters, as done by \citet{kacprzak12}. A significant improvement 
in the current analysis lies in
our modeling of additive systematic errors as proportional to PSF ellipticity,
which we also apply as a calibration (cf.~\S\ref{sec:im3shape:noisebias}).

A different strategy to account for noise bias (although not model bias)
is to include the known distribution of
intrinsic galaxy shapes as Bayesian prior information
and fully sample the posterior likelihood surface.
\citet{Miller2007} proposed a first order approximation to this, and a more
rigorous treatment was given by \citet{BernsteinArmstrong2014}. 
%MJ: I thought to also cite Anze, but he doesn't seem to have published his even 
%    more rigorous treatment yet.
For the \ngmix\ shear measurements, we follow the approach of \citet{Miller2007} (cf.~\S\ref{sec:ngmix:shear}).

Each part of the sky in a weak lensing survey is generally observed multiple times.
Most commonly, the shape measurements are made on 
\coadd ed images of these multiple exposures \citep[e.g.][]{Wittman00, LVW00, Heymans05, Leauthaud07, Fu08}.
While \coadd ed images reduce the total data volume, making data handling easier,
differences in the PSFs between the epochs complicate the modeling of the \coadd ed PSF
and often introduce spurious effects that are problematic for the most
sensitive shear probes.  Multi-epoch methods \citep{Tyson08, Bosch11, Miller2013}
instead simultaneously use all individual exposures of a galaxy
with the corresponding single-epoch PSF models and weights,
thereby avoiding these problems.

The current state-of-the-art weak lensing shear measurement comes from
the Canada-France-Hawaii Telescope Lensing Survey \citep[CFHTLenS;][]{Heymans12}, 
which observed 154 square degrees of sky and measured 7.6 million galaxy shapes.
They discovered that the previous CFHTLenS analysis \citep{Fu08}, using
\coadd\ images, had significant systematic errors and that switching to
a multi-epoch method \citep{Miller2013} was superior.  
We use similar multi-epoch algorithms in this work (cf. \S\ref{sec:shear:multiepoch}).

For removing problematic data, the CFHTLenS analysis 
trimmed the survey area to only those fields
in which the shape catalogues passed certain systematic tests. 
We use a somewhat different strategy in our analysis.  We
blacklist \SE\ images that fail tests of the image quality, the astrometric solution,
or the PSF model, and exclude them from the multi-epoch fitting process (cf. \S\ref{sec:meds:blacklists}).

\begin{figure*}
\centerline{\includegraphics[width=14cm]{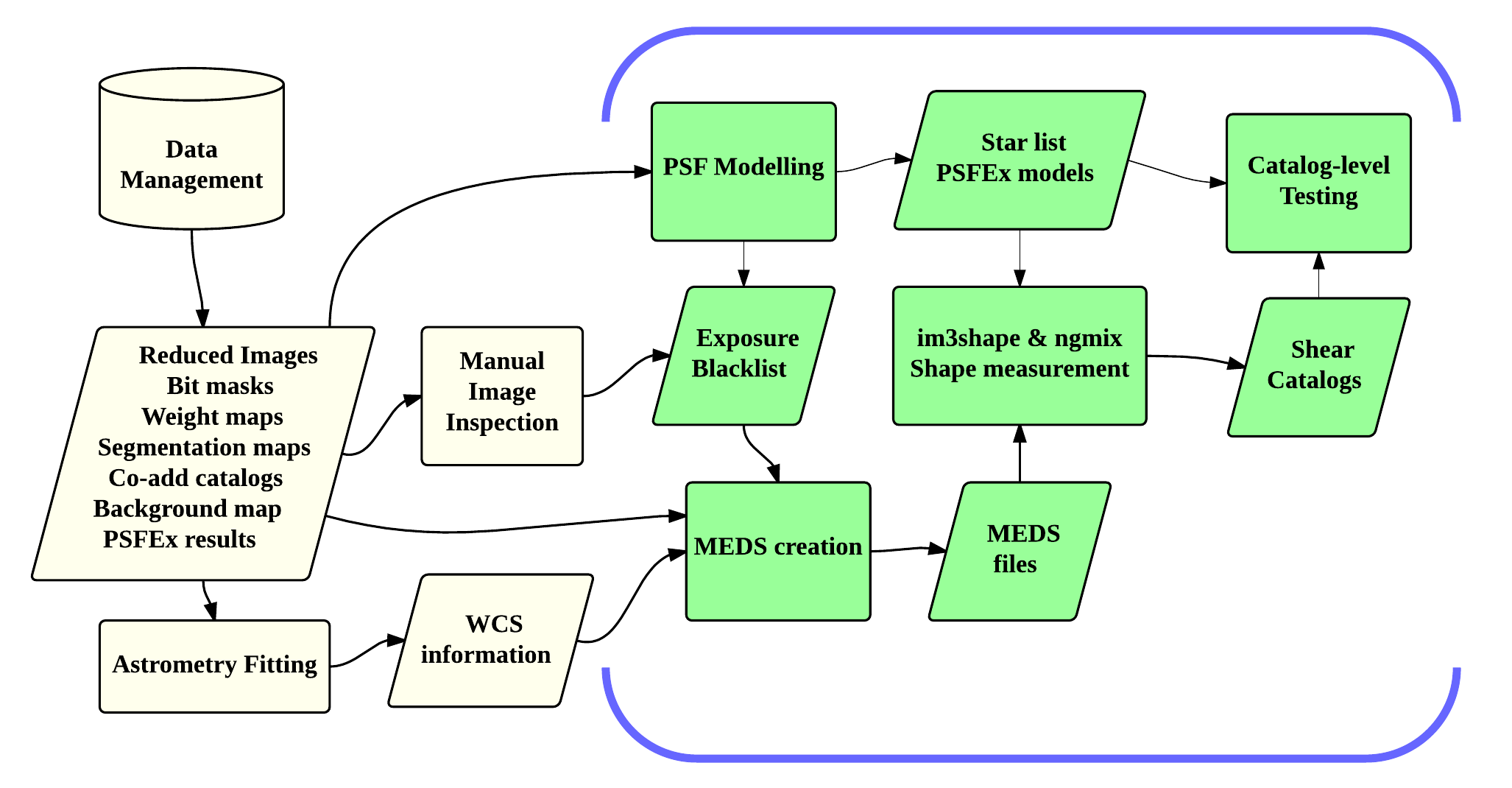}}
\caption{A flowchart showing the main stages in the production of the shear catalogues.  The items
inside the blue bracket are done by the weak lensing group in DES and are the principal subject of this
paper.
\label{fig:flowchart}
}
\end{figure*}

In this paper we present the shear catalogue for the DES Science Verification (SV) data,
described in \S\ref{sec:data}.  We derive requirements for our systematic
uncertainties in \S\ref{sec:req}.  The PSF model is described and tested in \S\ref{sec:psf}.
To facilitate multi-epoch shear measurements,
we developed \medsfull\ (\meds), which we describe in \S\ref{sec:meds}.
Two sets of simulations that we used for calibration and testing are presented in \S\ref{sec:sims}.
We present our two shear estimation codes, \imshape\ and \ngmix\ in \S\ref{sec:shear}.
Then we submit our catalogues to a suite of null
tests, described in \S\ref{sec:tests}, which constitutes the
main results of this paper.
Finally, we describe our final shear catalogues in \S\ref{sec:cats} and conclude in
\S\ref{sec:conclusions}.  Appendices provide more information on the data
structures and catalogue flags.  A flowchart outlining the main stages in the
production of the shear catalogues is shown in \fig{fig:flowchart}.

\section{Data}
\label{sec:data}

\assign{Erin}

The Dark Energy Camera
\citep[DECam;][]{FlaugherDECam2015,DiehlDECam2012,HonscheidControl2012}
was installed on the 4m Victor M. Blanco Telescope
at the Cerro Tololo Inter-American Observatory (CTIO) in Chile from June, 2011, to September, 2012 \citep{Diehl14}.
The first light ceremony was September 12, 2012.

DECam holds sixty-two $2048\times4096$ science CCDs, four $2048\times2048$ guider CCDs, and eight
$2048\times2048$ focus and alignment CCDs, for a total of 570 megapixels covering a
roughly hexagonal footprint.  The CCDs were fabricated at Teledyne 
Dalsa\footnote{\url{https://www.teledynedalsa.com}},
further processed by Lawrence Berkeley National Laboratory (LBNL),
and assembled and tested at 
% Steve says this should go as a footnote the first time Fermilab is mentioned, as an affil or otherwise.
Fermilab\footnote{Fermilab 
is operated by Fermi Research Alliance, LLC under Contract No. De-AC02-07CH11359 with the United States Department of Energy.
}. Each CCD is 250
microns thick and fully depleted, with two amplifiers per CCD. 

The DECam field of view has a diameter of 2.2 degrees on the sky.  Unfortunately, one the
62 science CCDs was damaged during commissioning, so we have only
61 working CCDs\footnote{One additional science CCD failed in the first year of the
DES main survey \citep{Diehl14}, but it was still functional for the work presented here.}.
The total usable footprint of an exposure, excluding the gaps
between the CCDs, totals \edit{2.7} square degrees.
Five filters are used during normal survey operations, \grizY, exchanged using
an automated shutter-filter system \citep{TarleFilterChanger2010}.

The Dark Energy Survey (DES) 
officially started taking survey data in August, 2013 \citep{Diehl14}.
It will cover about 5000 square degrees in the South Galactic Cap
region, with $\sim$\nomepochs\ visits per field in the $g$, $r$, $i$ and
%SB: Josh told me we have to put "sim" i.e. ~ 10 epochs for the full survey not 10 without the ~
$z$ bands (two visits per year), for a $10 \sigma$ limiting magnitude of about
\nomdepth\ in the $i$ band.  In addition to the
main survey, the DES supernova survey contains smaller patches optimized for time-domain science, which
are visited more often, and which are useful as a deeper dataset observed with
the same instrument.

Before the start of the main survey, a small Science Verification (SV)
survey was conducted from November 2012 to February 2013.  The strategy was to
observe the SV area at \nomepochs\ different epochs, mimicking the number of
visits and total image depth planned for the full 5-year DES survey.
\edit{The dither pattern matches that of the main survey, which uses large dithers
to minimize the impact of any systematic errors related to the location on the field of view.
Each tiling is typically observed on different nights to vary the observing conditions
as much as possible.}
Significant depth variations exist in the SV data due to weather, issues with the telescope,
and no data quality checks to ensure uniformity
\citep[cf.][]{Leistedt15}.

For the current study we restricted our measurements to the
largest portion of the SV area, known as SPT-East (\spte\ for
short), an area of approximately \sptearea\ square degrees contained within the
eastern part of the
region observed by the South Pole Telescope \citep[SPT; ][]{CarlstromSPT2011}.

The SV data were reduced by the DES Data Management (DESDM) system
\citep{DESDM2012, Desai12}, resulting in calibrated and background-subtracted images.
Catalogues 
%of detected objects %in these images % fluff text to get the para to wrap not at a hyperlink.
were produced using the software package Source Extractor
\citep[\sex;][]{BertinSExtractor1996,BertinPSFEx2011}.  
The point spread
function was characterized using the \psfex\ package (\citealp{BertinPSFEx2011}; for more details,
see \S\ref{sec:psf}).

On a set of pre-defined areas of sky, all overlapping \SE\
images were registered and combined into a \coadd\ image using the \scamp\ and
\swarp\ packages \citep{BertinSWarp2002,BertinSCAMP2006}.

For weak lensing we used these \coadd\ images only for object detection, deblending,
fluxes (for use in
photometric redshift measurements, see \citealp{Sanchez14, Bonnett15}),
and for the detailed informational
flags which were important for determining a good set of galaxies to use for
shear measurement.  

In contrast to previous work on DES data by \cite{Melchior14},
we performed object shape measurement directly on all
available \SE\ images in which an object was observed, using \ME\
fitting techniques.  See \S\ref{sec:meds} for more details of how we
repackaged the data for \ME\ fitting and \S\ref{sec:shear:multiepoch} for 
a description of the multi-epoch measurement process.

\subsection{Object Catalogue}
\label{sec:gold}
\assign{Erin}
\contrib{Erin,Peter}

The starting point for our object catalogue was
the ``SVA1 Gold Catalogue''
\footnote{\url{http://des.ncsa.illinois.edu/releases/sva1}},
which excludes regions of the data that were found to be problematic in 
some way, due to imaging artefacts, scattered light, failed observations, etc.
The selection criteria for the Gold Catalogue included the following:
\begin{itemize}
\item Required object to have been observed at least once in each of the $g, r, i,$ and $z$ bands.
\item Required Declination to be north of $61^\circ S$ to avoid the Large Magellanic Cloud and R Doradus,
where the photometric calibration was found to have severe problems.
\item Removed regions with a high density of objects with ``crazy colours'', 
i.e. those with any of the following:
$g-r < -1,~ g-r > 4,~ i-z < -1,$ or $i-z > 4$.
Such regions are usually due to satellite trails, ghosts, scattered light, etc.
\item Removed regions with a density less than $3\sigma$ below the mean density.
\item Removed regions near bright stars.  We eliminated a circular region around all
stars detected in the 2-micron All Sky Survey \citep[2MASS;][]{Skrutskie2MASS} brighter than $J_M = 12$ with a mask radius of
$r = (-10 J_M + 150)$ arcseconds up to a maximum radius of 120 arcseconds.
\item Removed regions with a concentration of objects with large centroid shifts between bandpasses.
Some of these objects are just dropout galaxies or large galaxies with complex, wavelength-dependent substructure,
but many are due to scattered light, ghosts, satellite trails, etc.  25\% of such objects
fall into 4\% of the total area, so we removed all objects in that 4\% on the assumption
that the other nearby objects probably have corrupted shapes and photometry.
\end{itemize}
The full \spte\ area observed during SV totals 163 square degrees.
Applying the above selection criteria brings this down to 148 square degrees for the Gold Catalogue.

\begin{figure}
\includegraphics[width=\columnwidth]{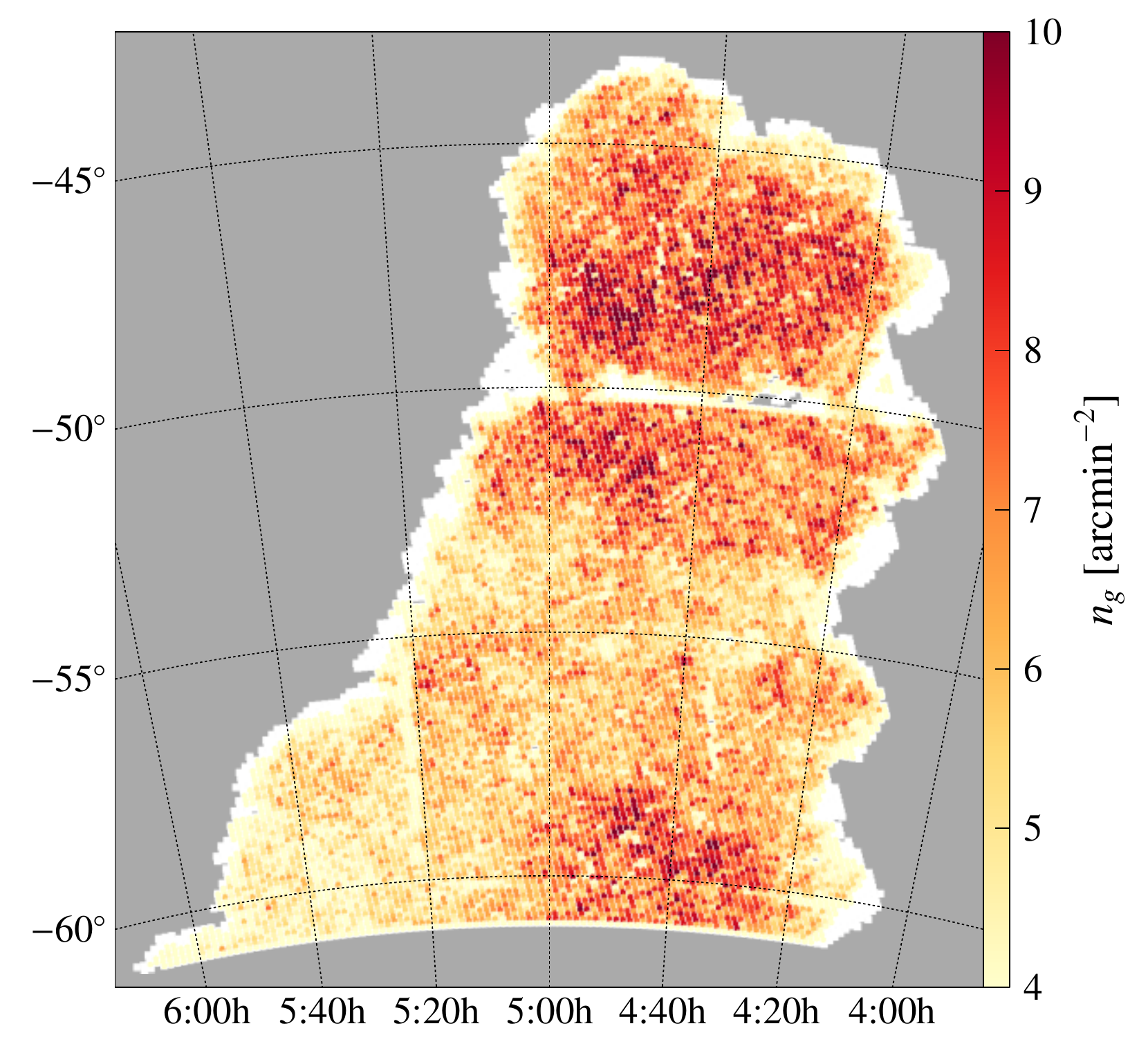}
\caption{A \healpix\ map of the \spte\ region.
The white background shows the full ``Gold'' area.
The colours show the galaxy density in the \ngmix\ shear catalogue.
(The map for \imshape\ is qualitatively similar, although about 40\% shallower.)
The map has \healpix\ resolution \code{nside} = 512.
\label{fig:sptemap}
}
\end{figure}

The selection criteria listed above removed galaxies in a non-random way that varied across
the sky.   We characterized this selection using a geometrical ``mask'',
implemented as a \healpix{} map \citep{healpix}.  The \healpix{} map for the DES
\spte\ region is shown in \fig{fig:sptemap}.  The white background represents
the Gold Catalogue area.  The coloured intensity represents the galaxy number
density in the \ngmix\ catalogue (cf. \S\ref{sec:ngmix}). 

The region used for the
weak lensing analysis is somewhat smaller than the full Gold Catalogue region,
because we additionally excluded
CCD images with poor astrometric solutions (cf. \S\ref{sec:astrometry}), poor PSF
solutions (cf. \S\ref{sec:psf:cuts}), and blacklisted CCDs containing bright
stars, ghosts, airplanes etc. (cf. \S\ref{sec:meds:blacklists}).  The
astrometric cuts in particular removed regions near the edge, since the
solutions were poorly constrained there, resulting in a final area for the
shear catalogues of \sptearea\ square degrees.  The intensity map for the
\imshape\  catalogue (cf. \S\ref{sec:im3shape}) looks qualitatively similar,
although it is about 40\% shallower (cf. \S\ref{sec:cats:neff}).

\subsection{Galaxy Selection}
\label{sec:modest}

The preliminary galaxy selection was performed using
standard
\sex\ outputs from the \iband\ detections in the Gold Catalogue.
The selection, in pseudo-code, was

{\footnotesize 
\begin{verbatim}
  bright_test = CLASS_STAR > 0.3
                AND MAG_AUTO < 18.0
  locus_test = SPREAD_MODEL + 
               3*SPREADERR_MODEL < 0.003
  faint_psf_test = MAG_PSF > 30.0
                   AND MAG_AUTO < 21.0

  galaxies = NOT bright_test
             AND NOT locus_test
             AND NOT faint_psf_test
\end{verbatim}
}

Within DES, this is called the ``Modest Classification'' scheme.
Bright stars were identified by the standard \sex\ classifier (\code{bright\_test}). 
Fainter objects were considered stars if they were near the stellar locus in the
\spreadmodel\ measure, introduced by \citet{Desai12}, \edit{which uses a model of
the local PSF to quantify the difference between PSF-like objects and
resolved objects} (\code{locus\_test}; see also \citealp{Bouy13}).

\edit{Objects whose best estimate of the total magnitude is much brighter
than their PSF magnitude (\code{faint\_psf\_test})} are often spurious detections
and were considered ``junk'' in this classification.  Our \edit{initial} galaxy
selection then included every object not classified as either a star or junk
\edit{by this scheme}.

\edit{Many faint stars and spurious detections remained in the catalog
at this stage.}
Later, further selection criteria, described in \S\ref{sec:cats:flags}, were
applied based on measurements from the shear pipelines.  \edit{The initial
selection was not intended to produce an accurate galaxy catalogue, but rather
to produce a superset of the objects that would eventually be trimmed based
on more stringent selection criteria.  All objects in this
preliminary galaxy catalogue were processed by both shear measurement algorithms
(cf. \S\ref{sec:shear}).}

\subsection{Astrometry}
\label{sec:astrometry}
\assign{Eli}
\contrib{Gary}

For each CCD image we must establish an astrometric solution, i.e.\ a map
from pixel coordinates $(x,y$) to celestial coordinates
$(\theta,\phi)$, known as the World Coordinate System (WCS). 
Since the determination of galaxy shapes is
done by a simultaneous fit to the pixel data for all \SE\ exposures
covering the galaxy, any misregistration of the exposures
will introduce spurious shear signals into the inferred galaxy shapes
and sizes.

We found that the astrometric solutions provided by DESDM were not sufficiently
accurate for our needs.  They included misregistrations of more than 150 milliarcseconds
on some CCDs, which induced
unacceptably high systematic errors in the galaxy shapes.  
Here we
describe the process we used to improve these solutions to the WCS.

Astrometric solutions for the SV exposures were assumed to take the
form
\begin{equation}
\label{eq:astrometric}
P(\theta,\phi) = E\left(C(x,y)\right), 
\end{equation}
where $P$ is a gnomonic projection from the (curved) sky on to a planar
coordinate system, using a chosen field coordinate for the pole of the
projection; $E$ is an affine transformation chosen to be distinct for
each CCD image of each exposure; and $C$ is a cubic polynomial mapping that
is common to all exposures in a given filter with a given CCD.  In the
nomenclature of the \scamp\
code\footnote{\url{http://www.astromatic.net/software/scamp}},
$C$ is the ``instrument'' solution, and $E$ is the ``exposure'' solution.

The instrument solution $C$ was derived as follows. We took
a series of $\approx 20$ exposures of a rich
star field in succession, with the telescope displaced by 
angles ranging from 10 arcseconds up to the field of view of the camera.
Coordinates of stars were determined in the pixel coordinates of each
exposure, and we adjusted the parameters of the map in \eqn{eq:astrometric} to 
minimize the internal disagreement between
sky coordinates of all the observations of each star.  The solution
also minimized the discrepancies between the positions of stars in the
2MASS \citep{Skrutskie2MASS} catalogue and our measurements of these stars,
thereby anchoring the absolute pointing and scale of our astrometric maps.

All 20 coefficients of the cubic polynomial $C$ were left free for
each of the 61 functional CCDs \citep[cf.][]{Diehl14,FlaugherDECam2015}.
While fitting the star field data, we
forced all CCDs in a given exposure to share a common affine map $E$,
so there were 6 additional free parameters in the fit for each 
exposure.  The instrument maps $C$ derived in this way were
assumed to apply to all SV exposures taken with the same CCD in the
same filter.  The process was repeated for each of the \grizY\ filters.

\edit{Repeating the above ``star flat'' procedure every few months revealed small changes
in the astrometric map, consistent with rigid motion of some CCDs relative
to the others at a level of $\approx 10$~milliarcseconds, probably occurring when the
camera was cycled to room temperature for occasional engineering tasks.}

\edit{To account for these small shifts,}
we determined an independent $E$ function (6 degrees of freedom)
for each CCD image in the SV data
in another round of fitting.  In this
second stage we minimized the disagreements between positions reported
for all CCDs that contribute to each DES \coadd\ image.  The
coefficients of the affine transformations $E$ were allowed to float, 
but the higher-order polynomials $C$
were held fixed at the values determined from the star field data.  These
solutions again minimized residuals with respect to matching sources from the
2MASS catalogue in order to fix the absolute position on the sky.

Note that the principal effects of differential chromatic refraction (DCR) are a 
shift and a shear along the direction toward zenith, which are both properly 
included as part of the affine transformation $E$ for each CCD.  We did not,
however, make any attempt to address the intra-band chromatic effects
related to DCR \citep[cf.][]{Plazas12, Myers15}.

The RMS disagreement between sky positions of bright stars inferred from
distinct DES exposures are consistent with errors in the astrometric maps of
10--20~milliarcseconds RMS in each coordinate.  We found these errors to
be coherent over arcminute scales in a given exposure, but
were uncorrelated between distinct exposures.  We interpret this to mean 
the remaining relative astrometric errors are dominated by stochastic
atmospheric distortions \citep[cf.][]{Heymans12b}.  Indeed, equation~8 of
\citet{Bouy13} predicts an RMS astrometric residual due to the atmosphere
of order 10~milliarcseconds for our field of view and exposure time.

We found some remaining astrometric errors that
were coherent over time and correlated with position on the detector
array, which are consistent with
small components of the electric fields transverse to the 
surface of the CCD in some places \citep{plazas14}.  These residuals are
at the few milliarcsecond level, which is small enough to be irrelevant for SV data
reductions.

\section{Requirements on Systematic Errors}
\label{sec:req}

In this section, we derive the requirements for systematic uncertainties on the shear
estimates for the DES SV data.  These requirements will be used to assess the
quality of the PSF and shear catalogues in subsequent sections.

Throughout this paper we will use the notation $e = e_1 + i e_2 = |e| \exp(2i \phi)$ as the complex-valued 
shape of each galaxy.  We define the shape $e$ such that the expectation value of the mean shape
for an ensemble of galaxies is an estimate of the mean reduced gravitational shear acting on those galaxies
\begin{equation}
\langle e \rangle = g \equiv \frac{\gamma}{1-\kappa},
\label{eq:req:edef}
\end{equation}
where $\gamma$ and $\kappa$ are the shear and convergence, respectively (see e.g.~\citealp{Hoekstra13} 
for a review of weak lensing concepts and terminology).

For a galaxy with elliptical isophotes, one finds that $|e| = (a-b)/(a+b)$ satisfies \eqn{eq:req:edef},
where $a$ and $b$ are the semi-major and semi-minor
axes of the ellipse.  However, galaxies do not in general have elliptical isophotes, so this definition is of
little practical value.  For the more general case, the estimator
\begin{equation}
e = \frac{I_{xx} - I_{yy} + 2i I_{xy}}{I_{xx} + I_{yy} + 2\sqrt{I_{xx}I_{yy} - I_{xy}^2}}
\label{eq:req:eseitz}
\end{equation}
has been proposed by \citet{Seitz97},
where the second moments of the intensity profile $I(x,y)$ are defined as
\begin{equation}
I_{\mu\nu} = \frac{\int dxdy I(x,y) (\mu-\bar\mu)(\nu-\bar\nu)}{\int dxdy I(x,y)}.
\end{equation}
But since neither shear algorithm in this paper uses \eqn{eq:req:eseitz} directly,
we consider
\eqn{eq:req:edef} to be the functional definition of what we mean by the shape
of an arbitrary galaxy.  See \S\ref{sec:im3shape:model} and \S\ref{sec:ngmix:model} for 
details about the \imshape\ and \ngmix\ estimators of $e$.

While \eqn{eq:req:edef} is our goal for the shape estimates in our catalogue, it is inevitable that there will be
systematic errors in the shape measurements.  A convenient parameterization, based on
one first proposed by \citet{step1},
uses a first-order expansion of the form,
\begin{equation}
\langle e \rangle = (1+m) g_\mathrm{true} + \alpha \epsf + c,
\label{eq:req:esys}
\end{equation}
where $g_\mathrm{true}$ denotes the value that would be obtained from an ideal error-free shape estimator,
$m$ quantifies the \emph{multiplicative error}, $\alpha$ measures the \emph{leakage} of the PSF shape
into the galaxy shapes, and $c$ represents other sources of \emph{additive error}. 

Note that $m$ can in principle be different for each of the two components
$e_1$ and $e_2$.  However, we find in practice that the two coefficients are generally very close to
equal when they can be measured separately,
so we simply take $m$ to be a single real value here.  Similarly, $\alpha$ could in principle have up to 4 components
if the leakage were anisotropic and involved cross terms\footnote{In the complex formulation we are using,
this would 
involve terms $\alpha \epsf + \alpha^\prime \epsf^*$.  In formulations that treat $[e_1, e_2]$ as a vector,
$\alpha$ would be a $2\times2$ matrix.}, but we do not see evidence for anything
beyond a real-valued $\alpha$ in practice.

The leakage term $\alpha \epsf$ is 
commonly \citep[e.g.][]{step1} implicitly folded into the general additive error term, $c$, 
but we have found it useful to retain it explicitly, since
PSF leakage can be one of the more difficult additive errors to correct.
Furthermore, \citet{great3results} found that the additive systematic errors for essentially
all of the methods submitted
to the GREAT3 challenge were well-described by $\alpha \epsf$, which motivates us to
include it as an explicit term in \eqn{eq:req:esys}.

\subsection{Shear Correlation Functions}
\label{sec:req:dxi}

We set our requirements on the various kinds of systematic errors according to how they propagate 
into the shear two-point correlation functions (defined as in \citealp{Jarvis03}):
\begin{align}
\xi_+(\theta) &= \left\langle e^*(\mathbf{x}) e(\mathbf{x} + \boldsymbol\theta) \right\rangle 
\label{eq:req:xip} 
\\
\xi_-(\theta) &= \left\langle e(\mathbf{x}) e(\mathbf{x} + \boldsymbol\theta) \exp(-4i \arg(\boldsymbol\theta)) \right\rangle,
\label{eq:req:xim}
\end{align}
where ${}^*$ indicates complex conjugation.
%SB: note to self: I get confused with this xi+ and xi- notation - I prefer it in terms of e1 and e2. I need to convince myself 
%   that the above equation is correct, in particular that the theta in the arg is indeed the angular separation and not the 
%   angle of the galaxy i.e. arctan2(e2, e1)]
%MJ: I didn't realize I was being so novel here.  My own papers seem to be the only references for this definition.
%   I find it convenient, since you don't need to define e1,e2 *relative to the line connecting them*, which is often confusing.
%   This projection is automatically accounted for by the arg(theta) factor.
%   Anyway, I added a reference to Jarvis et al 2003 as prior art for this definition.

Substituting
\eqn{eq:req:esys} into these equations and assuming the three types of systematic errors are uncorrelated
(which is not necessarily true in general, but is a reasonable assumption for setting requirements), we find
\begin{equation}
\delta\xi_i(\theta) \simeq 2m \xi_i(\theta)
+ \alpha^2 \xi_i^{pp}(\theta)
+ \xi_i^{cc}(\theta)
\label{eq:req:dxi}
\end{equation}
to leading order in each type of systematic, 
where $i\!\in\!\{+,-\}$, $\delta\xi_i$ are the systematic errors in the two correlation functions, 
$\xi_i^{pp}$ are the auto-correlation functions of
the PSF shapes, and $\xi_i^{cc}$ are the auto-correlation functions of the additive error, $c$.

To set requirements on $\delta\xi_i$, we consider how the errors will affect our estimate of the 
cosmological parameter $\sigma_8$,  the present day amplitude of the (linear) matter power spectrum 
on the scale of $8 h^{-1}$ Mpc.
Our requirement is that the systematic errors change the estimated value of $\sigma_8$
by less than 3\%, $\delta \sigma_8/ \sigma_8<0.03$. 
This value was chosen to be about half of the expected statistical
uncertainty on $\sigma_8$ for the DES SV survey.

Propagating this limit to  the shear correlation functions, we obtain the requirement
\begin{equation}
\dximax_i = \frac{\partial \xi_i}{\partial \sigma_8} \delta \sigma_8.
\label{eq:req:dxi_sigma8}
\end{equation}
This constraint assumes that errors are fully correlated across $\theta$; assuming independent errors
would be less restrictive.  

\begin{figure}
\includegraphics[width=\columnwidth]{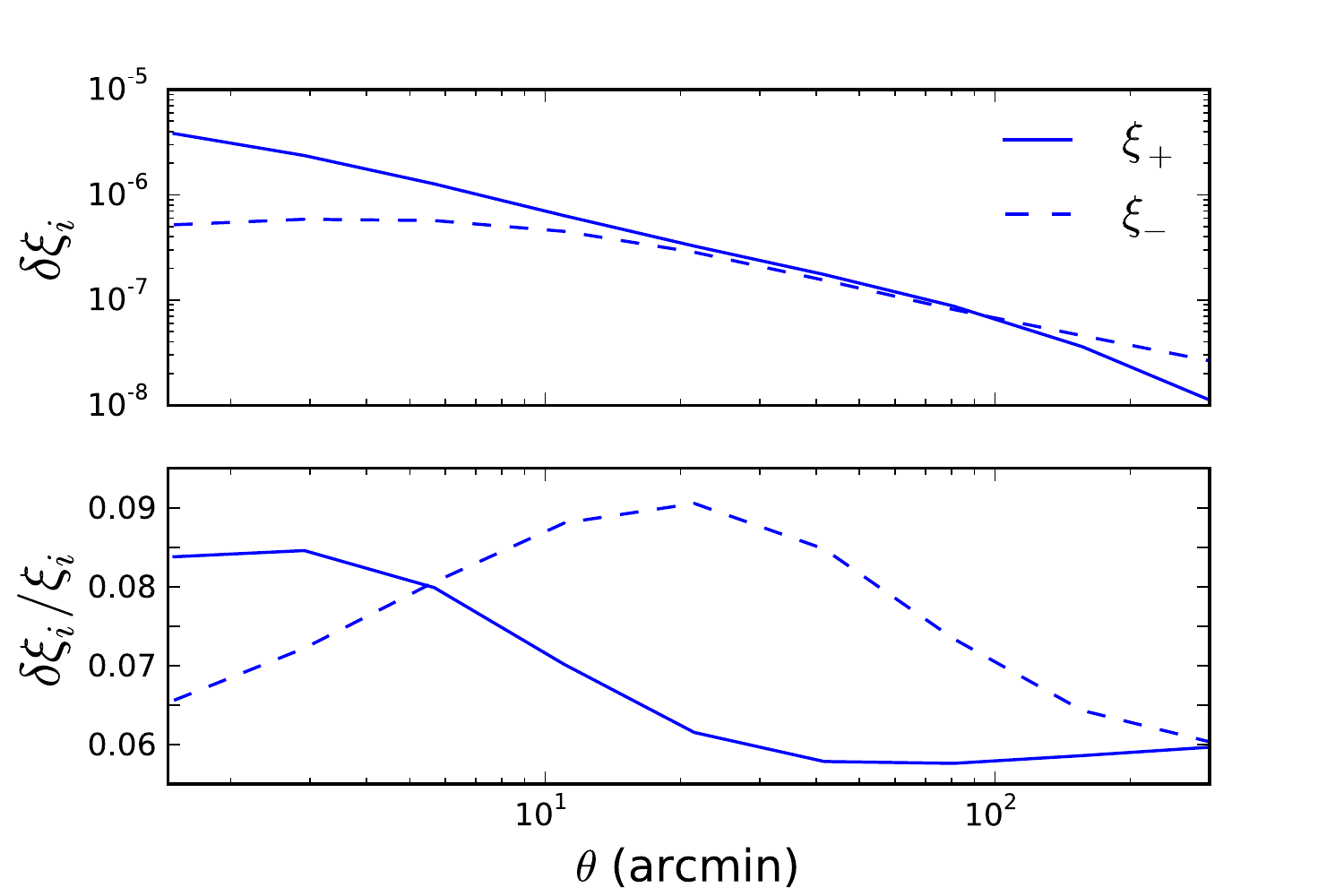}
\caption{Requirement for the maximum systematic error contribution to the shear correlation functions. The blue lines 
correspond to $\delta \sigma_8/\sigma_8=0.03$ for each of the correlation functions $\xi_+$ (solid) and 
$\xi_-$ (dashed).  The top and bottom panels shows the requirement for the absolute and relative error
in the correlation functions.
\label{fig:req:dxi}
}
\end{figure}

\fig{fig:req:dxi} shows the resulting requirements for $\delta{\xi_i}$
derived for a flat $\Lambda$CDM central cosmological model with $\sigma_8=0.82$, $\Omega_b = 0.047$ 
and $\Omega_c= 0.2344$, $h=0.7$ and $n_s=0.96$. 

\edit{We will apply this requirement to a number of different potential sources
    of systematic error.  If each of them just barely pass the requirement,
    this would be a problem, since the total net systematic error would then
    exceed the requirement.  We attempt to quantify the total realized
    systematic error in the shear measurements in \S\ref{sec:tests:summary}; it
    is this total error that must be propagated into the next stage(s) of the
    analysis, along with any other non-measurement sources of systematic error
(e.g. photometric redshift errors and intrinsic alignments) that may be
relevant for each specific science application.}

\subsection{Multiplicative and Additive Errors}
\label{sec:req:mc}

From \eqn{eq:req:dxi}, we find that the requirement on the multiplicative bias, $m$, is
\begin{equation}
|m| < \frac{1}{2} \left| \frac{\dximax_i}{\xi_i} \right|.
\end{equation}
As can be seen from the lower panel in \fig{fig:req:dxi}, the most stringent requirement on $\delta \xi_i/\xi_i$ is about 0.06,
yielding a requirement on the multiplicative error of
\begin{equation}
|m| < \mrequirement.
\label{eq:req:m}
\end{equation}

The requirement on the additive systematic error is somewhat more complicated, since it is the correlation function
of the additive systematic that matters.  For a systematic error that is coherent over small spatial scales (less
than $\sim 1$ arcminute), the requirement comes from the zero-lag value of $\delta\xi_+$ in \fig{fig:req:dxi},
$\langle c^2 \rangle < \dximax_+(0)$, or 
\begin{equation}
    c_\mathrm{rms} < \crequirement.
\label{eq:req:c}
\end{equation}

For additive errors that have longer correlation lengths, we will need to be more careful about calculating 
the correlation function of the systematic error.  The requirement in this case is
\begin{equation}
|\xi_i^{cc}(\theta)| < \dximax_i(\theta)
\label{eq:req:xic}
\end{equation}
using the function shown in \fig{fig:req:dxi}.  The most notable example of this will be systematic effects
due to the PSF: both leakage and modeling errors, which will be discussed in the next two sections.

Note that we do not need to satisfy these requirements for
all values of $\theta$.  The statistical uncertainties on $\xi_{+,-}(\theta)$ become much larger at large scales,
so such scales are not as important for constraining cosmology as smaller scales.  In practice,
\eqn{eq:req:xic} should ideally be satisfied for scales $\theta < 100$ arcminutes, where $\xi_{+,-}(\theta)$ are
relatively well-measured.

We note that these results are broadly consistent with those of \cite{AmaraRefregier08}, who 
derived requirements for a tomographic weak lensing survey, performing joint constraints on 
the set of cosmological parameters for a $w$CDM model.  They found requirements of 
$|m| < 4.0 \times 10^{-2}$ and $c_\mathrm{rms} < 2.1 \times 10^{-3}$  for DES SV survey parameters, 
which are in rough agreement with the requirements quoted above.

\subsection{PSF Leakage}
\label{sec:req:alpha}

The requirements for the PSF leakage term in \eqn{eq:req:esys} can be obtained from the general requirement
on additive errors, \eqn{eq:req:xic}.  
\begin{equation}
\alpha^2 \xi_i^{pp}(\theta) < \dximax_i(\theta),
\end{equation}
which can be solved for $\alpha$ as
\begin{equation}
|\alpha| < \left( \frac{\dximax_i(\theta)}{\xi_i^{pp}(\theta)} \right)^{\frac{1}{2}}.
\label{eq:alpha_theta}
\end{equation}

\begin{figure}
\includegraphics[width=\columnwidth]{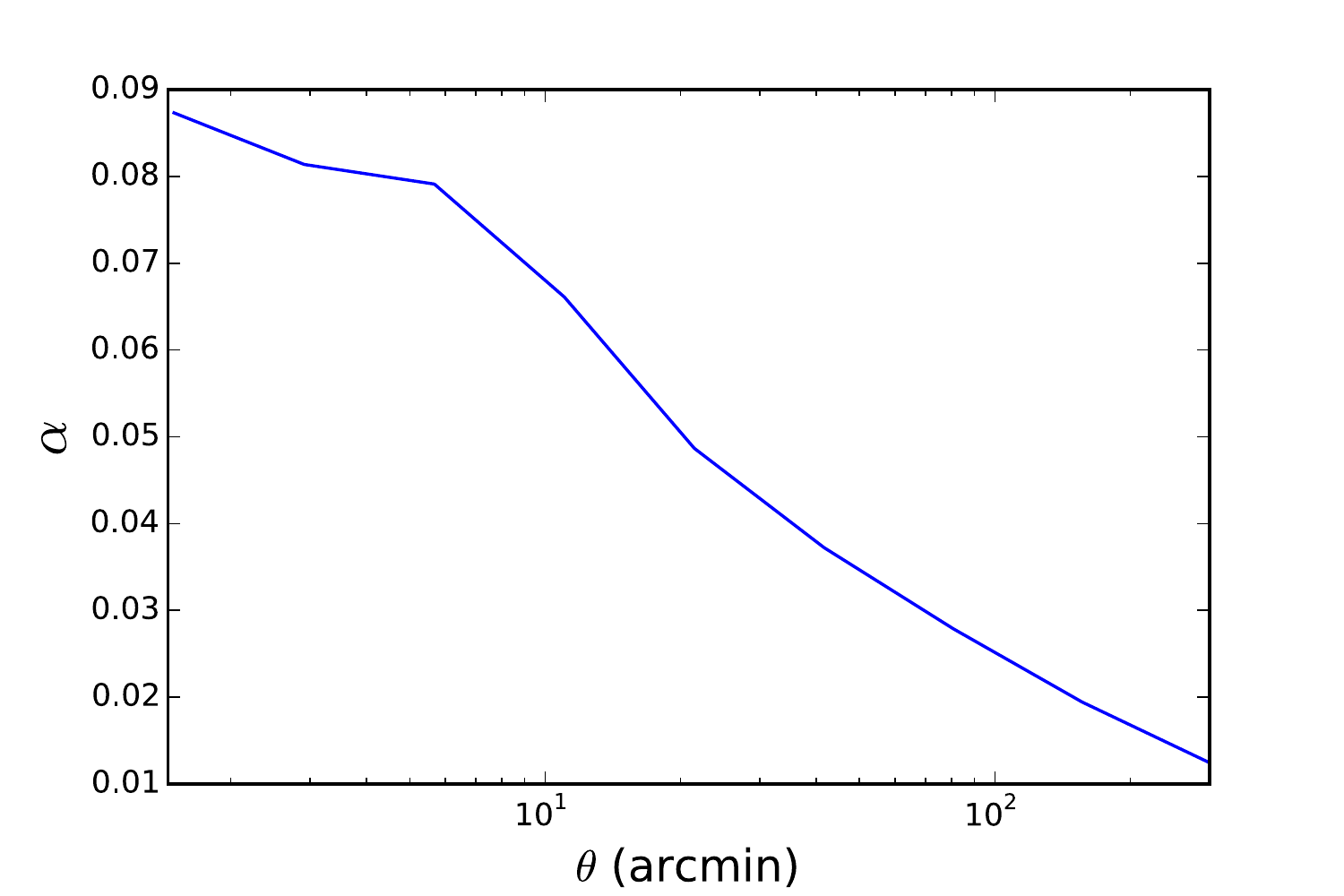}
\caption{Requirement for the PSF leakage factor $\alpha$ based on the relative error in $\sigma_8$ being
less than 3\%.
\label{fig:req:alpha}
}
\end{figure}

\fig{fig:req:alpha} shows this requirement on $\alpha$ as a function of $\theta$
using the observed $\xi_+^{pp}$ for DES SV data\footnote{See \fig{fig:tests:alpha}, top panels.
We use the \imshape\ measurement of $\xi_+^{pp}$ here.}.
The requirement arising from
$\xi_-^{pp}$ is always larger than 0.05 and is not shown. 

In general, the amount of leakage of PSF shapes into galaxy shapes from an imperfect correction scheme is not
expected to vary with scale.  Rather, we can use \fig{fig:req:alpha} to determine a conservative requirement
for $\alpha$
that would be applicable for scales $\theta<100$ arcminutes:
\begin{equation}
|\alpha| < 0.03.
\label{eq:req:alpha}
\end{equation}
We will estimate $\alpha$ from the data in \S\ref{sec:tests:psf}.

\subsection{PSF Model Errors}
\label{sec:req:rho}

We now consider errors in the modeling of the PSF itself.  The previous section dealt with the possibility
of the galaxy shear estimation algorithm imperfectly accounting for the PSF convolution and letting
some of the PSF shape leak into the galaxy shape.  However, even a perfect PSF 
correction scheme can suffer systematic biases if the PSF model itself is biased.

As our starting point, 
we use the unweighted moments approximation of \cite{PH08}, who
give the bias on the measured galaxy ellipticity in terms of errors in the PSF model
(their equation 13)\footnote{The
\citet{PH08} formalism is based on $\epsilon = (a^2-b^2)/(a^2+b^2)$ rather than our $e$
shape measure,
so there are factors of $O(1)$ differences that we are neglecting.  Similarly, they derive
their formula for unweighted moments, which are also not directly applicable to real
shear estimation algorithms, differing again by factors of $O(1)$.  Despite these 
possible shortcomings, we feel this is nonetheless a useful model for describing 
PSF modeling errors.
}:
\begin{equation}
\delta e_\mathrm{sys} = \left(e - \epsf \right) \left(\frac{\Tpsf}{\Tgal}\right) \frac{\delta \Tpsf}{\Tpsf}
- \left( \frac{\Tpsf}{\Tgal} \right) \delta \epsf,
\end{equation}
where $T \!\equiv\! I_{xx} \!+\! I_{yy}$ is the intensity-weighted second moment of the radius
(written as $R^2$ in their paper).  $\Tgal$ refers to the intrinsic galaxy size, unconvolved by the PSF.

Constructing the shear correlation function with this model, we find that the 
systematic error in $\xi_+$ is 
\begin{align}
\delta \xi_+(\theta) &= 2 \left\langle\frac{\Tpsf}{\Tgal} \frac{\delta\Tpsf}{\Tpsf}\right\rangle \xi_+(\theta)
+ \left\langle\frac{\Tpsf}{\Tgal} \right\rangle^2 \rho_1(\theta) 
    \nonumber\\& \quad
- \alpha \left\langle\frac{\Tpsf}{\Tgal} \right\rangle \rho_2(\theta)
+ \left\langle\frac{\Tpsf}{\Tgal} \right\rangle^2 \rho_3(\theta) 
    \nonumber\\& \quad
+ \left\langle\frac{\Tpsf}{\Tgal} \right\rangle^2 \rho_4(\theta) 
- \alpha \left\langle\frac{\Tpsf}{\Tgal} \right\rangle \rho_5(\theta),
\label{eq:req:psfmodel}
\end{align}
where $\rho_1(\theta)$ and $\rho_2(\theta)$ are defined as \citep[cf.][]{Rowe10}
\begin{align}
\label{eq:req:rho1def}
\rho_1(\theta) &\equiv \left \langle \delta \epsf^*(\bfx) \delta \epsf(\bfxpt) \right \rangle \\
\label{eq:req:rho2def}
\rho_2(\theta) &\equiv \left \langle \epsf^*(\bfx) \delta \epsf(\bfxpt) \right \rangle,
\end{align}
and we introduce three new statistics defined as\footnote{We
note that \citet{Melchior14} proposed a slightly different $\rho_3$ statistic,
\begin{equation}
\nonumber
\rho_3^\prime(\theta) = \left \langle \left(\frac{\delta\Tpsf}{\Tpsf}\right)\!(\bfx)
   \left(\frac{\delta\Tpsf}{\Tpsf}\right)\!(\bfxpt) \right \rangle,
\end{equation}
pulling the $\epsf$ factors out of the ensemble average. 
We believe it is more appropriate to leave them in, since errors in the size estimates
could easily be coupled to the PSF shapes.}
\begin{align}
\label{eq:req:rho3def}
\rho_3(\theta) &\equiv \left \langle \left(\epsf^* \frac{\delta\Tpsf}{\Tpsf}\right)\!(\bfx)
   \left(\epsf \frac{\delta\Tpsf}{\Tpsf}\right)\!(\bfxpt) \right \rangle \\
\label{eq:req:rho4def}
\rho_4(\theta) &\equiv \left \langle \delta\epsf^*(\bfx)
   \left(\epsf \frac{\delta\Tpsf}{\Tpsf}\right)\!(\bfxpt) \right \rangle \\
\label{eq:req:rho5def}
\rho_5(\theta) &\equiv \left \langle \epsf^*(\bfx)
   \left(\epsf \frac{\delta\Tpsf}{\Tpsf}\right)\!(\bfxpt) \right \rangle.
\end{align}

There are corresponding terms for $\delta\xi_-$,
which are negligible in practice and thus uninteresting as requirements.

The first term in \eqn{eq:req:psfmodel} is a multiplicative systematic, so the relevant requirement comes from \eqn{eq:req:m}.  
We approximate the ensemble average as
a product of two averages to set a requirement on the mean error in the PSF size
\begin{equation}
\left|\left\langle \frac{\delta\Tpsf}{\Tpsf}\right\rangle\right| < 0.03 \left\langle\frac{\Tpsf}{\Tgal} \right\rangle^{-1} . 
\label{eq:req:dpsfsize}
\end{equation}
This represents an error due to improperly accounting for the ``dilution'',
the amount by which the blurring of the PSF makes objects rounder than they originally were.
Estimating the wrong PSF size will lead to a systematic multiplicative bias in the inferred galaxy shapes.

The other terms are additive errors, contributing to $\xi_+^{cc}(\theta)$, so the requirements from \eqn{eq:req:xic}
are that each term be
less than $\dximax_+(\theta)$:
\begin{align}
\label{eq:req:rho1}
|\rho_{1,3,4}(\theta)| &<  \left\langle \frac{\Tpsf}{\Tgal}\right\rangle^{-2}  \dximax_+(\theta) \\
\label{eq:req:rho2}
|\rho_{2,5}(\theta)| &<  |\alpha|^{-1} \left\langle \frac{\Tpsf}{\Tgal}\right\rangle^{-1} \dximax_+(\theta).
\end{align}
We will test these requirements for our PSF model below in \S\ref{sec:psf:rho}.

For our data, we compute the factor $\langle\Tpsf/\Tgal\rangle$ that appears in these requirements to be
1.20 for \imshape\ and 2.42 for \ngmix; the latter is larger
because the final galaxy selection for the \ngmix\ catalogue keeps more small galaxies than
the \imshape\ selection.
We use the \ngmix\ value in \S\ref{sec:psf:rho}, as it gives the more stringent requirement.
For $\alpha$, we conservatively use the value 0.03.  We will find in \S\ref{sec:tests:alpha2}
that both codes estimate $\alpha$ to be consistent with zero; however, it is not estimated much more 
precisely than this value.

\section{PSF Estimation}
\label{sec:psf}
\assign{Mike}

The principal confounding factor that must be addressed in order to measure
accurate shears is the convolution of the galaxy surface brightness profiles by the 
point-spread function (PSF).
The net PSF is due to quite a number of physical processes including
atmospheric turbulence, telescope and camera aberrations, guiding
errors, vibrations of the telescope structure,
and charge diffusion in the CCDs, among other more subtle effects.
Furthermore, this PSF is not constant, but varies both spatially over the focal plane
and temporally from one exposure to the next.
The atmospheric component varies approximately according to a 
Kolmogorov turbulent spectrum.  The optical aberrations have 
characteristic patterns due to features in the telescope optics.

Fortunately, we do not need to have a complete physical model of all the contributors
to the PSF in order to accurately characterize it.  Instead, we
build an empirical model, based on observations of stars, which we interpolate to
obtain an estimate of the PSF at any location on the focal plane.  
In this section, we describe how we select appropriate stars
to use, and then build and test the PSF model.

\subsection{Initial Identification of Stars}
\label{sec:psf:selection}
\assign{Mike}
\contrib{Bob}

We first selected the stars to be used to constrain the PSF model.  As stars are
point sources, observations of them provide a sample
image of the PSF at the location of each star.  We desired a high-purity sample of
fairly bright stars to make sure we did not erroneously consider images of
small, faint galaxies to be images of the PSF, as that would bias the resulting
PSF model.

\edit{We found that, for some CCD images, the sets of objects identified as
    stars by the Modest Classification scheme\footnote{Stars were identified as
    \code{(bright\_test OR locus\_test)} in terms of the pseudo-code presented
in \S\ref{sec:modest}} included a relatively high number of galaxies, and in
other cases too few stars were identified.  The cause of these failures is
dependent on many factors, but may be partly related to the use of \coadd\ data
for the classification.  The \coadd\ PSF can change abruptly at the locations
of chip edges in the original single-epoch images, which may have affected the
stellar classification near these discontinuities.}

\edit{Ultimately, the problems with the modest classifier were common enough
    that we decided to develop a new algorithm tailored specifically to the
    identification of a pure set of PSF stars.}
    Our algorithm works on each CCD
    image separately, using a size-magnitude diagram of all the objects
    detected on the image.  For the magnitude, we use the \sex\ measurement
    \magauto.  For the size, we use the scale size, $\sigma$, of the
    best-fitting elliptical Gaussian profile using an adaptive moments
    algorithm.  We found that these measures produce a flatter and tighter
    stellar locus than the \texttt{FLUX\_RADIUS} value output by \sex, and is
    thus better suited for selection of stars. As a further improvement, we
    initialize the algorithm with some stars identified by \sex\ to have
    \classstar\ between 0.9 and 1.0.  This was found to give a decent estimate
    of the size of the PSF, providing a good starting guess for the location of
the stellar locus.

\begin{figure}
\includegraphics[width=\columnwidth]{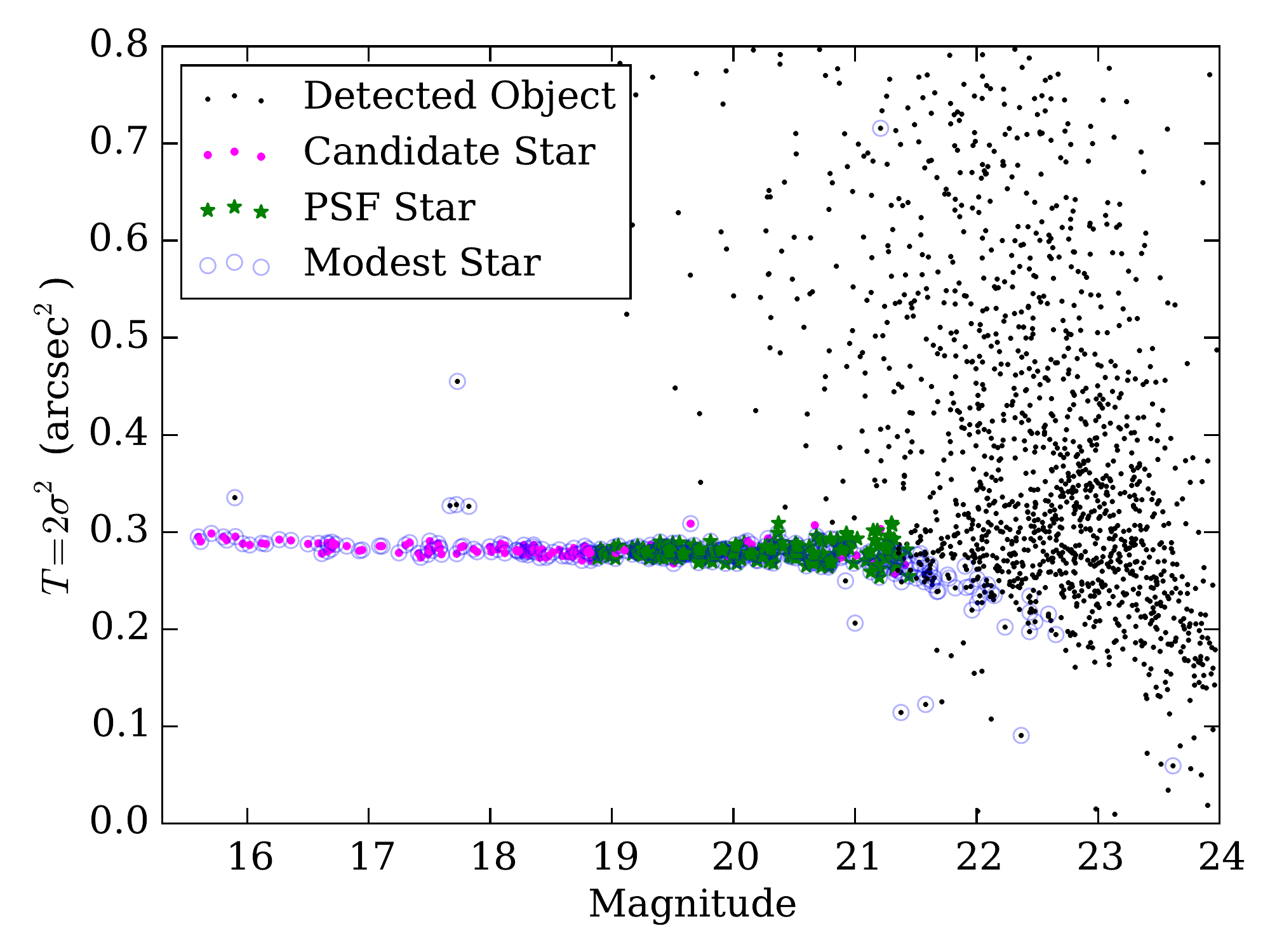}
\caption{An example size-magnitude diagram for a single CCD image, 
used to identify stars.  The size $T=2\sigma^2$ is based on the scale size
of the best-fitting elliptical Gaussian.
The pink and green points are the objects initially identified as stars.
The green points are the ones that pass our selection criteria outlined in \S\ref{sec:psf:cuts}, 
most notably the
magnitude cut to avoid objects contaminated by the brighter-fatter effect.  
These objects are then used to constrain the PSF model.
The blue circles show an an alternate star classification, called the Modest Classification
within DES, which was found not to work as well for our specific purpose.
\label{fig:psf:findstars}
}
\end{figure}

The stars are easily identified at bright magnitudes as a locus of points 
with constant size \edit{nearly} independent of magnitude.  The galaxies have a 
range of sizes, all larger than the PSF size.
Thus, the algorithm starts with a tight locus at small size for the stars
and a broad locus of larger sizes for the galaxies for objects in the 
brightest 5 magnitudes (excluding saturated objects).
Then the algorithm proceeds to fainter magnitudes, building up both loci, until the
stellar locus and the galaxy locus start to merge.
The precise magnitude at which this happens is a function of the 
seeing as well as the density of stars and galaxies in the particular
part of the sky being observed.  As such the faint-end magnitude 
of the resulting stellar sample varies among the different exposures.

\fig{fig:psf:findstars} shows such a size-magnitude diagram for a 
representative CCD image.  The stellar locus is easily identified by eye, and
the stellar sample identified by our algorithm is marked in pink and green.
The pink points are stars that are removed by subsequent steps in the
process outlined below, while the green points are the stars that 
survive these cuts.  The blue circles show the objects identified
as stars according to the Modest Classification, which includes more outliers
and misses some of the objects clearly within the stellar locus.

While the algorithm we currently use is found to work well enough for
the SV data, we plan to investigate whether the neural net star-galaxy
separator recently developed by \citet{Soumagnac15} is more robust
or could let us include additional stars.

\subsection{Selection of PSF Stars}
\label{sec:psf:cuts}
\assign{Mike}

Some of the stars in this sample are not appropriate to use for
PSF modeling, even ignoring the inevitable few galaxies that get misidentified
as stars.  The CCDs on the Dark Energy Camera each have six spots where
100 micron thick spacers were placed
behind the CCDs when they were glued to their carriers \citep[cf.][]{FlaugherDECam2015},
which affects the electric field lines near each 2mm $\times$ 2mm spacer.
These features, which we call tape bumps, 
distort the shapes in those parts of the CCDs, so the stellar images
there are not accurate samples of the PSF.  We exclude any star whose
position is within 2 \edit{PSF} FWHM separation of the outline of a tape bump.  The tape
bumps are relatively small, so this procedure excludes less than 0.1\% of the 
total area of the CCD, but removes a noticeable bias in the PSF model
near the bumps.

Another problem we addressed with regards to star selection is the
so-called ``brighter-fatter effect'' \citep{Antilogus14, Guyonnet15}.  As charge builds
up in each pixel during the exposure, the resulting lateral electric fields and
increased lateral diffusion push newly incoming charges slightly away from
the existing charge. This makes bright objects appear a bit larger than fainter
objects.  In addition, an asymmetry in the magnitude of the effect between rows and columns can
make bright stars more elliptical. The galaxies we used for weak lensing
are generally faint, so the brightest stars do not accurately sample the PSF
that we need to measure.  Furthermore, the brighter-fatter effect does not manifest as a
convolution of the signal, so the bright stars do not even provide an estimate
of the correct PSF to be used for bright galaxies.

\begin{figure}
\includegraphics[width=\columnwidth]{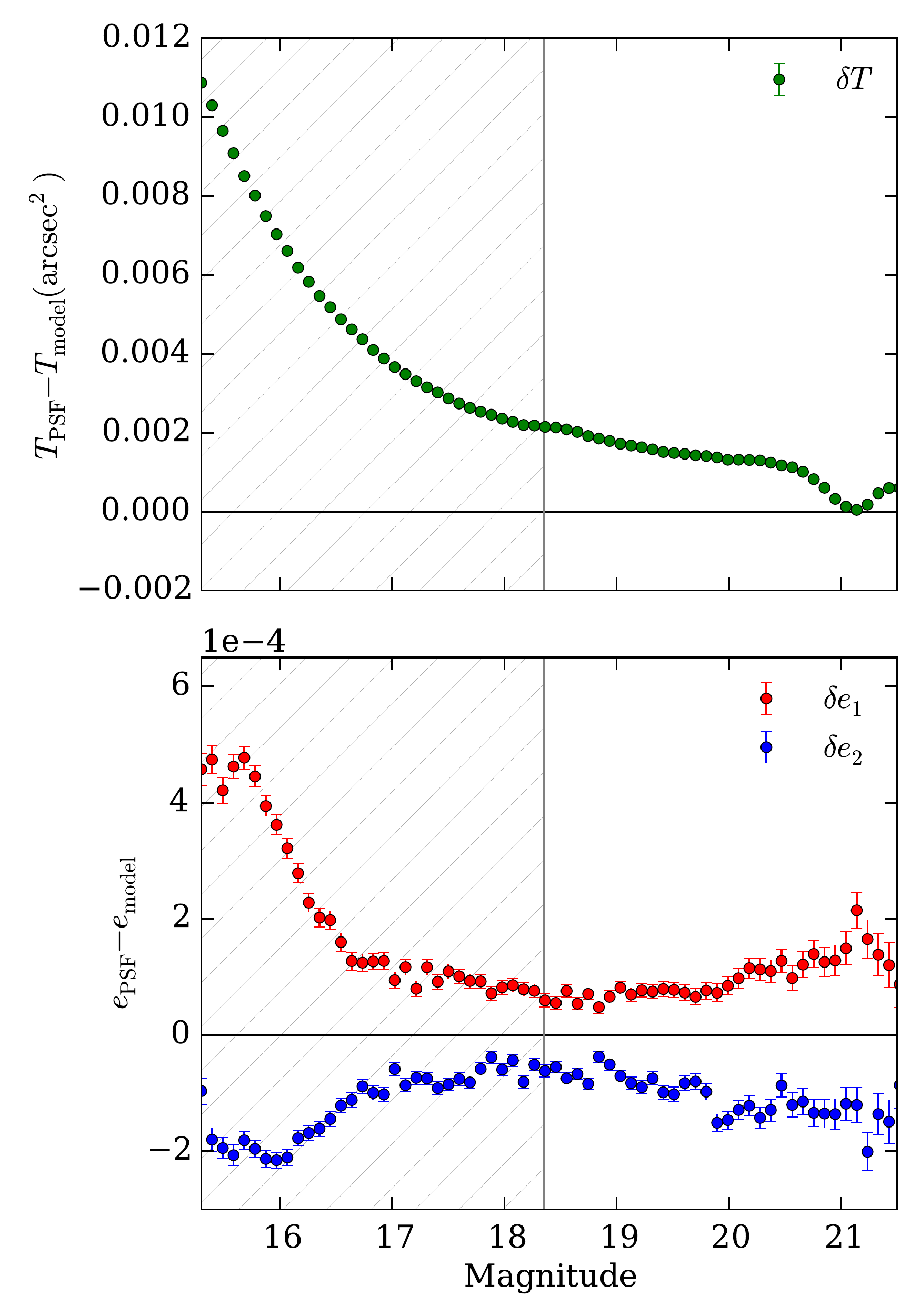}
\caption{The residual size (top) and shape (bottom) of stars relative to that of the 
PSF model as a function of magnitude.
The hatched region on the left shows the magnitude range of the stars we exclude
from the sample to reduce the impact of the brighter-fatter effect.
\label{fig:psf:bfe}
}
\end{figure}

The appropriate solution is to move the shifted charge back to where
it would have fallen in the absence of this effect. This will be
implemented in future DES data releases \citep{Gruen15}.
For the current round of catalogues, we
instead partially avoided the problem by removing the brightest stars from
our sample.  Specifically, we removed all stars within 3 magnitudes of the
saturation limit for the exposure.  That is, in our
final selection of PSF stars we required that the brightest pixel in the 
stellar image be less than 6\% of the pixel full well.  Since the brighter-fatter effect scales 
approximately linearly
with flux, this reduces the magnitude of the effect by a factor of 16.
We were left with stars of lower \snr,
so it is not the ideal solution, but it is an acceptable interim
measure (as we demonstrate below) until the more sophisticated solution can be implemented.

In \fig{fig:psf:bfe} we show the mean difference between the measured 
sizes of observed stars and the size of the PSF model at their locations,
using the model described below in \S\ref{sec:psf:measurement}.
For the measurements of the sizes and shapes described here, we used
the implementation of the HSM \citep{HSMa,HSMb} algorithm included
in the \galsim\ software package.
The hatched region marks the range we excluded to avoid the spurious
increase in PSF size from the brighter-fatter effect.  In \fig{fig:psf:bfe} 
we have also shown the mean difference in ellipticity due to the brighter-fatter
effect; it affects the shapes of the stars as well as the size.

We do not yet understand why the residual sizes and shapes shown in 
\fig{fig:psf:bfe} do not level off to zero at fainter magnitudes where
the brighter-fatter effect is negligible.  
The requirement on this residual value is given by \eqn{eq:req:dpsfsize}.
We calculate $\langle \delta\Tpsf/\Tpsf\rangle$ to be $0.0044$,
which is well below our requirement of 0.013 for the SV data.
However, this residual will not be acceptable
for future DES analyses, so we will need to investigate what is causing 
the problem and fix it.

\begin{figure}
\center\includegraphics[width=\columnwidth]{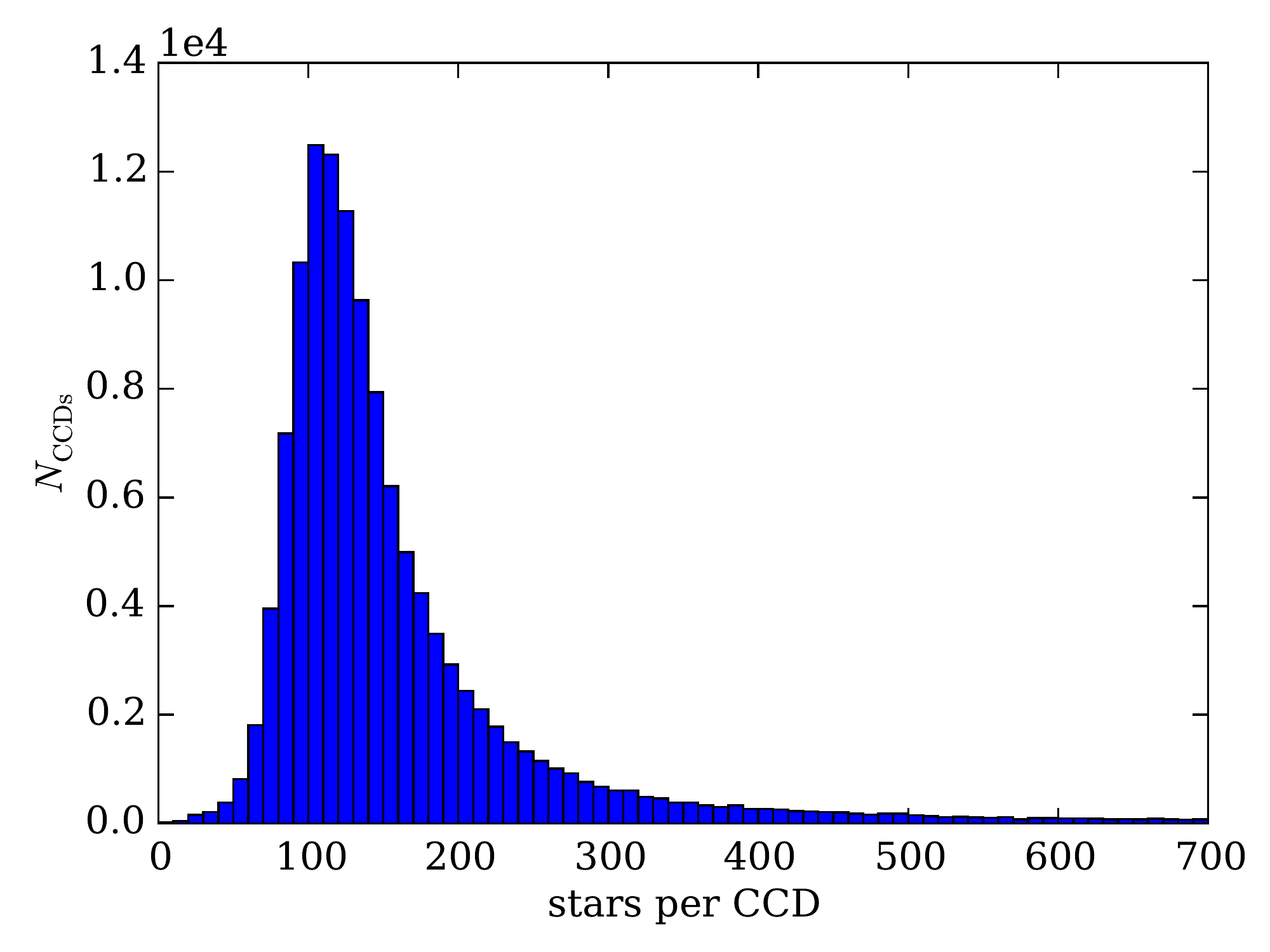}
\caption{The distribution of the number of stars per CCD image used for
constraining the PSF model.
\label{fig:psf:nstars}
}
\end{figure}

\begin{figure}
\center\includegraphics[width=\columnwidth]{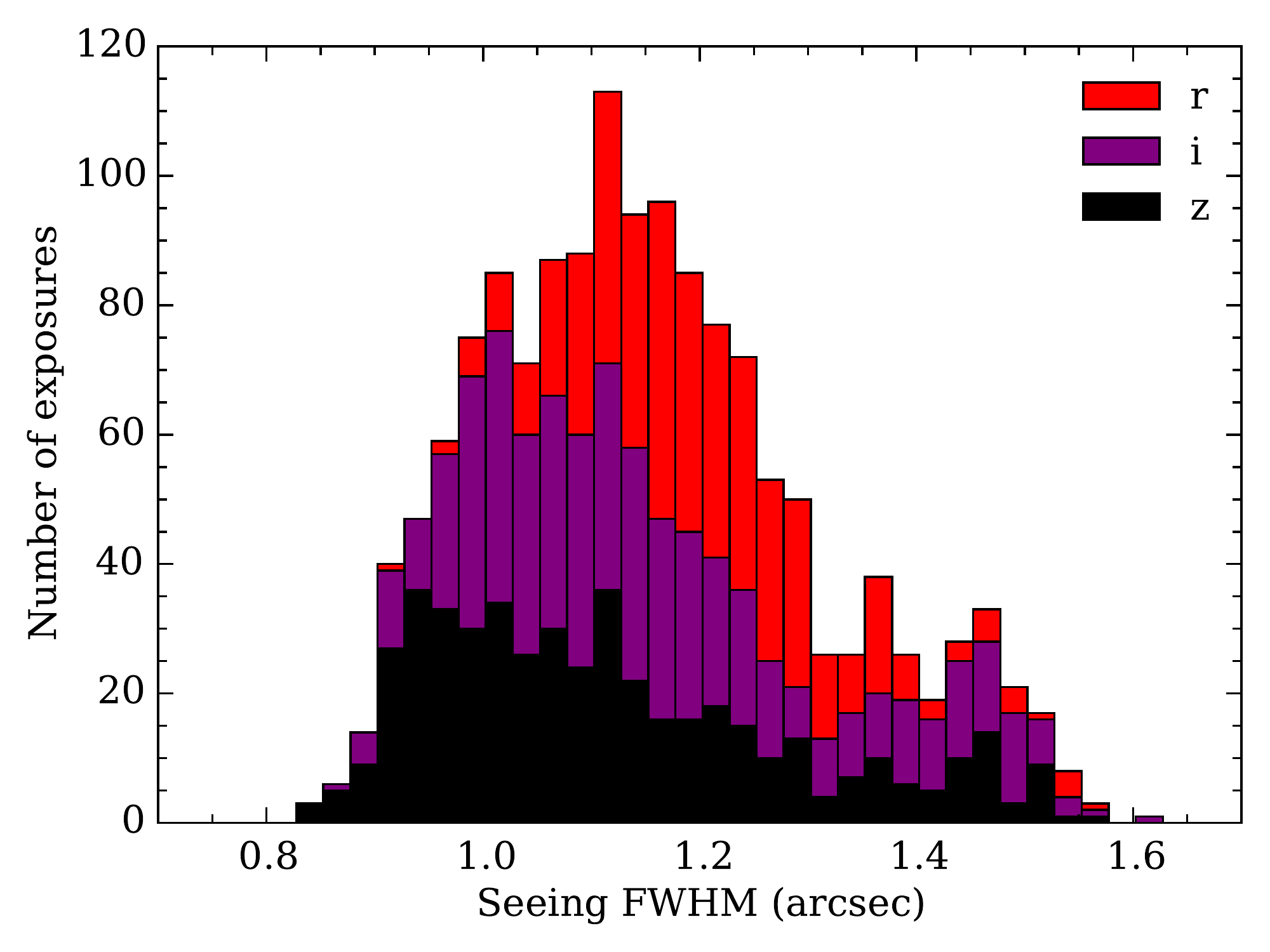}
\caption{\edit{The distribution of the median FWHM of the PSF stars
in the non-blacklisted exposures (cf. \S\ref{sec:meds:blacklists}).
The $r$, $i$ and $z$-band exposures are indicated from top to
bottom within each bar in red, purple, and black.}
\label{fig:psf:seeing}
}
\end{figure}

In the complete process described above we find a median of 130 useful
stars per CCD image, which we use to constrain the PSF model.  The
distribution is shown in \fig{fig:psf:nstars}.

\edit{In \fig{fig:psf:seeing} we show the distribution of the median measured full-width half-max (FWHM)
for the PSF stars used in our study, restricted to the exposures used
for shear measurements.  The overall median seeing of these exposures
was 1.08 arcsec.  The $r$, $i$ and $z$-bands had median
seeing of 1.11, 1.08, and 1.03 arcsec respectively.  This was somewhat worse
than expected (0.9 arcesec) and reflects the fact that a number of problems
related to the instrument, telescope, and control software
were being diagnosed and fixed concurrently with the observations.
The realized seeing has significantly improved in the subsequent main survey observations
\citep{Diehl14}.}

\edit{For some CCD images no stars passed our final selection criteria}, usually
because the initial 
stellar selection could not find any stars or no stars survived the magnitude cuts.
For instance, this can happen when there is a very bright object in the 
image that essentially masks out the entire image, leaving zero or very few
objects detected.  In less extreme cases, a bright object
can sufficiently contaminate the fluxes and sizes of the other detections
that the stellar locus is either difficult to find or merges with the 
galaxy locus at a fairly bright magnitude, such that the brighter-fatter cut
excludes the entire sample.

Whenever the process failed for any reason on a given CCD image,
we flagged the image and excluded it from being used in
subsequent shear estimation.  We also flagged images with less than 20 identified
PSF stars, since it is difficult to accurately interpolate the PSF model with
so few stars.  These flagged images were added to the set of blacklisted images
described in \S\ref{sec:meds:blacklists}.

\subsection{PSF Measurement and Interpolation}
\label{sec:psf:measurement}
\assign{Mike}

\begin{figure*}
\includegraphics[width=0.48\textwidth]{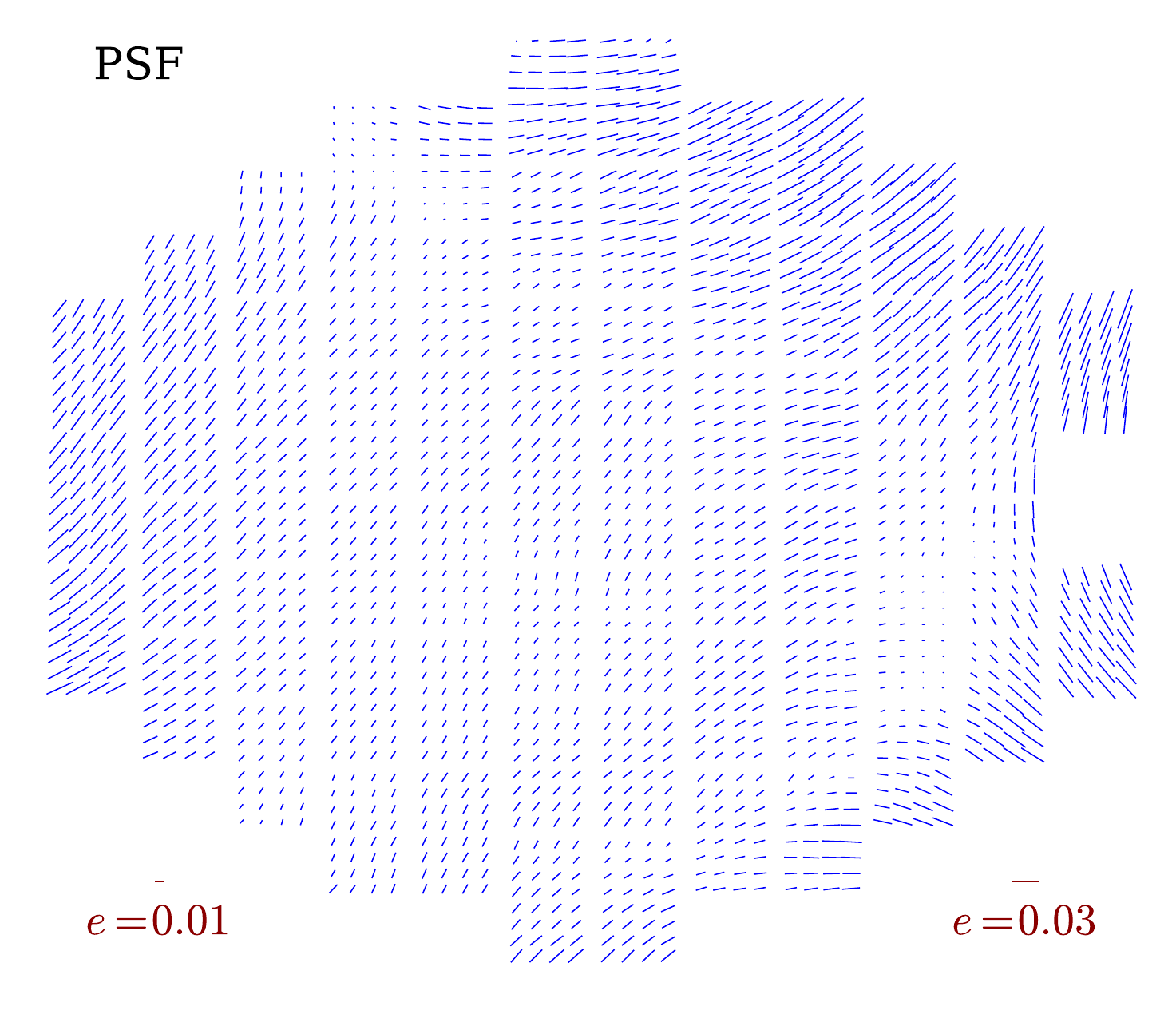}
\includegraphics[width=0.48\textwidth]{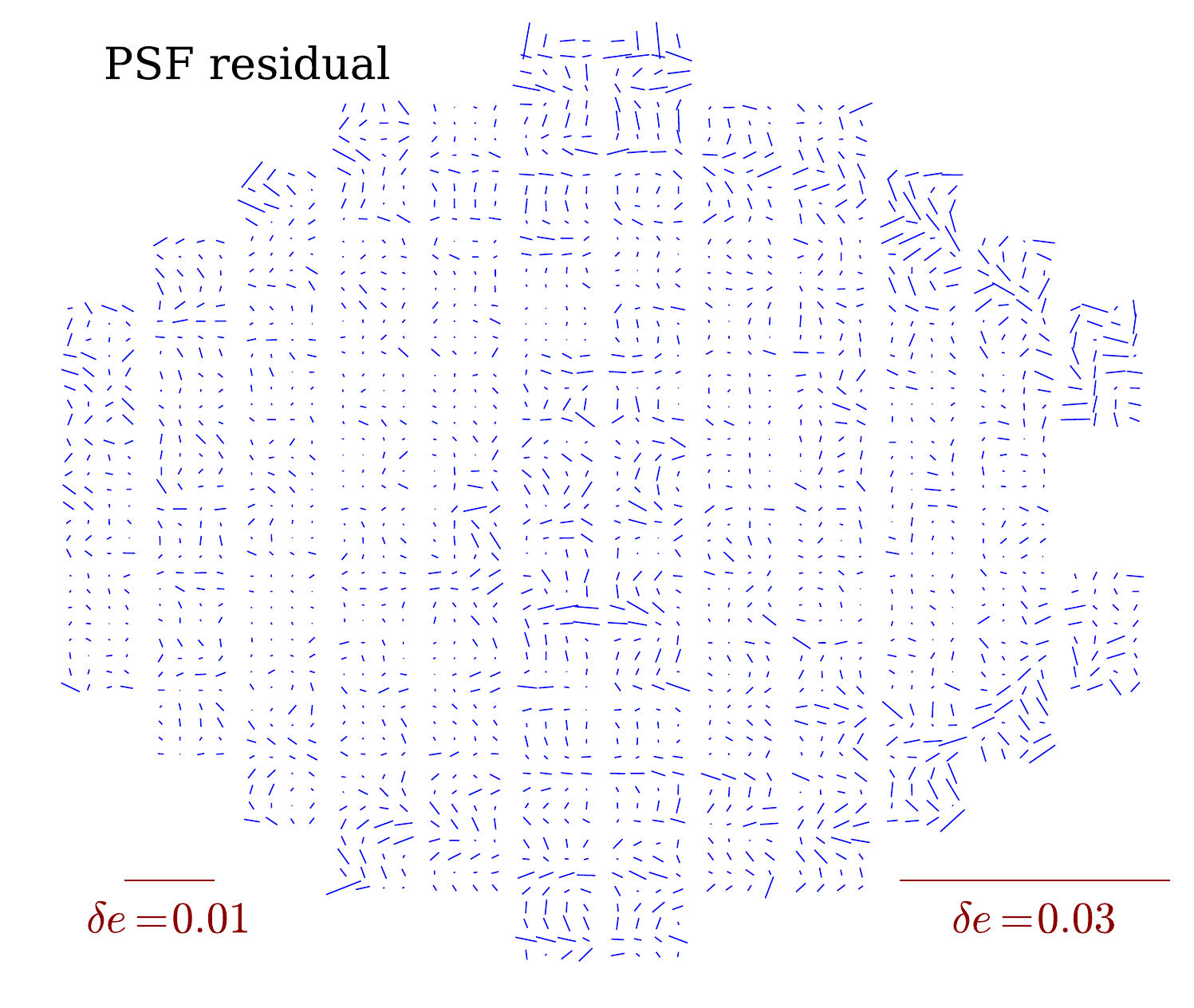}
\caption{Whisker plots of the mean PSF pattern (left) 
and of the mean residual after subtracting off the model PSF (right) 
as a function of position in the focal plane.
The length of each whisker is proportional to the measured ellipticity, and the orientation is
aligned with the direction of the ellipticity.  There is still some apparent structure in the plot
of the residuals, but the level is below the requirements for SV science. 
Reference whiskers of 1\% and 3\% are shown at the bottom of each plot, and
we have exaggerated the scale on the right plot by a factor of 10 to make the residual structure
more apparent.
\label{fig:psf:fov}
}
\end{figure*}

To measure the PSF and its spatial variation on each CCD, we used the software package
\psfex\ \citep{BertinPSFEx2011}.  Normally, \psfex\ takes as input the full list of objects detected
by \sex\ and finds the bright stars automatically.  However, as described in 
\S\ref{sec:psf:cuts}, we
removed some of the stars in the catalogue to avoid the brighter-fatter effect and the tape bumps.
This edited catalogue of stars was then passed to \psfex.

We used the \code{BASIS\_TYPE = PIXEL\_AUTO} option, which uses pixelated images to 
model the PSF profile, rather than fitting to some functional form.  
In \citet{Kitching13}, for undersampled PSFs a fixed oversampling was
found to perform better than the default \psfex\ choice; 
therefore, we forced the oversampling of these images to be a factor of 2 finer than the original
pixel size with \code{PSF\_SAMPLING = 0.5}.
The basis images are set to be $101 \times 101$ in the resampled pixels, or approximately 
13 arcseconds on a side.  

For the interpolation, we used a second order polynomials in chip coordinates,
interpolated separately on each CCD.  Specifically, we use the following 
parameters:
\begin{verbatim}
PSFVAR_KEYS = XWIN_IMAGE,YWIN_IMAGE
PSFVAR_GROUPS = 1,1
PSFVAR_DEGREES = 2
\end{verbatim}
%SB: not sure how to punctuate the above... I stuck a full stop at the end out of pedantry. but some 
%   pedants may say it shouldn't be there...
%MJ: The period looks wrong to me.  This is more like a code snippet, which shouldn't include
%    punctuation from the enclosing sentence.  I think with the "the following parameters:" intro,
%    it may be more acceptable to not include further punctuation. (?) 
We found that there was not much gain in using higher order polynomials than this and some evidence 
that they were overfitting the noise for some CCDs.  So we decided to use second order in all cases.

To assess the quality of the PSF interpolation, we first examined the differences between the 
measured shapes (using the HSM algorithm again) of actual stars on the image and the corresponding values for the
\psfex\ model at the locations of these stars. 
In \fig{fig:psf:fov} we show the whisker plots
of the PSF and the residuals as a function of position on the focal plane.
The residual whiskers are small, but not quite zero.  The impact of
these spatially correlated residuals are investigated below in \S\ref{sec:psf:rho},
and we will show that they meet
the requirements for science with SV data.

We believe the remaining structure
seen in the residual plot is largely due to the fact that the PSF modeling and interpolation is done in
pixel coordinates rather than sky coordinates.  Therefore, the interpolation must also include the
effects of the non-uniform WCS.  In particular, the distortion due to the telescope optics is a fifth order radial 
function,
but we fit the PSF with only a second order polynomial on each CCD.  This is most markedly
seen in the CCDs near the edges of the field of view where the residuals look
consistent with a fifth order radial function after the local second order approximation has been
subtracted off.  One of our planned improvements to the analysis is
to interpolate the PSF in sky coordinates rather than pixel coordinates,
so that WCS variations can be modeled separately from real PSF variations.  We expect this change
to remove most of the remaining residual PSF pattern.

\subsection{PSF Model Diagnostics}
\label{sec:psf:rho}
\assign{Mike}

Errors in the PSF model, and particularly errors in the interpolation, will directly affect the
shear estimates of galaxies, since they would be accounting for the effect of the PSF 
convolution incorrectly, as discussed in \S\ref{sec:req:rho}.
If the PSF errors were random, independent values for each galaxy,
they would constitute merely an additional contribution to the shear measurement uncertainty,
which would be highly subdominant to other sources of statistical noise, such as the 
unknown intrinsic shapes of the galaxies.  
However, this is not the case.  Because the PSF is interpolated between stars, the errors in the
PSF estimate are correlated among nearby galaxies.  The two-point correlation function of these
errors will directly impact the two-point correlation function of the shear estimates, which
means they would be a systematic error, as quantified in \eqn{eq:req:psfmodel}.

\citet{Rowe10} describes two diagnostic functions to quantify the level of interpolation errors
in the PSF model using the measured shapes of stars and the interpolated 
value of the model at the locations of these stars.  As we already introduced in \S\ref{sec:req:rho},
\begin{align}
\rho_1(\theta) &\equiv \left \langle \delta \epsf^*(\bfx) \delta \epsf(\bfxpt) \right \rangle \\
\rho_2(\theta) &\equiv \left \langle \epsf^*(\bfx) \delta \epsf(\bfxpt) \right \rangle,
\end{align}
where $\delta \epsf$ represents the difference between the measured ellipticity of the observed stars and 
the ellipticity of the \psfex\ models at the same locations, 
which is an estimate of the systematic uncertainty
in the shape of the PSF model at those locations.

In addition, we test three other statistics that appear at the same order of the expansion
of PSF model errors, involving errors in the PSF size, $\Tpsf$, which we
call $\rho_3$, $\rho_4$, and $\rho_5$.  They are defined in \eqnc{eq:req:rho3def}{eq:req:rho5def}
and are generally smaller than the two described by \citet{Rowe10}.

\begin{figure*}
\includegraphics[width=0.48\textwidth]{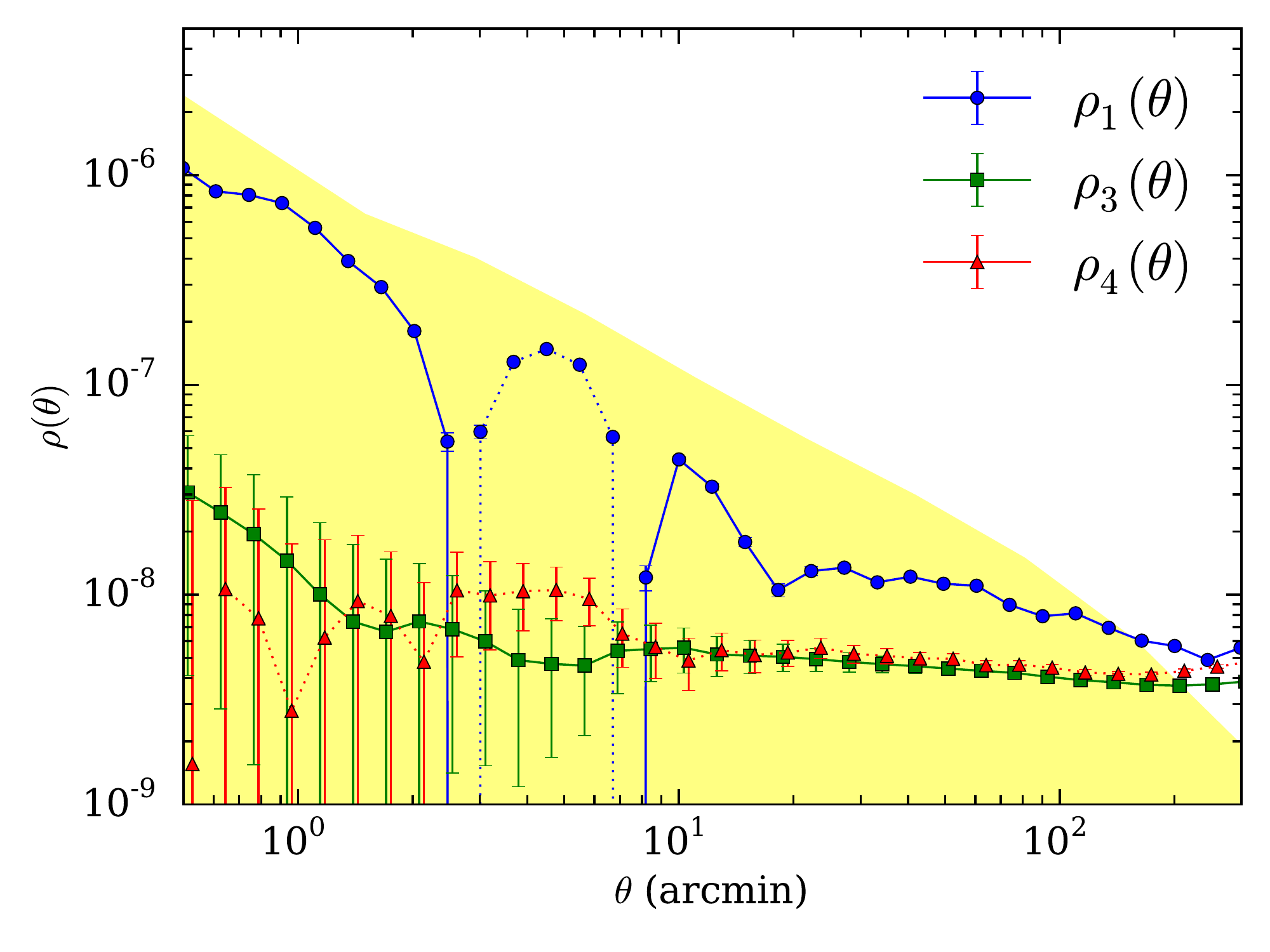}
\includegraphics[width=0.48\textwidth]{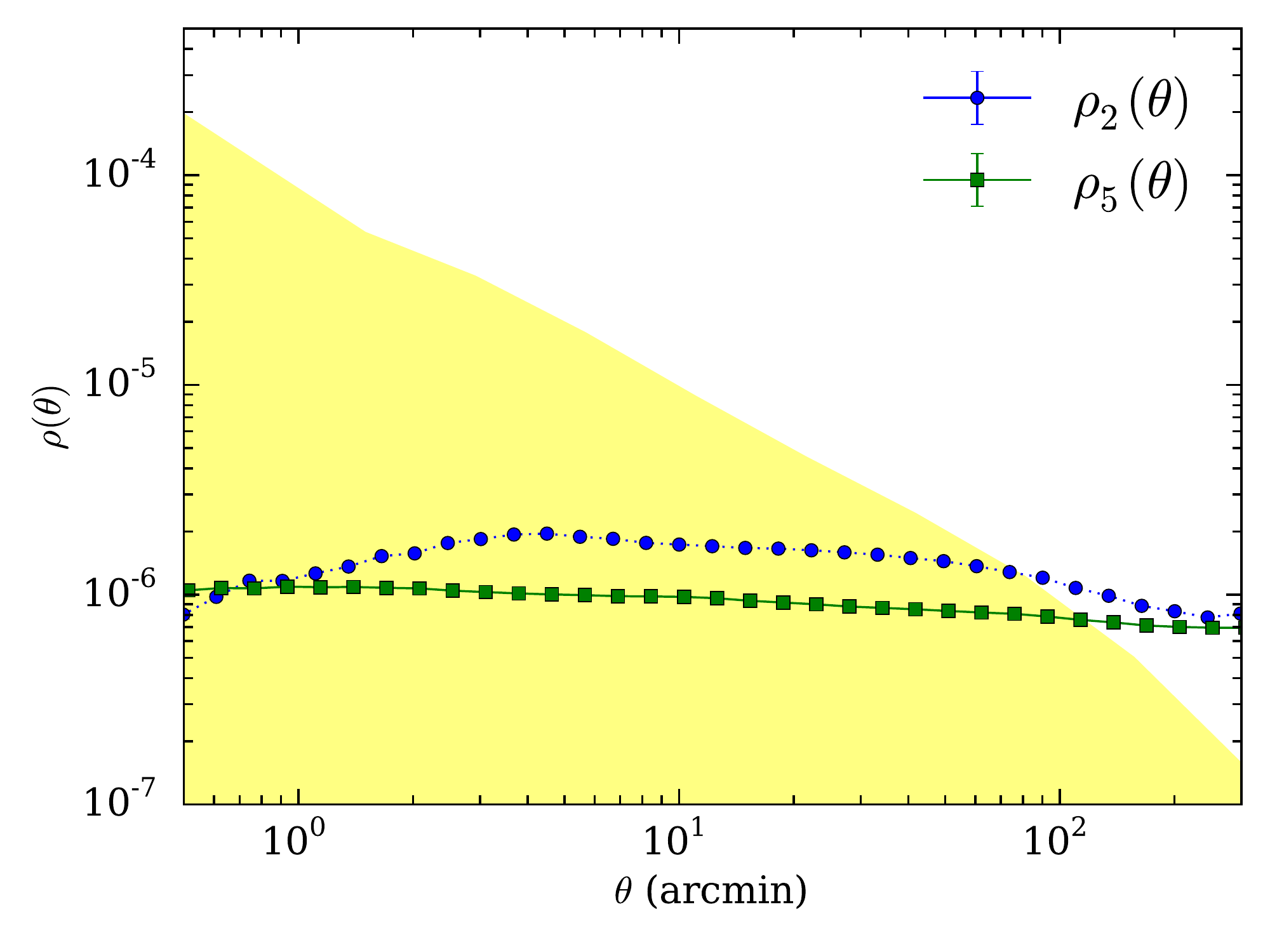}
\caption{The $\rho$ statistics for the PSF shape residuals.  
Negative values are shown in absolute value as dotted lines.
The shaded regions are the requirements for SV data.
\label{fig:psf:rho}
}
\end{figure*}

\fig{fig:psf:rho} shows the results for these five statistics.  The shaded regions show
the requirements, from \eqnb{eq:req:rho1}{eq:req:rho2}.
In all cases, the results are seen to be passing our requirements 
for scales less than about 100 arcminutes.
We see in \fig{fig:psf:rho} that $\rho_1$ changes sign twice and is below the requirement line
by only a factor of $\sim$ 2.
However, our requirements make the conservative assumption that additive errors are fully correlated
across scales.
So we have directly propagated the measured $\rho_1$ through to the bias on $\sigma_8$ and 
found the influence on $\sigma_8$ to be much less than one percent.

Of course, the \psfex\ model describes the full surface brightness profile of the PSF, not just 
its shape $\epsf$ and size $\Tpsf$.  However, these
are the dominant ways that errors in the PSF model could affect the galaxy shapes,
so these statistics are the most important checks of the model accuracy.

\section{\medsfull}
\label{sec:meds}
\assign{Erin}

As outlined in \S\ref{sec:data}, we used the \coadd\ images for object
detection and deblending.  For shear measurement we worked directly with the
pixel data from the original \SE\ images (cf.~\S\ref{sec:shear:multiepoch}).  
To simplify the bookkeeping we
developed a new data storage format, which we named \medsfull\ 
(\meds)\footnote{\url{https://github.com/esheldon/meds}}.

We created a \meds\ file corresponding to each \coadd\ image.  
In these files we stored a postage stamp for each observation of every object
detected in the \coadd\ image along with the corresponding weight maps, 
segmentation maps, and other relevant data.  
The postage stamps for each \coadd\ object were stored contiguously in the file, 
making sequential access of individual objects efficient.
The files are quite large, so loading the whole file into memory is
not generally feasible, but it is also not necessary.

The postage stamps from the original \SE\ images
were sky-subtracted and then scaled to be on a common
photometric system, which simplified the model fitting using these images.  We
also stored the local affine approximation of the WCS function, evaluated at the
object centre, so that models could be made in sky coordinates and constrained
using the different image coordinates for each postage stamp.

See \app{app:medsappendix} for details about how we build 
and store the \meds\ files.

\subsection{Exposure Selection}
\label{sec:meds:blacklists}
\assign{Alex}
\contrib{Steve, Mike}

We did not use all \SE\ images for measuring shapes.  We excluded a small fraction
of the CCD images that had known problems in the original data
or in some step of the data reduction and processing.
We created simple ``blacklist'' files, in which we stored information
for CCD images we wished to exclude, and that information was incorporated
into the \meds\ files as a set of bitmask flags.
Postage stamps from blacklisted images were then easily excluded from the analysis when
measuring shears.
Here we list some
of the reasons that images were blacklisted.

Some of the astrometry solutions (cf.~\S\ref{sec:astrometry}) provided a poor map
from CCD coordinates to sky coordinates.  This happened primarily near the edges of the
\spte\ region where there are not enough overlapping exposures to constrain the fit.

Some of the PSF solutions (cf.~\S\ref{sec:psf}) provided a poor model of the
PSF across the CCD.  In some cases there were too few stars detected to
constrain the model; occasionally there was some error when running either the star
finding code or \psfex.

A small fraction of the SV images were contaminated by bright scattered-light
artefacts.  Scattered-light artefacts fall into two broad categories: internal
reflections between the CCDs and other elements of the optics, known as
``ghosts''; and grazing incidence reflections off of the walls and edges of the
shutter and filter changer mechanism.  Ghosts primarily occur when a
bright star is within the field of view, while grazing incidence scatters
occur predominantly for stars just outside the field-of-view.  Using the
positions of bright stars from the Yale Bright Star Catalogue
\citep{YaleBSC} and knowledge of the telescope optics, it is possible to predict
locations on the focal plane that will be most affected by scattered light.  We
identified and removed a total of 862 CCD images (out of 135,481) from the single-exposure SV
data set in this manner.
In April 2013, filter baffles were installed to block some of this scattered light,
and non-reflective paint was applied to the filter changer and shutter in March 2014 \citep{FlaugherDECam2015}.
These modifications have greatly reduced the occurrence of
grazing incidence reflections in subsequent DES seasons.

It is common for human-made objects to cross the large DECam field of view
during an exposure.  The brightest and most impactful of these are low-flying
airplanes (two Chilean flight paths pass through the sky viewable by
the Blanco telescope).  Airplane trails are both bright and broad, and cause significant
issues in estimating the sky background in CCDs that they cross.  We identified these airplane
trails by eye and removed a total of
56 individual CCD images due to airplane contamination
(corresponding to 4 distinct exposures).  This rate of airplane contamination
is expected to continue throughout the DES survey.

In addition to airplanes, earth-orbiting satellites are a common occurrence in DES images.
During the 90 second exposure time of a DES survey image, a satellite in low-earth-orbit can traverse 
the entire focal plane, while geosynchronous satellites travel approximately 1.25 CCD lengths.
The impact of these satellite streaks is significantly less than that of airplanes; however, 
because they only occur in a single filter, they can 
introduce a strong bias in the colour of objects that they cross.
For SV, the ``crazy colours'' cut mentioned in \S\ref{sec:gold} removes most of the 
contaminated objects.
At the end of Year 1, an automated tool was developed by DESDM for detecting
and masking satellite streaks using the Hough transform \citep{Hough1959,DudaHart72}.
This should greatly reduce the impact of satellite streaks in upcoming seasons of DES observing 
and will be retroactively applied to reprocessing of earlier data.

\subsection{Masks}
\label{sec:meds:masking}
\assign{Erin}
\contrib{Niall}

The user can construct a ``mask'' for each postage stamp in the \meds\ files in a variety of ways.
For this analysis, we used what we call an ``\uberseg'' mask,
constructed from the weight maps, segmentation maps and locations of nearby objects.

To create the \uberseg\ mask,
we started with the \sex\ segmentation map from the \coadd\ image, mapping
it on to the corresponding
pixels of the \SE\ images.  We prefer this map to the segmentation map
derived for each \SE\ image because the \coadd\ image is less noisy, and thus
has more object detections and more information for determining the extent of
each object.

We then set pixels in the weight map to zero if they were either associated
with other objects in the segmentation map or were closer to any other object
than to the object of interest. The result was a superset of the information 
found in the weight maps and segmentation maps alone, hence the name \uberseg.

\begin{figure}
\centering
\includegraphics[width=1.0\columnwidth]{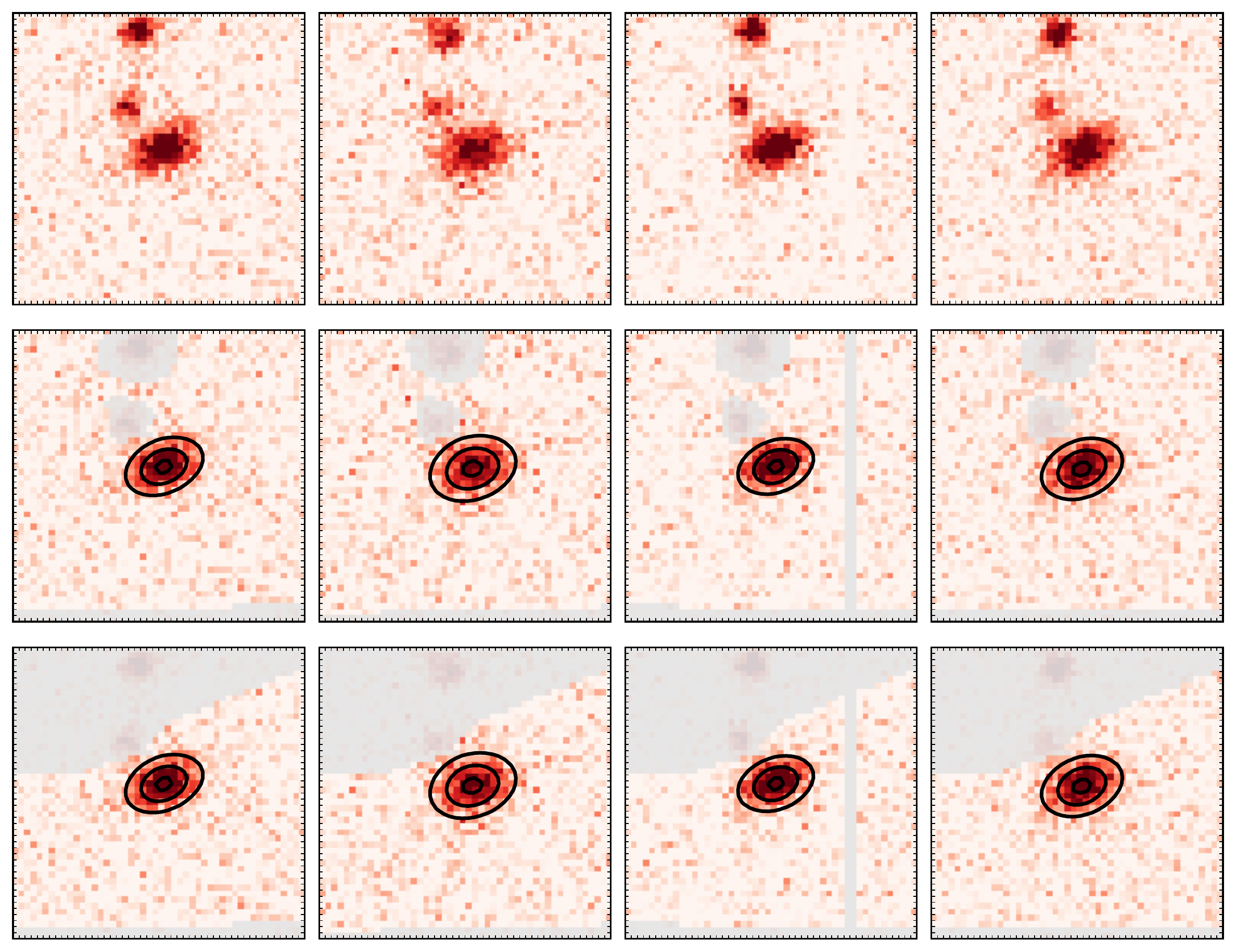}
\caption{Example galaxy image demonstrating two masking strategies.  
The top row shows the original postage stamps in the \meds\ file.  
The second row shows the result
when only the \sex\ segmentation map was used to mask neighbors.  
The third row shows the result when the \uberseg\ algorithm was used
to mask neighbors, as described in the text.
\label{fig:uberseg}
}
\end{figure}

An example set of images and \uberseg\ maps are shown in \fig{fig:uberseg}.
In tests on a simulation with realistically blended galaxies (cf. \S\ref{sec:sims:endtoend}),
we found a large reduction
in the shear biases when using the \uberseg\ masking as compared to the
ordinary \sex\ segmentation maps.  In particular, when using
ordinary segmentation maps we found a significant bias of the galaxy
shape in the direction toward neighbors.
With the \uberseg\ masking, such a bias was undetectable.

\section{Simulations}
\label{sec:sims}

Simulations were a crucial part of our shear pipeline development process,
providing data with known values of the applied shear for testing the shear
estimation code. There is no such absolute calibration source in the real data.
In addition to many small targeted simulations designed to answer particular
questions about the algorithms, we developed two general purpose simulations
that we used extensively to test the shear pipelines.

The first, which we call \greatdes, was modeled on the GREAT3 challenge.  We
used the \greatdes\ simulation to test the accuracy of the shear estimation
codes on realistic space-based galaxy images with a realistic range of noise
levels and galaxy sizes.  As with the GREAT3 challenge, the galaxies were
placed on postage stamps, so there were no blending or object detection issues
to consider.

The second, which we call the end-to-end simulation, was a high \snr\
simulation with analytic galaxy models with elliptical isophotes.  The
motivation with these simulations was to test that various bookkeeping details
were implemented correctly, such as the file conventions used by \psfex, the
application of the WCS transformations, and conventions about the origin of the
postage stamps in the \meds\ files.  These are all details that are easy to get
wrong, but which can be difficult to notice on noisy data.  In these simulations
we also tested the efficacy of the \uberseg\ masking (c.f. \ref{sec:meds:masking}).

We have found the \galsim\ \citep{GalSim} image simulation software 
to be invaluable for this purpose.  
In particular, its ability to accurately render sheared versions of space-based
images using their reimplementation of the SHERA
algorithm \citep[SHEar Reconvolution Analysis;][]{shera},
correctly accounting for the original HST PSF \citep{BernsteinGruen14},
was particularly important for making the \greatdes\ simulation.
The end-to-end simulation relied on \galsim's ability to generate multiple epochs of the
same scene and accurately handle non-trivial WCS transformations for the
various exposures.

\subsection{\greatdes}
\label{sec:sims:greatdes}
\assign{Tomek}

We used the \greatdes\ simulation to test the precision and accuracy of our
shear measurement codes, using DES-tuned sampling of both the population of
galaxies (size, shape, morphology) and the observing conditions (PSF
ellipticity, noise level).  The simulation consists of individual $48 \times
48$ pixel postage stamp images.  We ignored issues of crowding, bad pixels, and
imaging artefacts, but we otherwise attempted to make the images a close
approximation to the DES SV data.

We built the \greatdes\ simulation using galaxies from the COSMOS 
survey \citep{COSMOS}, made available for use with 
\galsim\footnote{\url{https://github.com/GalSim-developers/GalSim/wiki/RealGalaxy\%20Data}}. 
\citet{Kannawadi2014} showed that this sample of galaxies is a good 
representation of galaxy properties, and can be used in shear calibration of lensing surveys to a 
precision level of $m\!=\!0.01$.

\begin{figure*}
\includegraphics[width=\textwidth]{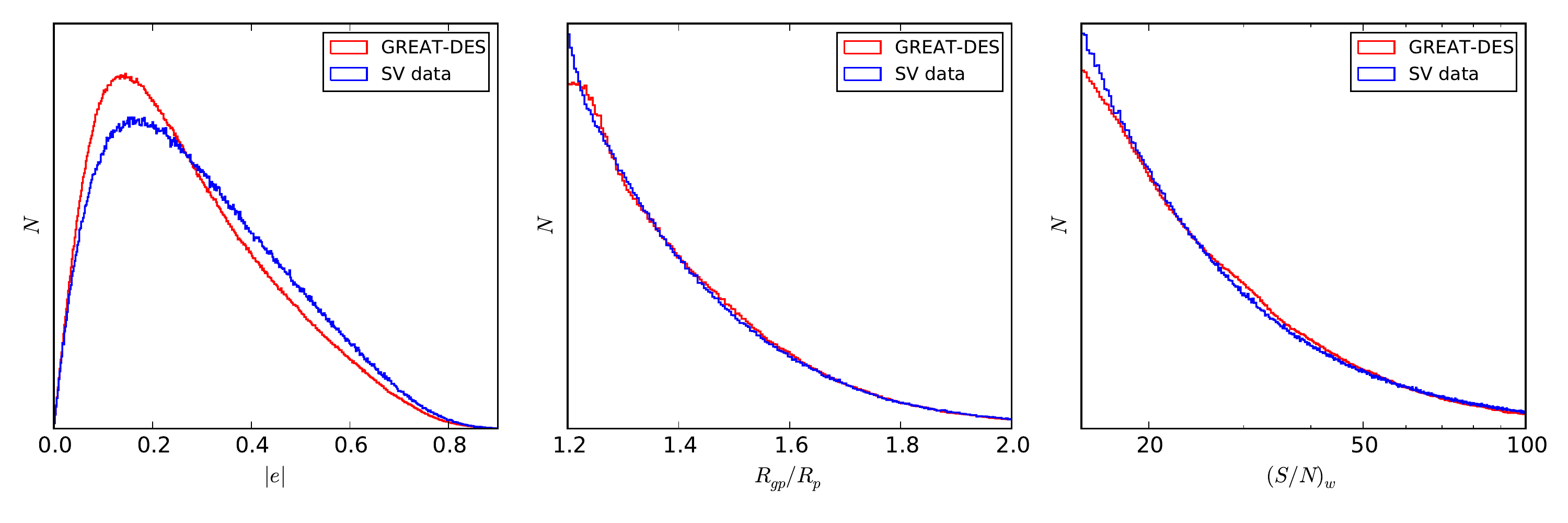}
\includegraphics[width=\textwidth]{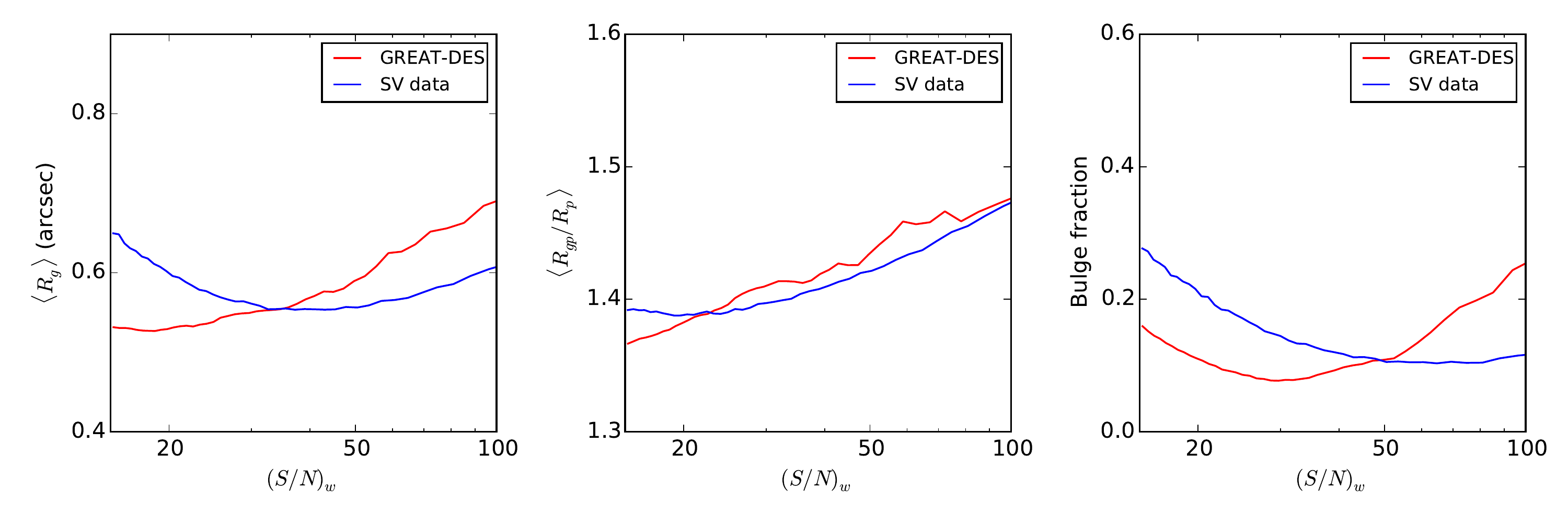}
\caption{A comparison of the galaxy properties in \greatdes\ (red) and the SV data (blue).
The top row shows histograms of $|e|$, \rgp, and \snr\ as measured by \imshape.
The bottom row shows the dependence of
$\langle R_g \rangle$, $\langle \rgp\ \rangle$, and bulge fraction as functions of \snr.
\label{fig:greatdes:histograms}
}
\end{figure*}

We started with the entire COSMOS sample distributed for use with \galsim\,
and discarded objects that were flagged as unusable in the GREAT3 challenge \citep{great3handbook},
which removed about 3\% of the objects and left more than
54,000 COSMOS galaxies available for use in the simulation.  
Next we selected individual galaxies from this set in such
a way as to mimic the distribution of galaxy properties
found in DES SV data.

For the PSF, we used a Kolmogorov profile with
sizes and ellipticities taken to match the range of values present in the SV data.
Specifically the PSF size took one of 6 values between 0.8 and 1.3 arcseconds FWHM,
and each component $e_1$, $e_2$ of the shape took one of 4 values from -0.02 to +0.02.
Thus, a total of 96 unique PSF images were used in the entire sample.
Gaussian noise was added based on the typical noise level observed in SV \coadd\ images.

We then applied a constant shear value within each simulation field, 
with a magnitude of $|g|=0.05$ and rotated at 8 evenly spaced position angles $\phi$.

Each of the COSMOS galaxies was used hundreds of times, with different noise
realizations, different random orientations and different centroid offsets.  We
did not use 90 degrees rotated galaxy pairs, as has commonly been done for the
GREAT challenges \citep[cf.][]{great10handbook} to reduce the number of
galaxies required to reach a given measurement precision.  We instead randomly
oriented each galaxy.  With this choice, we retained the ability to select
subsets of the galaxies according to their measured characteristics.  Such
selections tends to break up the pairs, which obviates any advantage from using
them.  More importantly, additive errors can cancel between the pairs of
galaxies, which would hide important systematic errors that would yet appear
real data. Therefore, we instead use a very large number of galaxies in each
field to get to the desired precision on the mean shear.  In total, we use 48
million rendered galaxy images.
%MJ: 8 shears x 600 files x 10K per file = 48 million objects total in the
%truth catalogs for nbc-sva1-006

We developed a DES-specific module for \galsim\ to store the simulated images directly
in \meds\ format, so we could run \imshape\ and \ngmix\ on the resulting \meds\ files with 
minimal modification compared to how we run the code on the data.
We did not actually create multiple epochs for this simulation, but it was helpful to 
use the same file format as the data.

To estimate the level of systematic errors well, we desired
the simulation to be closely representative of the data \citep[see e.g.][]{Berge2012,Bruderer2015}. 
To check that we have achieved this goal,
we compared the \imshape\ measurements of relevant galaxy properties
in \greatdes\ to those in the SV data.
In \fig{fig:greatdes:histograms} we show that the distributions of ellipticity, size and \snr\
are well-matched between the simulation and the data, as is the 
dependence of $R_g$ and \rgp\ as functions of \snr.
The bulge fraction, estimated by which model \imshape\ chooses as the better fit,
shows systematic disagreement as a functioin of \snr, although
the overall bulge fraction matches very well:
0.161 and 0.167 for the simulation and the data respectively.

Note that the choice to show \imshape\ measurements in \fig{fig:greatdes:histograms} is arbitrary;
the analogous plot of \ngmix\ measurements shows similar agreement, except that there is no estimate of bulge
fraction from the \ngmix\ exponential \disk\ model.
Also, the \imshape\ ``bulge fraction'' should not be considered an estimate of the actual \sersic\ index of the galaxies;
it is merely a diagnostic measure
related to the concentration of the galaxies.

Since \ngmix\ uses an exponential \disk\ model (cf.  \S\ref{sec:ngmix:model}),
and thus has worse model bias for bulge-like galaxies, the discrepancy in the
bulge fraction limits our ability to infer what the model bias would be in real
data (cf.~\S\ref{sec:tests:sims}). The \imshape\ shear estimate is based on a
bulge-or-\disk\ galaxy model (cf.~\S\ref{sec:im3shape:model}), which has less
overall model bias \citep{Kacprzak14}, so the discrepancy may have less impact.
However, the ability to accurately choose bulge vs. disk is dependent on \snr, so
our ability to test this aspect of the fitting is also somewhat limited
in \greatdes.

\subsection{End-to-end Simulation}
\label{sec:sims:endtoend}

The end-to-end simulation is of an entirely different nature from the \greatdes\ simulation.  
It is a high \snr\ simulation used to test various mundane coding details that are easy to
mix up, but which can be difficult to verify in noisy data.
For the galaxies we used simple exponential \disk\ profiles, which have elliptical isophotes when sheared,
and the images were rendered with relatively little pixel noise.  The fundamental shape estimation
problem is thus straightforward for both algorithms.

The starting point for this simulation was one of the actual \meds\ files from the data, along
with the corresponding \coadd\ catalogue, the list of \SE\ images that contributed to the \coadd\ image,
and the WCS solutions and estimated background maps for each \SE\ image.

Next we built new versions of these \SE\ images using exponential \disk\ galaxies
with the same size, flux, ellipticity, and celestial position as the measurements of the real galaxies.
We used variable elliptical Gaussian profiles for the PSF, using different parameters for each
\SE\ image.  The convolved images were rendered at the correct position on each image
using the original WCS.  With \galsim\ we applied the Jacobian of the WCS to the surface brightness profile
as well, so this important detail was handled correctly.
Objects that were deemed to be stars in the original catalogue (based on the \sex\ 
\code{SPREAD\_MODEL} being less than 0.003) were drawn as a PSF profile,
with the same flux as the original object.  Finally, we added the original background sky level
to the image, but with relatively small noise so that the faintest galaxies had $\snr > 200$.

We then ran these images through the full weak lensing pipelines, starting with
\sex\ and \psfex\ to estimate the PSF model, then building a \meds\ file,
and finally running \imshape\ and \ngmix.  
The resulting measured shape estimates were then compared
to the true shapes of the simulated galaxies, which were
expected to match to quite high precision, given the nature of the simulation.

The end-to-end simulation was successful in finding several bugs in various parts of the 
shear pipeline.  However, the most notable result from this process was the 
development of the \uberseg\ mask (cf.~\S\ref{sec:meds:masking}).  
These tests revealed significant biases from the
masking procedure we had been using, involving just the \sex\ segmentation maps.
When a galaxy had a bright neighbor on the same postage stamp,
light from the neighbor that was just outside the segmentation map was being included
as part of the fit, thus significantly biasing the inferred shapes in the direction of 
the neighbor.

Switching to the \uberseg\ mask made a big difference; we found the measured
shapes were then much closer the true values.  We found there was still a small
effect from neighbors, which amounted to a slight increase in the effective
shape noise for such objects, but we no longer detected any systematic bias in
the shape estimates due to unmasked flux from neighboring objects.

\section{Shear Measurement} 
\label{sec:shear}

We used two different shear measurement codes for this study: \imshape\ and
\ngmix, both of which are based on model-fitting.
\imshape\ performs a maximum likelihood fit using a bulge-or-\disk\ galaxy model 
(cf \S\ref{sec:im3shape:model}).
\ngmix\ uses an exponential \disk\ model, exploring the full $N$-dimensional
posterior likelihood surface
with an informative prior applied on the ellipticity (cf.~\S\ref{sec:ngmix:model}).

With both shear methods we used the \psfex\ models of the PSF detailed in \S\ref{sec:psf:measurement},
although the way the PSF model was used differed.  \psfex\ produces a 2-D image of the 
PSF profile at the location of each galaxy.  With \imshape\ we resampled the PSF image
to a higher resolution grid and performed the convolution with the galaxy model
via FFT.  With \ngmix, we fit 3 free elliptical
Gaussians to the PSF image and performed an analytic convolution with the galaxy model,
which was also approximated as a sum of Gaussians, resulting in very fast
model creation.

Finally, with both shear codes we used the \meds\ files described in
\S\ref{sec:meds} to constrain the galaxy models, using pixel data from the
original \SE\ images rather than using the \coadd\ image, which we only
used for object detection.

We discuss the details of the multi-epoch fitting process in the next section,
\S\ref{sec:shear:multiepoch}. In \S\ref{sec:shear:snr}, we define what we mean by signal-to-noise.
The details of the \imshape\ and \ngmix\ algorithms are given in \S\ref{sec:im3shape}
and \S\ref{sec:ngmix}.  Finally, our strategy for blinding the shear estimates is described
in \S\ref{sec:blinding}.

\subsection{Multi-epoch Fitting}
\label{sec:shear:multiepoch}

The typical method for dealing with multiple exposures of a particular patch
of sky is co-addition of images (also known as ``stacking''; cf.~\citealp{Fu08}).  
However, 
co-addition can be problematic, since it necessarily loses information
and imparts non-trivial, spatially correlated noise into the final image.  
Furthermore, as each CCD covers a finite region of sky,
addition of a finite number of CCD images results in discontinuities in the
PSF at image boundaries (cf.~\citealp{Jee13} for a discussion of this effect).
A more optimal method for fitting a
collection of images is to simultaneously fit all independent pixel data,
as also advocated by \citet{Heymans12}.  We
call this process multi-epoch fitting.

Multi-epoch fitting requires some additional complexity in the fitting
process, as we must use the correct PSF and WCS information for each image,
rather than a single function for each as would be sufficient to process a
\coadd\ image.

In order to simplify the bookkeeping to process the multi-epoch and multi-band
DES data, we used the \medsfull\ described in \S \ref{sec:meds}.
Each observation of a particular galaxy experiences a different PSF, and the
local image coordinates are related to celestial coordinates via a different
WCS transformation.  This information was stored in the \meds\ file and used
during modeling.

For both codes (\ngmix\ and \imshape) the model for a given set of galaxy
parameters was generated in celestial coordinates.  For \ngmix, we modeled the
PSF in celestial coordinates as well, and convolved it with the galaxy model
analytically.  We then compared this model to the observed data using the WCS
transformation.  For \imshape\ we modeled the galaxy and PSF in image
coordinates and convolved via fast Fourier transform (FFT).

\subsection{Signal-to-noise Ratio}
\label{sec:shear:snr}

Before we describe the algorithms we used for measuring shapes, it is worth
describing in detail what we mean by the signal-to-noise ratio (\snr).
This will be relevant both in the 
next section (in particular \S\ref{sec:im3shape:noisebias}, where we discuss how
\imshape\ calibrates the shear bias) and in later sections such as \S\ref{sec:tests:gal},
where we test that the shear is independent of \snr, and \S\ref{sec:cats:flags},
where we use \snr\ while selecting galaxies for the final shear catalogues.

There is no single definition for the \snr\ of an image or a surface brightness
profile.  Rather, a \snr\ is only well defined for a single measured value -- some statistic
calculated from the image or profile.  Given some such statistic $x$, the \snr\ is
typically defined as that value (either the measurement or the true value) divided
by the square root of its variance
\begin{equation}
\snr \equiv \frac{x}{\sqrt{\textrm{Var}(x)}} .
\end{equation}

One of the standard \snr\ measures is the so-called ``optimal'' \snr\ estimator.
One can show that among all statistics that are linear in the pixel values $I_p$,
\begin{equation}
\hat I_w = \sum_p w_p I_p,
\end{equation}
the one with the highest expected \snr\ has weights 
$w_p = \langle I_p \rangle / \sigma_p^2$,
where $\sigma_p^2$ are the estimated variances in each pixel.\footnote{The
proof involves finding $w_p$ values such that expectation of the \snr\ is
stationary with respect to any infinitesimal changes $\delta w_p$.}

In practice, one does not know the true expectation value of the surface brightness profile,
$\langle I_p \rangle$,
so typically one uses the best-fitting model of the galaxy, which
we call $m_p$, as part of the weight.
The \snr\ of this statistic is thus estimated as
\begin{equation}
\snrw = \frac{ \sum_p m_p I_p / \sigma_p^2} { \left(\sum_p m_p^2 / \sigma_p^2 \right)^{1/2}} . 
\label{eq:snrw}
\end{equation}
This is the \snr\ measure used by GREAT3 \citep{great3handbook}, for example.

A drawback of this estimator is that it is not independent of an applied
shear.  Galaxies that look similar to the PSF will have a higher measured \snrw\ than galaxies
with a different size or shape.  The PSF essentially acts as a matched filter
for these galaxies.  This means that \snrw\ is not
invariant under an applied gravitational shear.

If the PSF is 
approximately round, as is the case for our data,
then more elliptical galaxies will have a lower estimated
\snrw\ than round galaxies (holding flux constant).
Thus if galaxies are selected according to their measured \snrw, the
resulting galaxy catalogue will have a selection bias towards round shapes, which will
bias the overall mean shear.

One solution to this potential systematic error is to use a \snr\ estimator that is 
not biased with respect to 
an applied shear.  There are a number of choices one could make for this.
We choose to calculate the \snrw\ that the galaxy would have had
\emph{if it and the PSF were round}.

That is, we take the model of the galaxy profile and apply a shear such that
its ellipticity becomes zero.  We do the same for
the PSF, convolve these two profiles together,
and then integrate over the pixels.
The resulting $m^r_p$ values are the intensities we predict would
have been observed if both the galaxy and the PSF had been round.
We then use these values for both the 
model $m_p$ and the intensity $I_p$ in \eqn{eq:snrw}, as the actual data are no longer appropriate
for this counterfactual surface brightness profile.  
The ``roundified'' \snr\ estimator is then
\begin{align}
\snrr &= \frac{ \sum_p m^r_p m^r_p / \sigma_p^2} { \left( \sum_p (m^r_p)^2 / \sigma_p^2 \right)^{1/2} } \nonumber \\
&= \left(\sum_p (m^r_p)^2 / \sigma_p^2 \right)^{\!\!1/2} .
\label{eq:snrr}
\end{align}

We find both measures of the signal-to-noise useful in different contexts.  For
\ngmix, we use \snrr\ for the reasons described here; we find significantly
smaller selection biases when we use \snrr\ to select galaxies for shear
measurement, as compared to using \snrw.  

For \imshape, we find that the noise
bias calibration (cf. \S\ref{sec:im3shape:noisebias}) is more accurate using \snrw\ than \snrr, presumably because
the noise bias is more directly related to the signal-to-noise of the actual
galaxy than to that of a counterfactual round version of the galaxy.  
Thus, the ``noise bias'' calibration in fact also approximately calibrates the selection bias resulting from using \snrw.
This is therefore the appropriate \snr\ measure to use for selecting galaxies for the final \imshape\ catalogue.

\subsection{Shear Measurements with \imshape}
\label{sec:im3shape}
\assign{Joe}

\imshape\ is a maximum-likelihood model-fitting code, which we used to fit \devauc\ bulge and
exponential \disk\ components to galaxy images.  The code was described in 
\citet{im3shape}, where its performance on GREAT08 and its known biases were characterized.

We have slightly modified the model described therein, improving both its stability and its performance
on the tests detailed in this paper (cf. \S\ref{sec:tests}).  Previously each galaxy was modeled as the sum of two components,
a bulge and a \disk.  In this paper, we fit each galaxy twice: once as a pure bulge and once as a pure \disk.
For shear estimation, we used the model with the higher likelihood, unless that model was flagged as a bad fit
(cf. \S\ref{sec:im3shape:diagnostics}).  If both models were flagged the galaxy was excluded from the catalogue.

\edit{We found this ``bulge or \disk'' scheme to be much more robust on
    simulations of realistic galaxies. This scheme produced good model fits in almost all
    cases, whereas with the previous ``bulge plus \disk'' scheme we frequently
    found the best fit model was unphysical, with highly negative-flux components.}

The parameters of the best-fitting model were found using the numerical optimizer 
\LevMar\footnote{\url{http://users.ics.forth.gr/~lourakis/levmar/}} \citep{LevMar}, which
is an implemention of the \levmar\ algorithm \citep{Levenberg1944, Marquardt1963}, iterating towards a
model image which minimizes the $\chi^2$ with the data image.

For the SV data, we ran \imshape\ on \rband\ images only.  With the future data we plan to test fitting multiple bands simultaneously with marginalized relative amplitudes.

We made a number of additions to the original code presented
in \citet{im3shape}; in this
section we briefly review the code and its methodology, with particular focus on these changes.

The complete code with all the changes described below is available for
download\footnote{\url{https://bitbucket.org/joezuntz/im3shape}}.
One particularly useful infrastructure improvement was the implementation of a
Python interface to the
existing C functions.  We used the Python interface to load data from MEDS files (via
the \code{meds} module), select exposures, mask images, and compute most of the
diagnostic information described in \S\ref{sec:im3shape:diagnostics} below.

The biggest change we made to \imshape\ was the addition of a new model which fits multiple
exposures of the same galaxy simultaneously.  
We now define our model parameters in a celestial coordinate system:
a local tangent plane centred at the nominal Right Ascension and Declination of the galaxy.
This model is then constrained by the pixel data from each epoch where the galaxy was
observed, as discussed in \S\ref{sec:shear:multiepoch}.

\subsubsection{Bulge or \disk\ model}
\label{sec:im3shape:model}

Each galaxy model was defined by six varied parameters: 
the amplitudes of either the bulge or \disk\ components ($A_b, A_d$),
a centroid relative to the nominal detection position ($\delta u, \delta v$), 
an ellipticity ($e_1, e_2$), and a half-light radius ($r$).

To compute the likelihood of a particular model for a given galaxy observed on a number 
of individual exposures, we used the local affine approximation to the WCS for each postage
stamp (stored in the \meds\ file) to transform these parameters into each image's local
pixel coordinate system.  Schematically,
\begin{align}
&\left\{ \delta u, \delta v, e^c_1, e^c_2, r^c \right\} \rightarrow \nonumber\\
&\qquad \bigl[
\left\{ \delta x, \delta y,  e^p_1, e^p_2, r^p \right\}_{\textrm{Image 1}}, \nonumber\\
&\qquad ~\left\{ \delta x, \delta y,  e^p_1, e^p_2, r^p \right\}_{\textrm{Image 2}},
... \bigr]
\end{align}
where ${}^c$ indicates the parameters in celestial coordinates ($u,v$) and ${}^p$ 
indicates the transformed parameters in pixel coordinates ($x,y$).
The amplitudes do not require any transformation, since the \meds\ files have already put the
postage stamps on the same photometric system.

Given the appropriate parameters for each postage stamp, 
we then built the galaxy models in pixel coordinates, each convolved by the correct PSF
for that stamp, and computed a $\chi^2$ of the model relative to the data, using the correct
pixel noise.  The total $\chi^2$ from all the postage stamps then gave us the final
likelihood to use for that set of model parameters.  We then iterated to find the maximum
likelihood parameters for each galaxy.
The maximum likelihood was typically found in less than 50 iterations.  At 150 iterations, we
stopped the algorithm and declared failure.

The Levenberg-Marquadt code that we used to find the maximum likelihood, 
\LevMar, does not directly handle 
problems where different weights are applied to each data point.  The 
straightforward fix for this is to scale both the observed intensity $I_p$
and the model $m_p$ by the standard deviation of the intensity
$\sigma_p$ before passing them to \LevMar:
\begin{align}
I^\mathrm{LM}_p &= I_p / \sigma_p \nonumber \\
m^\mathrm{LM}_p &= m_p / \sigma_p.
\end{align}
This maintains the $\chi^2$ per pixel that the \LevMar\ algorithm uses as its objective function.
The estimates of $\sigma_p$ came from the weight map (as $\sigma_p^{-2}$) provided with the images.

% NOTE: If I put this any later, it shows up on the next page.  With this here, it shows up on the same page as its reference
%    in section 7.3.2
\begin{figure*}
\includegraphics[width=0.48\textwidth]{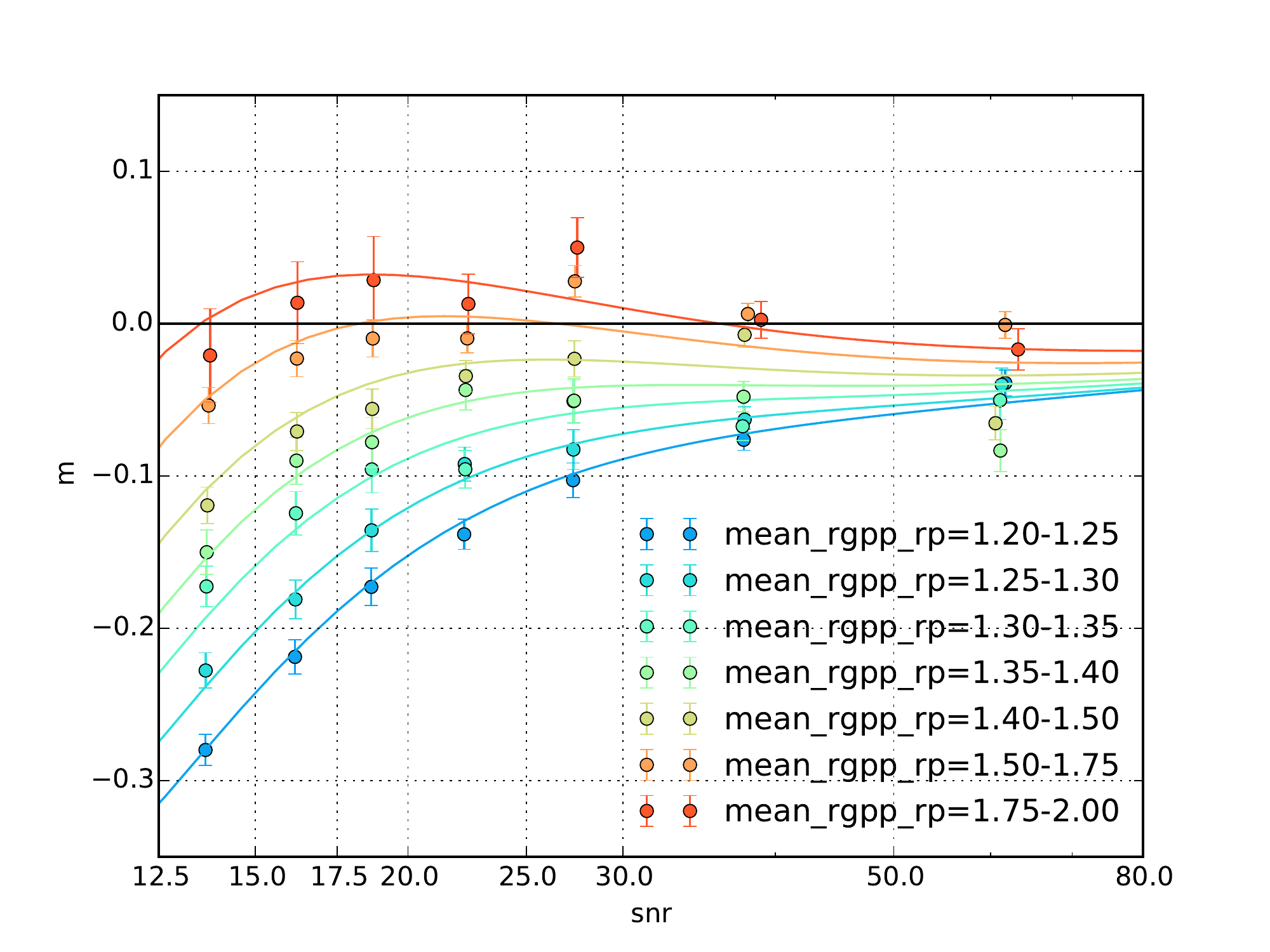}
\includegraphics[width=0.48\textwidth]{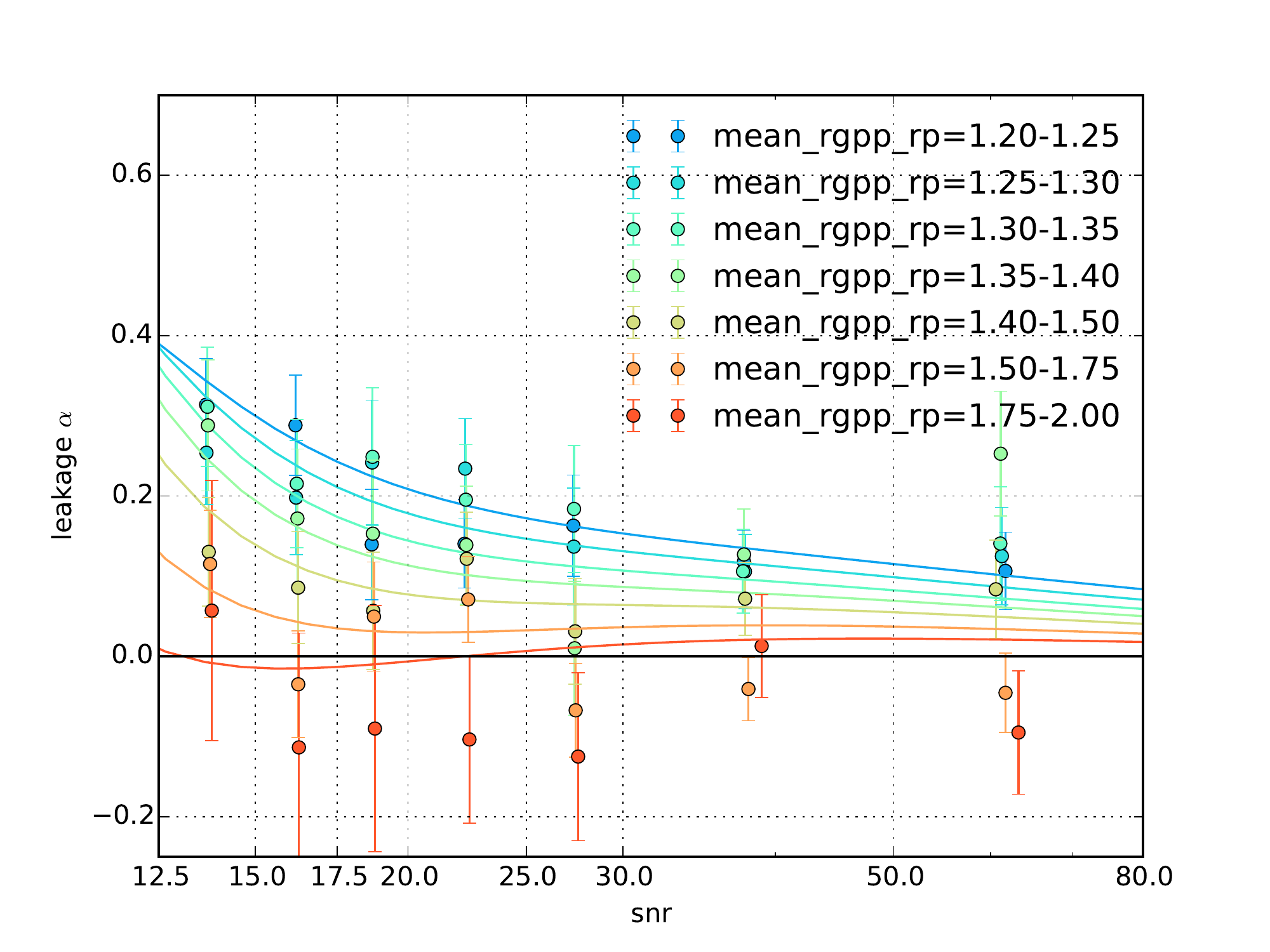}
\caption{Shear bias for \imshape\ measurements on the \greatdes\ simulation: multiplicative bias (left) and PSF
leakage (right), as functions of the measured \snrw\ and \rgp.
The fits, which are used to calibrate the shear estimates on the data, are smooth functions in both of these variables.
Solid lines show the fits vs \snrw\ at particular choices of \rgp.
\label{fig:tests:calibration_factors}
}
\end{figure*}

\subsubsection{Shear calibration}
\label{sec:im3shape:noisebias}
\assign{Tomek}

A significant problem with maximum likelihood shear estimators is that the peak of the likelihood
distribution is not an unbiased estimator of the shear in the presence of noise \citep{refregier12, kacprzak12}. 
The fitted model parameters are a non-linear function of pixel intensities affected by 
Gaussian noise, resulting in \emph{noise bias} in the estimated shear values.
The \imshape\ algorithm, being a maximum likelihood estimator, is known to suffer from this effect.

In addition, we found a small \emph{selection bias}, which is introduced by using recommended \imshape\ flags 
(cf.~\S\ref{sec:im3shape:diagnostics}) and the selection based on galaxy size and \snr\ (cf.~\S\ref{sec:cats:flags}).
We also expect a small amount of \emph{model bias} due to realistic galaxies not always being well fit by our
bulge-or-\disk\ model.
This model bias is expected to be small compared to the requirements \citep{Kacprzak14}.

To account for all of these sources of error in our shape measurements, we calculated bias 
corrections of the form shown in \eqn{eq:req:esys}. Specifically, we fit for $m$ and $\alpha$
as functions of \snrw\ (defined in \eqn{eq:snrw}) and \rgp\
(the FWHM of the PSF-convolved galaxy divided by the FWHM of the PSF)
on simulated data from the \greatdes\ simulation (cf. \S\ref{sec:sims:greatdes}).
We ran \imshape\ on the simulated data in the same way as we do on the DES data, including the 
same choices of input parameters.  

In principle, the two multiplicative terms, $m_1$ and $m_2$ should be treated as independent biases. 
In practice, however, when averaged over many galaxies we find virtually no difference between the two. 
As such, we correct both $e_1$ and $e_2$ by the average $m=(m_{1}+m_{2})/2$. 

We fit both $m$ and $\alpha$ as two-dimensional surfaces in the \snr\ and size parameters. 
Due to the complicated structure of this surface, we fit $m$ with 15 terms 
of the form $\snrw^{-x} (\rgp)^{-y}$, where $x$ and $y$ are various powers ranging from 1.5 to 4.
To control overfitting, we used a regularization term in the least-square fit and optimized it such that the 
fitted surface has a reduced $\chi^2=1$.
A similar procedure was applied to $\alpha$, where we used 18 parameters in the fit.
In \fig{fig:tests:calibration_factors} we show these fits as curves in \snrw\ in bins of \rgp.  
However, the actual functions are smooth in both parameters.

We checked if our calibration is robust to the details of this model
by (1) varying the number of terms in the basis expansion and (2) splitting the training data into halves. 
For both tests the changes in the mean multiplicative and additive corrections applied to
the SV data did not vary by more than 1\%.

In \S\ref{sec:shear:snr}, we mentioned that \snrw\ is a biased measure of \snr\ with respect to shear,
so if it is used to select a population of galaxies, it will induce a selection bias on the mean shear.
\rgp\ similarly induces such a bias.
Thus, when we bin the shears by these quantities to construct the calibration functions, 
there is a selection bias induced in every bin.  
The scale of selection bias reaches $m\simeq-0.05$ for the most populous bins.
This is not a problem for the correction scheme so long as the overall selection is also made
using these same quantities.  In that case, the shear calibration automatically accounts for the 
selection bias in addition to the noise bias.

We tried using \snrr\ in the calibration model rather than \snrw\ to help reduce the level of the 
selection bias in each bin, but we found that it does not perform as well as using the standard \snrw. 
Perhaps not surprisingly, the noise bias seems to be more related to the \snr\ of the actual galaxy than 
it is to the counterfactual round version of the galaxy used for \snrr.  
In future work, it would be interesting to seek an effective shear calibration scheme
that disentangles noise and selection biases, but we have not found one yet.

We used these fits to estimate the multiplicative and additive corrections to use for every galaxy in the 
\imshape\ catalogue.  However, it should be stressed that this bias estimate is itself a noisy 
quantity, being based on noisy estimates of the size and \snr.  Therefore one should not
directly apply the correction to each galaxy individually.  Rather, the mean shear of an ensemble
of galaxies should be corrected by the mean shear bias correction of that same ensemble
(cf. \S\ref{sec:cats:calib}).

Note that a selection bias can appear whenever a subset of galaxies is selected from a larger sample.
In the cosmological analysis, we apply recommended \imshape\ flags, 
cut on \rgp\ and \snrw, and then typically split the galaxies into redshift bins.
The redshift selection in particular is not used in the shear calibration process, so it is 
possible for there to be uncorrected selection biases in the different redshift bins.
In \S\ref{sec:tests:sims}, we test that the shear calibration nevertheless performs well in this scenario
by applying the same selection procedure to the \greatdes\ simulation.
There we demonstrate that all biases are removed to the required tolerance level
in all redshift bins.

\subsubsection{Diagnostics}
\label{sec:im3shape:diagnostics}
\assign{Joe}
\contrib{Troxel}

After performing the shape measurement, we generated a large suite of diagnostic information
based on the results of the fits to help identify objects that potentially should not be used for weak lensing.
Many objects showed evidence of imaging artefacts or some other problem that violates the assumptions 
we have made in the model, so we wanted to be able to remove these objects from the final shear catalogue.

We distinguished two types of flags: ``error'' flags, which identify objects that
should definitely be removed from any analysis, and ``info'' flags, which identify objects
that may be somewhat contaminated, but which may have some value depending on the science 
application.  Most of the 
info flags are derived by examining histograms of the relevant parameters and flagging extreme tails.

The full listing of \imshape\ flags is given in \app{app:im3shapeflags}.
In \S\ref{sec:cats:flags}, we will detail the final selection criteria that we recommend for
the \imshape\ catalogue,
which will include both \code{ERROR\_FLAG==0} and \code{INFO\_FLAG==0}.
Moving to a less restrictive selection should only be done after carefully testing for the possibility of
increased systematic errors.

\subsubsection{Galaxy weights}
\label{sec:im3shape:weights}

We assigned a weight to each shear measured by \imshape\ based on an estimate of the total
shear uncertainty including both shape noise $\SN$ (the standard deviation of the intrinsic ellipticities)
and measurement uncertainty $\sigma_e$:
\begin{equation}
w = \frac{1}{\SN^2 + \sigma_e^2}.
\end{equation}

\edit{The \LevMar\ Levenberg-Marquardt implementation we used produces an estimate of the
parameter covariance for each galaxy as a by-product of optimization,
but we did not use this estimate to give us weights, for two reasons.
First, we found it to show rather wide scatter when compared to MCMC tests,
often showing spurious parameter covariances.  Second, our physical parameters
are a non-linear function of the numerical parameters in some regimes,
as discussed in \citet{im3shape}.}

To estimate the appropriate weight for each galaxy, we \edit{instead}
used the measured shears from the
\greatdes\ simulation.
We grouped galaxies in bins of \snrw\ and \rgp.
We then measured the width of the distribution of ellipticities in each bin, 
both by fitting a Gaussian to a histogram of the distribution
and by measuring the sample variance directly.
The larger of the two variance estimates was taken, and the weight was then given by the inverse variance.
%MJ: This seems to imply the opposite of the reason Tomek did this, so I think better to just remove it.
%This twin scheme handles both the high 
%signal-to-noise regime, where the distributions are close to Gaussian, and the low signal-to-noise regime,
%where they are close to uniform across the ellipticity plane.

We also imposed a maximum weight set by the mean variance of all high-\snr\ bins.  Otherwise
spuriously low variance estimates in some sparsely populated bins 
resulted in very high weight values for those bins.

\subsection{Shear Measurements with \ngmix}
\label{sec:ngmix}
\assign{Erin}

The code \ngmix\ is a general tool for fitting models to astronomical images
\citep{Sheldon2014}.  The code is free
software\footnote{\url{https://www.gnu.org/philosophy/free-sw.html}}, and is
available for download\footnote{\url{https://github.com/esheldon/ngmix}}.

In \ngmix\, both the PSF profile and the galaxy are modeled using mixtures of
Gaussians, from which the name \ngmix\ is derived.  Convolutions are thus
performed analytically, resulting in fast model generation as compared to
methods that perform the convolution in Fourier space.

\subsubsection{Exponential \disk\ model}
\label{sec:ngmix:model}

For the galaxy model, \ngmix\ supports various options including exponential
\disk s, \devauc\ profiles \citep{devauc1948}, and \sersic\ profiles
\citep{Sersic63}, all of which are implemented approximately as a sum of
Gaussians using the fits from \citet{HoggLangGMix}.  Additionally, any number
of Gaussians can be fit, either completely free or constrained to be co-centric
and co-elliptical.  For the DES SV catalogues, we used the exponential \disk\ 
model for galaxies.

Using this simple \disk\ model resulted in detectable model bias (cf.
\S\ref{sec:tests:sims}). In simulations, we found this model bias was reduced
when using a more flexible model, but the more flexible model was not
implemented for real survey data in time for this release.  We will explore
improved modeling in detail for future DES analyses.

We constructed the model in celestial coordinates and fit
it to multiple epochs and bands simultaneously (cf.~\S\ref{sec:shear:multiepoch}).
The centre, size and ellipticity were set to be the same
for all bands and epochs, but the flux was allowed to vary between bands.  For
this study we combined bands \ngmixbands, resulting in eight free parameters:
\begin{itemize}

    \item $u_c,v_c$, the object centre in celestial coordinates, relative to the fiducial
        centre from the \coadd\ object catalogue. The units are arcseconds.

    \item $e_1,e_2$, the ellipticity.

    \item $T$, the area of the object, defined in terms of the unweighted
        moments of the Gaussian mixture $T = \langle x^2\rangle + \langle y^2
        \rangle$.  The units are arcseconds squared.

    \item $F_k$, the flux in each of the \ngmixbands\ bands.

\end{itemize}

\subsubsection{Image fitting}
\label{sec:ngmix:fitting}

The \ngmix\ code supports multiple paradigms, all of which were used in the
current analysis:
\begin{itemize}

    \item Exploration of the full likelihood surface for a given set of model
        parameters with a Markov Chain Monte Carlo (MCMC) scheme, using either the standard
        Metro\-polis-Hastings algorithm \citep[MH;][]{Metropolis53} or the
        recently introduced affine invariant method
        \citep{GoodmanWeare10,Mackey13}.  The model can be fit directly to the
        pixel data, or it can include convolution by a point-spread function
        (PSF).

    \item Maximum-likelihood fitting using
        any of a variety of function minimizers.
        We used \levmar\ \citep[LM;][]{Levenberg1944, Marquardt1963}
        as well as the method of \citet{NelderMead1965} (NM) in this work.
        The model can be fit directly to the pixel data, or it can include
        convolution by a PSF.

    \item Expectation Maximization \citep[EM; ][]{Dempster1977}, fitting
        directly to the pixels only. This method is used for PSF fitting.

\end{itemize}

For PSF measurement, the EM code was used, with three completely
free Gaussians.  EM is a good choice for PSF
measurement, since it is extremely stable even with many components.  By
allowing all components to be completely free, the off-centre PSF components that
are occasionally found in the SV PSF images were fitted without instability.  

We chose to handle the WCS information by projecting each pixel into celestial
coordinates and building both the galaxy and PSF models in that coordinate system.
%This is different from how \imshape\ handles the WCS information, but the
%two approaches are mathematically equivalent.

Our procedure for fitting the galaxy shapes involves a number of
steps:\footnote{We
tried using the affine invariant fitter, and found it to be very robust, 
but the burn-in period was too slow for large-scale processing.  This hybrid approach
using both maximum-likelihood fitters and MH is significantly faster and 
sufficiently accurate.}
\begin{enumerate}

    \item Estimate a flux for the object by fitting the PSF model to the galaxy
        with a single free parameter, which is the overall normalization
        (keeping the centroid fixed at its fiducial value).

    \item 
        Run NM to find the maximum likelihood model,
        guessing the flux from the result of step 1, and guessing
        the size to be the typical seeing size.  We find
        NM to be more robust than LM for this fit.
        
    \item 
        Run LM starting from the maximum likelihood model to
        estimate the covariance matrix, since NM does not produce one. 
        Relatively few evaluations are made in this step.

    \item 
        Run an MCMC chain with MH
        using the maximum likelihood position as a starting guess and the
        covariance matrix as a proposal distribution. 
        We run a few thousand burn-in steps, followed by
        a few thousand post-burn-in evaluations.  If the acceptance rate is
        outside the range $[0.4,0.6]$ we reset the proposal distribution based
        on the covariance matrix from previous MH run, and run a new burn-in
        and post-burn in.  If the acceptance rate remains outside the desired
        range we try again up to four times.  These bounds on the
        acceptance rate are somewhat arbitrary, but for our problem 
        we found that rates above 0.6 result in highly correlated chains, and
        lower than 0.4 can result in a poorly sampled peak.

\end{enumerate}

\subsubsection{Shear estimation}
\label{sec:ngmix:shear}

Multiple methods are supported for shear measurement, but for this study we
adopted the ``\lensfit''-style method, based on the work of \cite{Miller2007}.
We found our implementation of this method to be sufficiently accurate for the
precision of our current data set; for this study \ngmix\ measurements were
instead limited by the use of an overly simple exponential \disk\ model for
galaxies (cf. \S\ref{sec:tests:sims}).

The \lensfit\ method involves multiplication by a prior on the distribution of
galaxy ellipticities when estimating the expectation value of the ellipticity
for each galaxy.
\begin{align}
    \langle e_\mu \rangle &=  \frac{\int \mathcal{L}(\vece) p(e) e_\mu \mathrm{d} \vece}{\int
    \mathcal{L}(\vece) p(e) \mathrm{d} \vece} \\
                          & \simeq  \frac{\sum_j p(e^j) e^j_\mu}{\sum_j p(e^j)}, 
\end{align}
where $\mathcal{L}(\vece)$ is the likelihood and $p(e)$ is the prior on the galaxy
shapes. We approximate the integral over the likelihood with the sum of points
from an MCMC chain. The index $\mu$ takes values 1,2 for each ellipticity
component such that $\vece = (e_1,e_2)$; the ellipticity magnitude is given by $e$.

Multiplying by an ellipticity prior reduces the effects of noise, which
broadens and distorts the likelihood surface. However, application of the prior
also biases the recovered shear, in effect reducing the ``sensitivity'' of the
shear estimate.  \cite{Miller2007} derive a measure of the sensitivity of this
estimator to a shear $\vecg$, which is approximately given for each component
by:
\begin{align} \label{eq:lensfitsens}
    s_\mu \equiv
      \frac{\partial \langle e_\mu \rangle}{\partial g_\mu} 
      & \simeq  1 - \left[ \frac{\int \left( \langle e_\mu \rangle - e_\mu \right) \mathcal{L}(\vece)
      \frac{\partial p}{\partial e_\mu} \mathrm{d} e}{\int \mathcal{L}(\vece) p(e) \mathrm{d} e} \right] 
      \nonumber \\
      & \simeq  1 - \left[ \frac{ \sum_j \left( \langle e_\mu \rangle - e^j_\mu \right)
      \frac{\partial p}{\partial e_\mu} }{\sum_j p(e^j)} \right].
\end{align}

No expression was formally derived by \cite{Miller2007} for the mean
of the shear field acting on
an ensemble of galaxies; however, it was proposed to use the same formula
as derived for a constant applied shear:
\begin{equation}
    g_\mu = \frac{ \sum_i \langle e^i_\mu \rangle }{\sum_i s^i_\mu},
\end{equation}
where the index $i$ runs over all galaxies in the measurement.  In practice
we also apply weights in both sums,
\begin{equation}
    w = \frac{1}{2\SN^2 + C_{1,1} + C_{2,2}},
\end{equation}
where $\SN$ is the shape noise per component, which we have calculated
to be \ngmixSN\ based on fits to COSMOS galaxies (cf. \S
\ref{sec:ngmix:priors}), and $C_{i,j}$ are elements of
the 2x2 ellipticity subset of the covariance matrix
produced by \ngmix.

The sensitivities in \eqn{eq:lensfitsens} do not transform as
polarizations.  Thus for practical shear measurements, such as tangential shear
or two-point functions, which require rotation of the ellipticities, we chose
to use a scalar sensitivity for each galaxy that is the mean of its two
components.

\subsubsection{Ellipticity prior}
\label{sec:ngmix:priors}
\assign{Erin}

The \lensfit\ method requires as input a prior on the shapes of galaxies, $p(e)$.
The prior must be continuous for $e_1,e_2$ in the unit circle in order to evaluate
the derivatives in equation \ref{eq:lensfitsens}.

In simulations, we found that the accuracy of the shear recovery was sensitive
to the details of the ellipticity prior.  For example, we ran the shear code on
the \greatdes\ simulation presented in \S\ref{sec:sims:greatdes} using a prior with
intrinsic variance in ellipticity 35\% higher than the true variance, and found
the multiplicative bias increased by $(1.3 \pm 0.2)$\%.

For application to real data, we based our prior on the ellipticities of \sersic\ model fits to
COSMOS galaxies, as released by the GREAT3 team \citep{great3handbook}.  We
fit the observed distribution to a simple model
\begin{align}
p(e) &= 2 \pi e A ~ \frac{ \left( 1-\mathrm{exp}\left[\frac{e-1}{a}\right] \right)}{(1+e) \sqrt{e^2 +
e_0^2}} ~ c(e) \\
c(e) &= \frac{1}{2} \left(1 + \mathrm{erf}\left[ \frac{e_{cut}-e}{\sigma_e}\right]\right).
\end{align}
This model is a modified version of that introduced in \cite{Miller2013}. Note
in particular the cutoff at high ellipticities achieved by using an error function. We found
this formula improved the fit to the distribution of ellipticities.

\begin{figure}
\center
\includegraphics[width=\columnwidth]{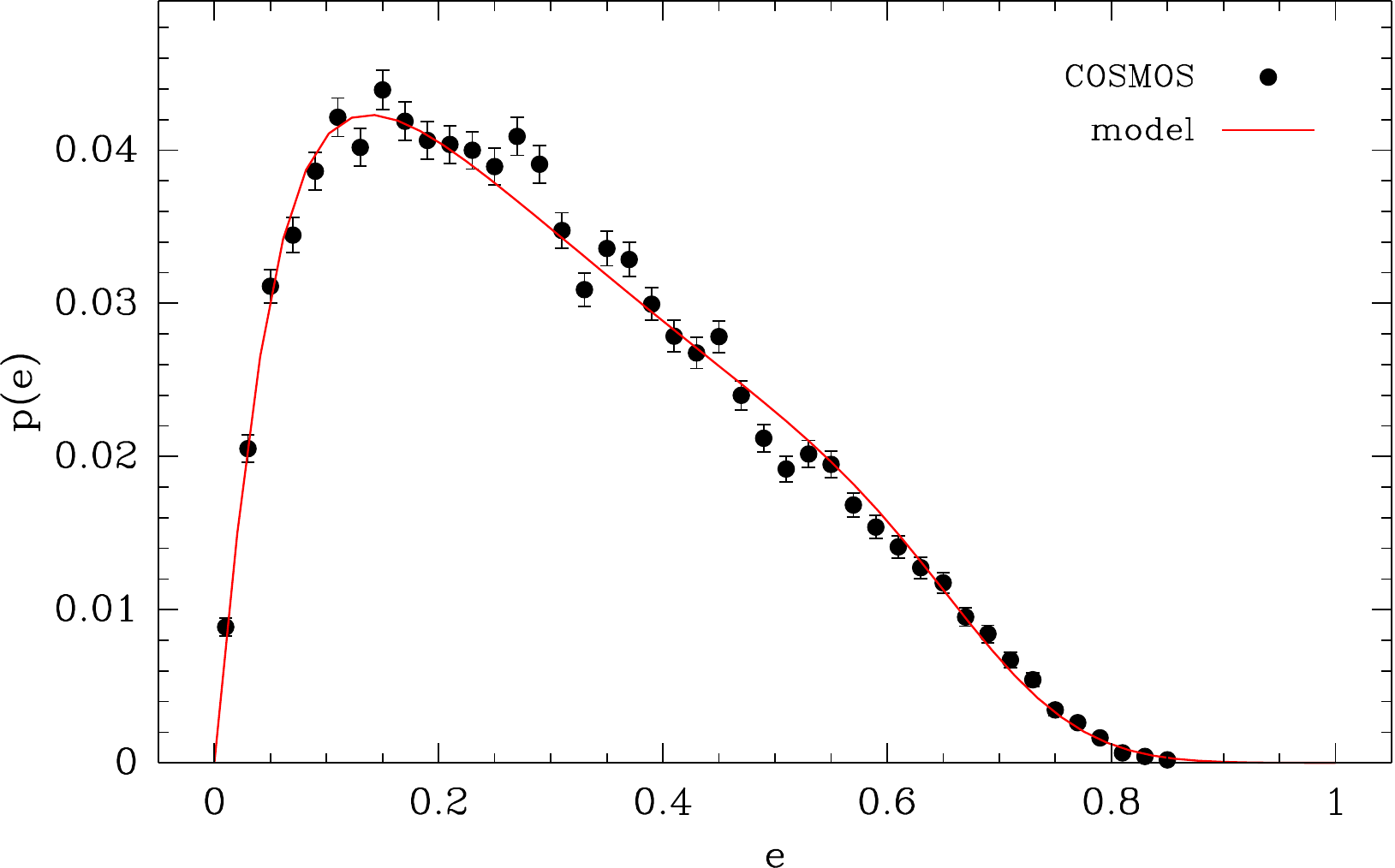}
\caption{Distribution of shapes for COSMOS galaxies, selected
as described in the text.  The model fit was used as a prior
for the \ngmix\ shear analysis.
\label{fig:ngmix:prior}
}
\end{figure}

We fit this model to the ellipticities of COSMOS galaxies selected to fall in a 
range of size and flux that corresponds to the galaxy population seen in our data.
A comparison between the measured $p(e)$ and the fit model is shown in
\fig{fig:ngmix:prior}.  The best-fitting parameters are given in
\tab{tab:ngmix:priorpars}.

\begin{table}
\centering
\begin{tabulary}{1\columnwidth}{C C}
\hline
Parameter & Fit Value \\
\hline
$A$ &  $0.025 \pm 0.002$ \\
$a$ &  $2.0 \pm 0.2$    \\
$e_0$ & $0.079 \pm 0.003$  \\
$e_{cut}$ & $0.706 \pm 0.004$ \\
$\sigma_e$ &  $0.125 \pm 0.006$ \\
\hline
\end{tabulary}
\caption{Parameters for the ellipticity prior used with the \ngmix\ 
shear code, with best-fitting values for the distribution
of shapes of COSMOS galaxies.
\label{tab:ngmix:priorpars}
}
\end{table}

We used this same prior for all galaxies, but the distribution of COSMOS galaxy
shapes depends on redshift.  However, for this study we found that the
uncertainties due to the redshift dependence of the shape distribution were sub-dominant
to model bias for \ngmix\ (cf. \S\ref{sec:tests:sims}).

\subsection{Blinding}
\label{sec:blinding}

We blinded the shape catalogues from both pipelines before they were used for any
tests or SV science papers.  We did this to prevent the 
\emph{experimenter bias} effect, wherein researchers work harder on finding
bugs, tuning methodology, etc.~when results are inconsistent with a previous
experiment, or otherwise do not match expectations, than when they do match
\citep[cf.][]{Klein05}.

We blinded the SV shear catalogues by scaling all measured shears by a secret
factor generated by an algorithmic, but unpredictable, process (using an MD5
hash of a code phrase) to be between $0.9$ and $1.0$.  This unknown scaling
meant that it was harder for DES members to, for example, accidentally tune
results to get the $\sigma_8$ value predicted by Planck.  We only unblined the
catalogues after the analysis for a given paper was finalized.

This was a gentle blinding approach that was appropriate for the relatively
loose statistical constraints that will come from SV data.  It has the useful
feature that, being linear, correlation tests on it such as those listed in
this paper remain valid. It has a significant downside in that it is asymmetric -
unblinding could only increase the measured $\sigma_8$, so the potential for
bias was still present.  We will consider new blinding methodologies for future data.

\section{Tests of the Shear Measurements}
\label{sec:tests}
\assign{Mike}

We developed an extensive test suite to check that the shear catalogues do not
have significant systematic errors that would adversely affect weak lensing science.
While there is no way to definitely prove that the shear catalogues are free of all
possible systematic errors, there are many tests that can reveal 
systematic errors that might be present in the data.  These tests were formulated
as ``null tests'', which should have zero signal in the absence of systematic errors.
Most of our null tests were similar to ones that have been performed in
previous analyses \citep[cf.~e.g.][]{Jarvis03, Schrabback10, Velander11, Heymans12, Jee13, Kuijken15}.

These null tests can be broken up into several broad categories.
\begin{enumerate}

\item \textbf{Spatial tests} to check for systematic errors that are connected to the physical
    structure of the camera.  Examples of these are errors in the WCS
    correction, including effects like edge distortions or tree rings
    \citep{plazas14}, and errors related to features on the CCDs such as the
    tape bumps.
    (\S\ref{sec:tests:spatial})

\item \textbf{PSF tests} to check for systematic errors that are connected to the PSF correction.
    This includes errors due to inaccurate PSF modeling as well as
    leakage of the PSF shapes into the galaxy shape estimates.
    (\S\ref{sec:tests:psf})

\item \textbf{Galaxy property tests} to check for
    errors in the shear measurement algorithm related to properties of the
    galaxy or its image.  This can include effects of masking as well, which
    involve the other objects near the galaxy being measured.
    (\S\ref{sec:tests:gal})

\item \textbf{B-mode statistics} to check for systematic errors that show up as a B-mode signal in the
    shear pattern.  The gravitational lensing signal is expected to be
    essentially pure E-mode.
    Most systematic errors, in contrast, affect the E- and B-mode
    approximately equally,
    so the B-mode is a direct test of systematic
    errors.
    (\S\ref{sec:tests:bmode})

\item \textbf{Calibration tests} to check for systematic errors that affect the overall
    calibration of the shears.  If all of the shear values are scaled by a
    constant factor, most null tests remain zero (if they were zero to start
    with).  Furthermore, there are no known absolute shear calibration sources
    that we can use to calibrate our results. For these reasons it can be hard
    to tease out errors in the calibration from the data.  However, we can use
    simulated data where the true shear is known to check that we recover the
    correct values.
    (\S\ref{sec:tests:sims})

\item \textbf{Cross-catalogue comparisons} to check that the two shear catalogues are consistent
    with each other.  Because we have two shear catalogues available for testing,
    we can check that the two give consistent results, thus potentially uncovering
    problems that may be in one shear catalogue but not the other (or have
    different levels in each).  Considering the large differences between the
    \ngmix\ and \imshape\ codes, these are non-trivial tests.
    (\S\ref{sec:tests:cross})
\end{enumerate}

%MJ: I wish LaTex were smarter about placing these figure* floats.  If I put this closer to where 
%    it actually gets referenced, LaTex puts it after figure 15.
\begin{figure*}
\includegraphics[width=0.48\textwidth]{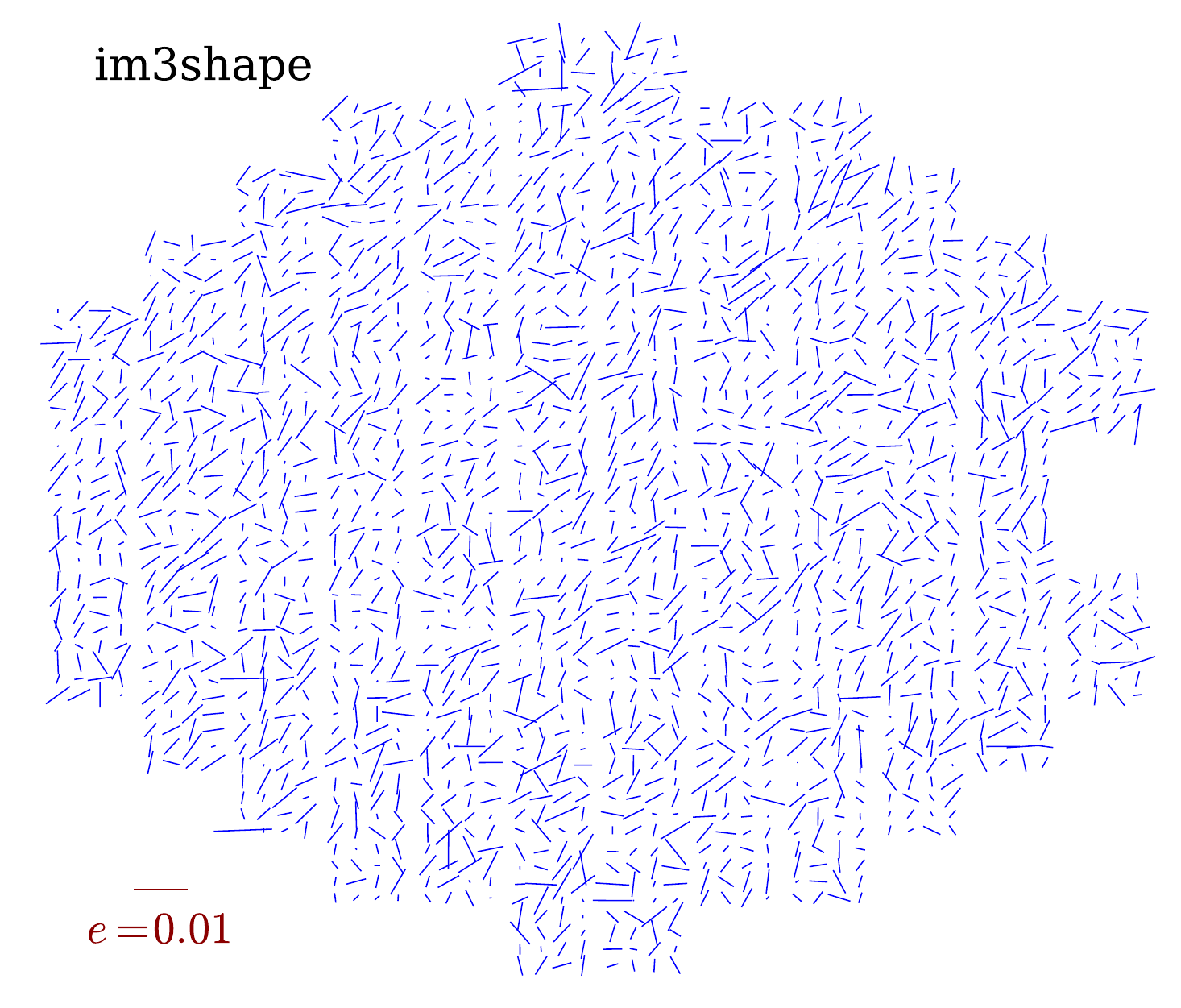}
\includegraphics[width=0.48\textwidth]{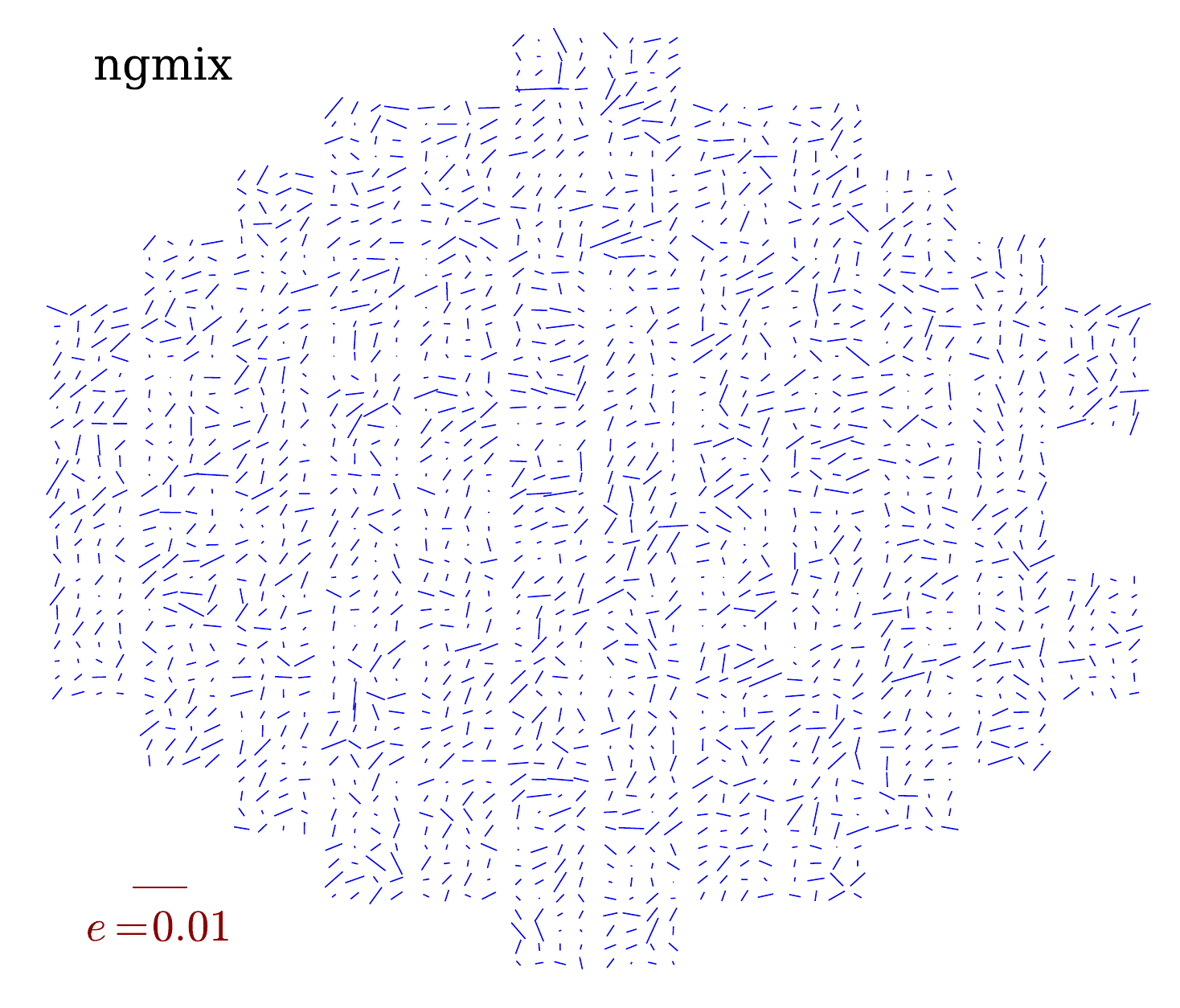}
\caption{Whisker plots of the mean shear binned by position in the focal plane for 
\imshape\ (left) and \ngmix\ (right).
As in \fig{fig:psf:fov}, the length of each whisker is proportional to the mean shear, and the 
orientation is aligned with the direction of the ellipticity.  A 1\% whisker is shown for scale
in the lower left.
\label{fig:tests:fov}
}
\end{figure*}

One caveat to keep in mind with the various null tests is that we do not necessarily expect
the overall mean shear to be precisely zero.  The SV region is small enough that the
rms value of the mean shear due to cosmic variance is expected to be about $4 \times 10^{-4}$.
In fact, the overall mean shear is measured to be
\begin{align*}
\imshape \quad&
\begin{array}{l}
\langle e_1 \rangle = 0.1 \times 10^{-4}\\
\langle e_2 \rangle = 6.8 \times 10^{-4}
\end{array}
\\
\ngmix \quad&
\begin{array}{l}
\langle e_1 \rangle = -0.4 \times 10^{-4}\\
\langle e_2 \rangle = 10.2 \times 10^{-4}.
\end{array}
\end{align*}
These values are both about $2\sigma$ from zero given the expected cosmic variance, so 
it may be due to an additive systematic error affecting both codes. 
However, the fact that they roughly agree with each other suggests at least the possibility that it could
be a real cosmic shear signal.
In any case, each of the 
null tests look for variations \emph{relative} to this overall mean shear to find
dependencies that may indicate systematic errors.

In \S\ref{sec:tests:spatial} -- \S\ref{sec:tests:cross} we show the results of our null tests in each of
the above categories.
In \S\ref{sec:tests:summary} we summarize these results and tries to quantify the total possible
systematic errors that may be present in the shear catalogues.

\subsection{Spatial Tests}
\label{sec:tests:spatial}
\assign{Mike}
\contrib{Rutu, Donnacha}

There are many potential sources of systematic error
related to the camera and telescope optics that can cause a
spatial dependence of the shear with respect to the camera's
field of view. The telescope
distortion pattern and some of the optical aberrations
are essentially static in time. The CCDs have bad
columns and other defects, including the tape bumps
mentioned in \S\ref{sec:psf:cuts}. There are also distortion
effects at the CCD edges due to the electric field lines
becoming non-parallel as well as tree ring distortion patterns due to
doping variations in the silicon \citep{plazas14}.

\subsubsection{Position in the field of view}
\label{sec:tests:fieldofview}

To check that we have adequately corrected for effects that are connected with the 
telescope and camera, or that they are small enough to ignore, we binned the shear spatially
with respect to the field of view.

In \fig{fig:tests:fov} we show the mean shear as a function of position on the focal plane for 
both \imshape\ (left) and \ngmix\ (right).
Each whisker represents the mean shear of all galaxies that were ever observed in that area of the 
focal plane.  As our shear measurements used information from multiple epochs,
each measurement contributed to this plot multiple times: once for each \SE\ observation of that galaxy.

This figure is similar to \fig{fig:psf:fov}, in which we showed the residual PSF pattern
as a function of position on the focal plane.  These plots are noisier due to
the shape noise of the galaxies, but there is a hint of the same radial patterns
that were seen for the PSF residuals, especially in the \ngmix\ results, which are slightly less
noisy due to the higher number of galaxies in the catalogue.  This is not surprising; we expected
these PSF interpolation errors to leak into the galaxy shapes.

\subsubsection{Position on CCD}
\label{sec:tests:ccd}

\begin{figure}
\includegraphics[width=\columnwidth]{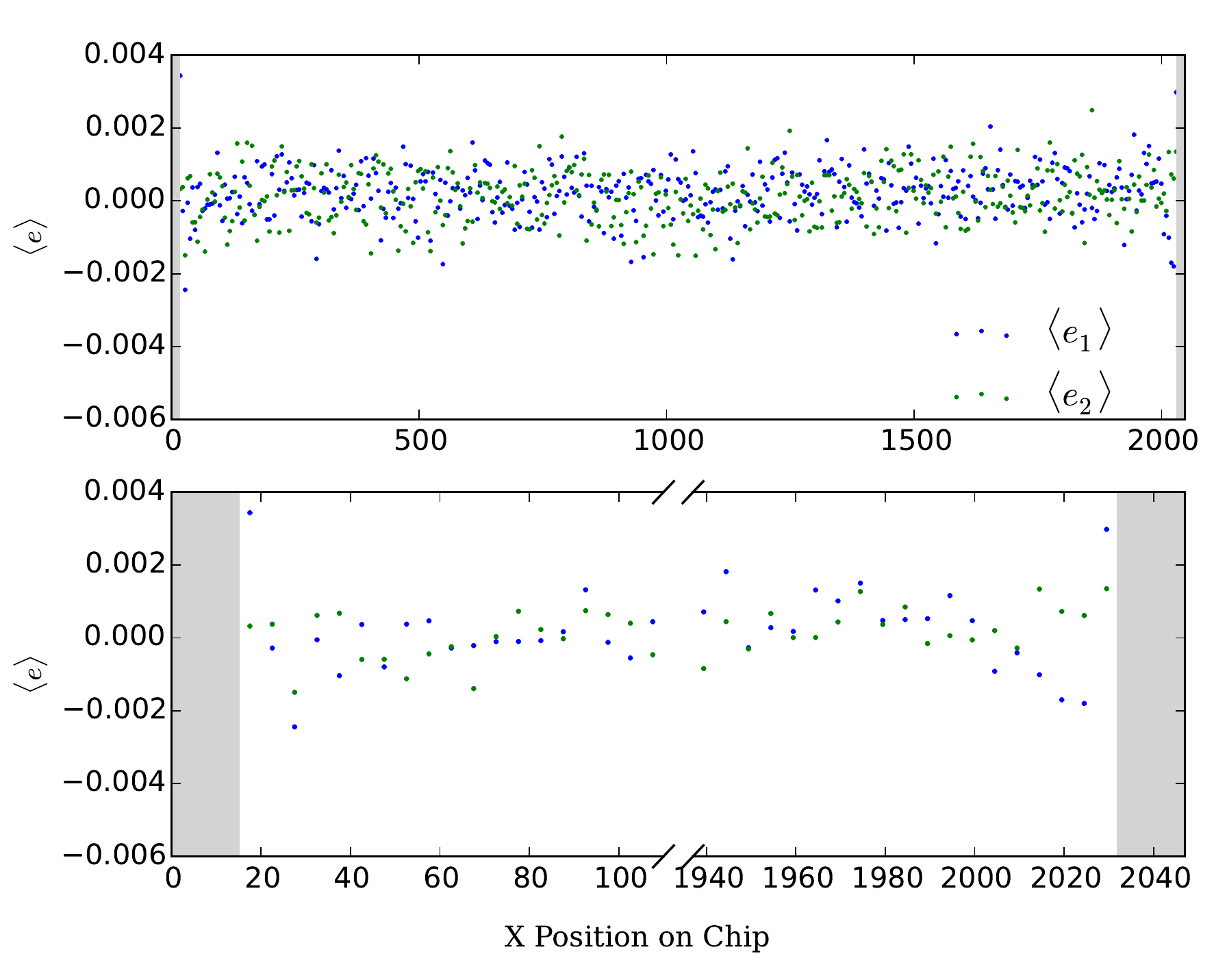}
\caption{The mean shear $\langle e_1\rangle$ (blue) and $\langle e_2 \rangle$ (green)
for \ngmix, binned by column number $X$; $X$ runs along the
readout rows of the CCDs.  (The corresponding
plot for \imshape\ looks similar but noisier.)
The bottom panel shows the same data blown up near the left and right edges to highlight
the effect of the edge distortion.
We mask the 15 columns along each edge where the distortion is strongest, but 
there is still a slight bias in the $e_1$ component of the shears up to 40 pixels from the edge.
\label{fig:tests:colnum}
}
\end{figure}

If we bin the shears by their column
number, irrespective of the CCD number, as shown in \fig{fig:tests:colnum}, we can see the
effect of something known as ``edge distortion'' \citep{plazas14}.  This is where the electric field lines in the
detector become slightly non-parallel near the edges of the CCDs.  The cross section of the pixels becomes
rectangular, elongated in the direction towards the edge of the CCD.
\citet[their Figure~6]{plazas14} showed that this effect led to photometric biases of $\sim20$ mmag
at $\sim30$ pixels from the edge of the CCDs.
Since flux and shape respond to the astrometric variation at the same order, this implies that we should
expect shape residuals of about $\delta e_1 \sim 0.02$ near the edge of the CCDs.

In \fig{fig:tests:colnum} we show the mean shape measured by \ngmix\ binned by column number.
As we do not measure the \SE\ shape, any effect on the shapes has been reduced by a factor of 
about 10, the number of \SE\ exposures of each galaxy.  So we might expect a signal of
$\langle e \rangle \sim 0.002$.
There does seem to be a slight effect visible in \fig{fig:tests:colnum} at this level
for $e_1$, although it is not highly significant.  
The effect of the edges is even less evident when binning by the row number (not shown).

To quantify how much this edge effect might impact the overall shear signal,
we estimated that the effect is only significant for about 20 pixels on any edge.  This is 
a fraction of $40/2048 + 40/4096 = 0.015$ of the area.  Galaxies have $\sim$ 10 chances to 
fall in this area, so about 15\% of the galaxies may have a spurious shear of $\sim 0.002$.
The net additive systematic shear
from this effect is thus about $c_\mathrm{rms} = 8 \times 10^{-4}$.  This is well below the requirements on
additive systematic errors given by \eqn{eq:req:c}, $c_\mathrm{rms} < \crequirement$;
however, 
it is not below the expected requirements for DES 5-year data.  Therefore, we
plan to remove this effect directly in the astrometry solution in future DES data analyses.

\subsubsection{Tangential shear around field centres}
\label{sec:tests:fieldcenters}

The telescope distortion pattern is approximately a fifth order radial function centred near the 
centre of the field of view.  If it is not corrected it can induce spurious shears oriented either radially
or tangentially relative to the field centres.  We looked for this effect by measuring the tangential
shear pattern around the set of field centres; essentially this is similar to a galaxy-galaxy lensing
measurement where the telescope pointings play the role of the lenses.  

\begin{figure}
\includegraphics[width=\columnwidth]{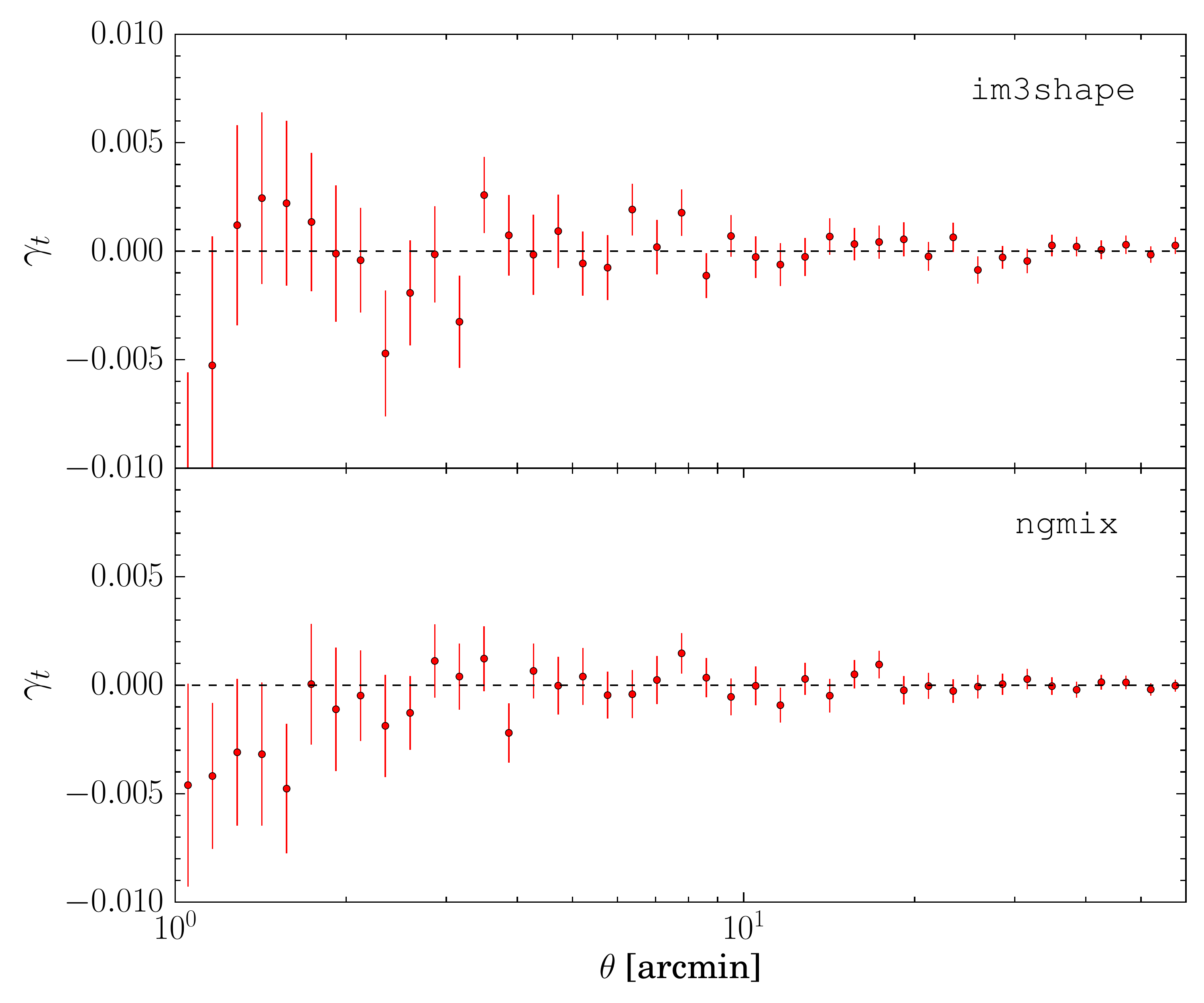}
\caption{The tangential shear of galaxies in \imshape\ (top) and \ngmix\ (bottom) around field 
centres. Both 
measurements are approximately consistent with zero, although at scales less than $10$ arcminutes, both
show a slight departure from the expected null signal. The magnitude of this
effect is well below our requirements in both cases.
\label{fig:tests:fieldcenter}
}
\end{figure}

In \fig{fig:tests:fieldcenter} we show the results of this test for both the
\imshape\ and \ngmix\ shear measurement pipelines.  Uncertainties are jackknife
estimates, made by splitting the total area into $152$ equal-area sub-fields.
At large scales, the measurements are consistent with zero, but at
scales less than about $10$ arcminutes there are a few consecutive bins with
$\sim1\sigma$ deviation from zero in both
cases.  The \imshape\ results show a slight oscillating pattern, and the \ngmix\
results are slightly negative (a radial shear pattern).

None of these features is highly significant, especially since the points are somewhat
correlated, so it may just be a noise fluctuation.
Also, since the telescope distortion is largest at the edge
of the field of view, we expected the absolute mis-estimation of the
distortion to be largest at a separation of around $1$ degree.  Furthermore,
\imshape\ and \ngmix\ use exactly the same WCS solution, since it is
incorporated into the MEDS files directly. So the fact the tangential shear
patterns are different in the two cases, and most significant near the centre,
indicates that this is probably not due to errors in the WCS solution, 
although we do not have a good hypothesis for a plausible cause.

We estimated the magnitude of this potential additive systematic error in the
same manner as we used above for the edge distortions.
The mean spurious shear in this case has a magnitude of at most
$0.005$ in both cases and occurs over a relative area
of about $(2\arcmin/62\arcmin)^2 = 0.001$. The net additive systematic
shear from this effect is thus at most $c_\mathrm{rms} = 2\times10^{-4}$, well below
our requirements for an additive systematic shear.

We also looked at the shear around the CCD corners.  While there was a very slight hint of a non-
zero signal at small scales, the magnitude was even smaller than the shear around the field centres.

\subsection{PSF Tests}
\label{sec:tests:psf}
\assign{Troxel}
\contrib{Niall, Vinu, Erin, Matt}

If the PSF interpolation is not sufficiently accurate or if the shear algorithm
does not fully account for the effects of the PSF convolution, the resulting shear estimates
will include a spurious additive error that is correlated with properties of the PSF.

We looked for such additive errors by examining: 
(1) the mean shear binned by PSF ellipticity and PSF size, 
(2) the PSF-shear two-point correlation function and derived quantities, and
(3) the tangential shear measured around stars.

\subsubsection{PSF leakage}
\label{sec:tests:alpha}

\begin{figure}
\includegraphics[width=\columnwidth]{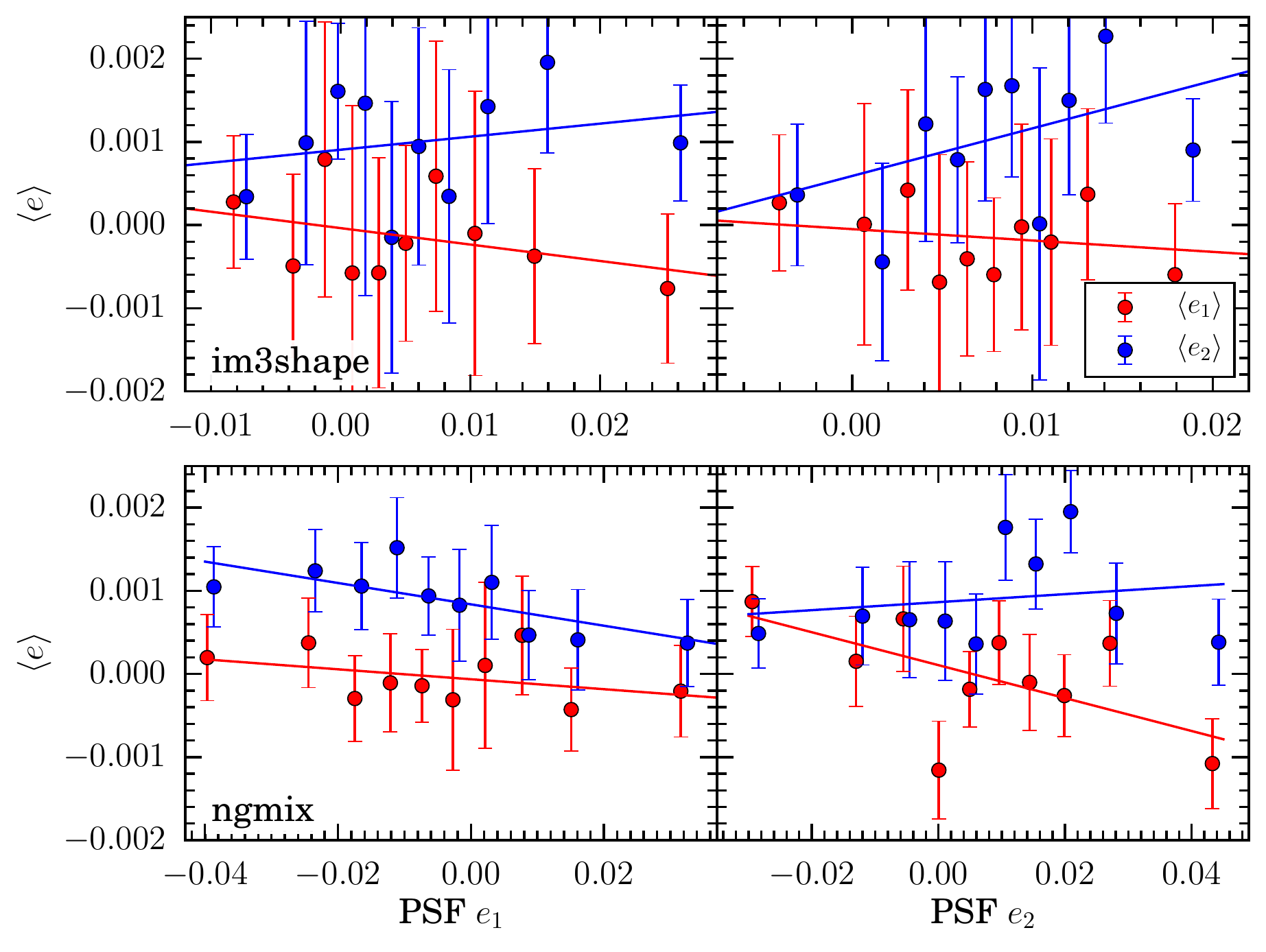}

\caption{The mean galaxy shear as a function of the input PSF ellipticity
    ($e_1$ left, and $e_2$ right) for \imshape\ (top) and \ngmix\ (bottom).
    The solid lines show the best linear fit without binning.  
    Note the range of the abscissa is different for the \ngmix\ and \imshape\
    plots.  The
    \ngmix\ measurements are averaged over $r$, $i$, and $z$-band images, while the
    \imshape\ measurements use $r$-band images, and different models are
    used for measuring the ellipticity, resulting in different PSF ellipticity
    ranges for the two catalogues.  
\label{fig:tests:psfe}
}
\end{figure}

As we introduced in \eqn{eq:req:esys},
we assumed that a component of the additive bias in the shear estimates 
comes from imperfect correction of the PSF, resulting in a term proportional to the PSF shape:
\begin{equation}
e_\mathrm{gal} = e_\mathrm{true} + \alpha \epsf + c. 
\label{eq:eobsalpha}
\end{equation}
A measured slope of galaxy ellipticity vs.~PSF ellipticity can be
identified as $\alpha$, where we use the mean PSF shape over all epochs.

The mean shear as a function of PSF
ellipticity is shown in \fig{fig:tests:psfe} for both \imshape\ and
\ngmix\ galaxies.  The points represent the
mean galaxy ellipticity in each of 10 equal-number bins of PSF ellipticity.
The line represents the best fit to the individual (unbinned) galaxy shapes.
The slopes of the linear fits range from -2.0\% to 5.7\% $\pm$ 3\% for \imshape\ and from
-2.0\% to 0.5\% $\pm$ 1\% for \ngmix.
The slopes are consistent with no PSF leakage for both catalogues.  

To obtain a more precise estimate of $\alpha$, we computed the (weighted) average of the slopes of the 
red lines on the left plots (i.e.~$\langle e_1 \rangle$ vs.~PSF $e_1$) and the blue lines on the
right plots (i.e.~$\langle e_2 \rangle$ vs.~PSF $e_2$).  For \imshape\ we found
$\alpha = 0.008 \pm 0.025$, and for \ngmix\ $\alpha = -0.001 \pm 0.007$.
There is no evidence for non-zero $\alpha$; however, for \imshape, we cannot definitively
confirm that $|\alpha| < 0.03$ (cf. \eqn{eq:req:alpha}) given the uncertainty in the estimate.

\begin{figure}
\includegraphics[width=\columnwidth]{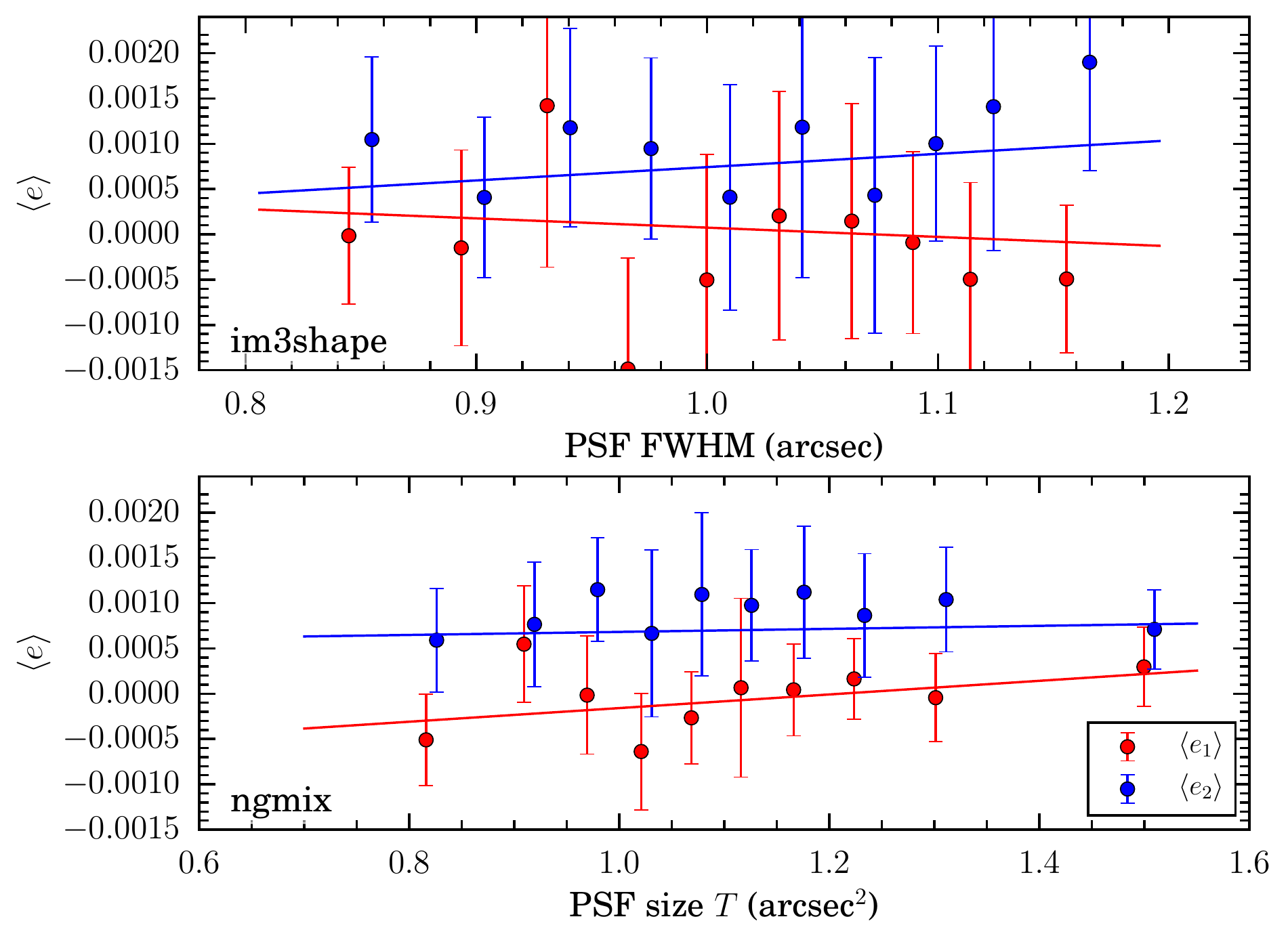}
\caption{The mean galaxy shear as a function of the input PSF size for \imshape\ (top) and \ngmix\
(bottom).  The solid lines show the best linear fit without binning.
\label{fig:tests:psfsize}
}
\end{figure}

We similarly plot the mean shear as a function of PSF size in \fig{fig:tests:psfsize} for both \imshape\ (left) and \ngmix\ 
(right). Linear best-fitting lines
are also included. The slopes here are also consistent with zero, being on 
the order of 0.1\% or less, which indicates negligible dependence of the mean shear on the PSF size. 

\subsubsection{Star/galaxy cross-correlation}
\label{sec:tests:alpha2}

Another estimate of the leakage factor $\alpha$ comes from the cross-correlation of 
the galaxy shapes with the PSF shapes, $\xi_+^{gp}$.  
Writing $\xi_+^{gp}$ in terms of
\eqn{eq:eobsalpha} and solving for $\alpha$, we find that
\begin{equation}
\alpha=\frac{\xi_+^{gp} - \langle e_\mathrm{gal} \rangle^* \langle \epsf \rangle}
{\xi_+^{pp} - |\langle \epsf \rangle|^2},
\label{eq:psfalpha}
\end{equation}
where $\xi_+^{pp}$ is the auto-correlation function of the PSF shapes, \epsf.

While this nominally gives us an estimate of 
$\alpha$ as a function of scale, $\alpha$ is not a scale-dependent quantity.  It is
a quantification of
a point process, the possible leakage of the PSF shape into the galaxy shape estimates.
Therefore, we expect this estimate of $\alpha$ to be consistent at all scales,
given the uncertainties in the estimate.

\begin{figure}
\includegraphics[width=\columnwidth]{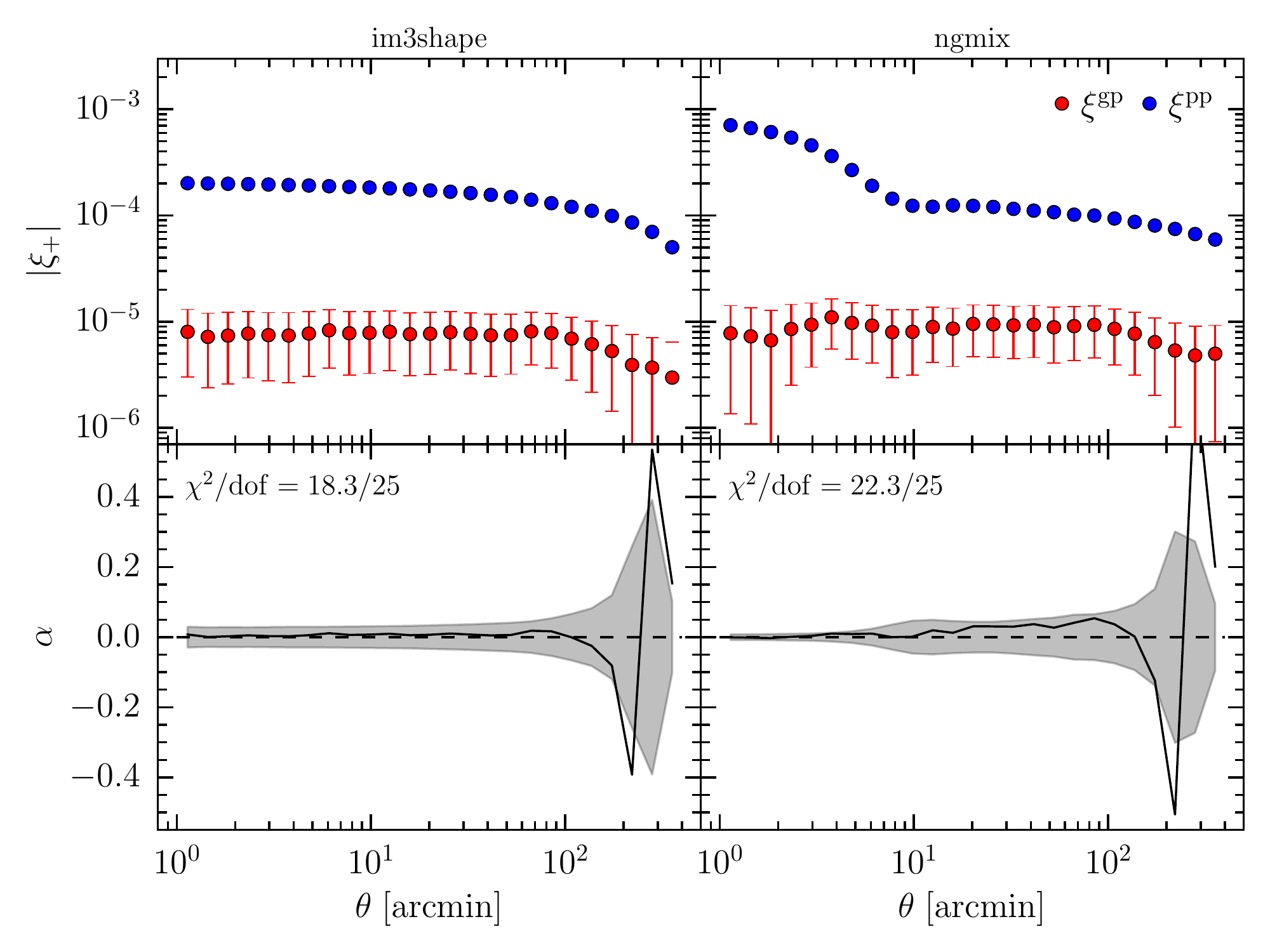}
\caption{The calculation of the PSF leakage parameter $\alpha$, which is given in 
\eqn{eq:psfalpha}.  The top plots show 
$\xi^{gp}$ (red), the cross correlation of the galaxy shapes with the PSF shapes, and 
$\xi^{pp}$ (blue), the auto-correlation of the PSF shapes, for \imshape\ (left) and \ngmix\ (right). 
The bottom plots show $\alpha$, which is a measure of the leakage of the PSF shapes into the 
galaxy shapes as a function of scale. The grey band shows the sample variance plus shape 
noise uncertainty for $\alpha$. The $\chi^{2}/{\rm d.o.f.}$ is given for $\alpha$ over all scales.
\label{fig:tests:alpha}
}
\end{figure}

The measured $\xi_+^{gp}$ and $\xi_+^{pp}$ correlation functions are shown in the top panels of
\fig{fig:tests:alpha} for \imshape\ (left) and \ngmix\ (right). $\alpha$ is then calculated based on 
these and shown in the lower panels. Due to sample variance, $\alpha$ can be non-zero in this test even if 
the measured shears have no PSF contamination. We used the mock catalogues described in \citet{Becker15}
to compute the total uncertainty for $\alpha$. These catalogues were populated with PSF shapes by using the 
PSF shape from the nearest observed galaxy to each mock galaxy. We then used
the full suite of 126 mock catalogues to compute the total uncertainty on $\alpha$ including both
shape noise and sample variance.

We found that both \imshape\ and \ngmix\ show no significant PSF contamination in this test, with a total 
$\chi^{2}/\mathrm{d.o.f.}$ of 18.3/25 and 22.3/25 for $\alpha$ computed over all scales. 
The best-fitting value for $\alpha$ in each case, properly taking into account the correlations
\citep{Avery96}, is
$\alpha = 0.010 \pm 0.023$ for \imshape\ and $\alpha = -0.008 \pm 0.006$ for \ngmix,
both below the requirement of $|\alpha| < 0.03$ from \eqn{eq:req:alpha}, 
although in the case of \imshape\ we are only able to constrain $|\alpha|$ to be less than 0.03
at about $1\sigma$.

\subsubsection{Tangential shear around stars}
\label{sec:tests:ggstars}

\begin{figure}
\includegraphics[width=\columnwidth]{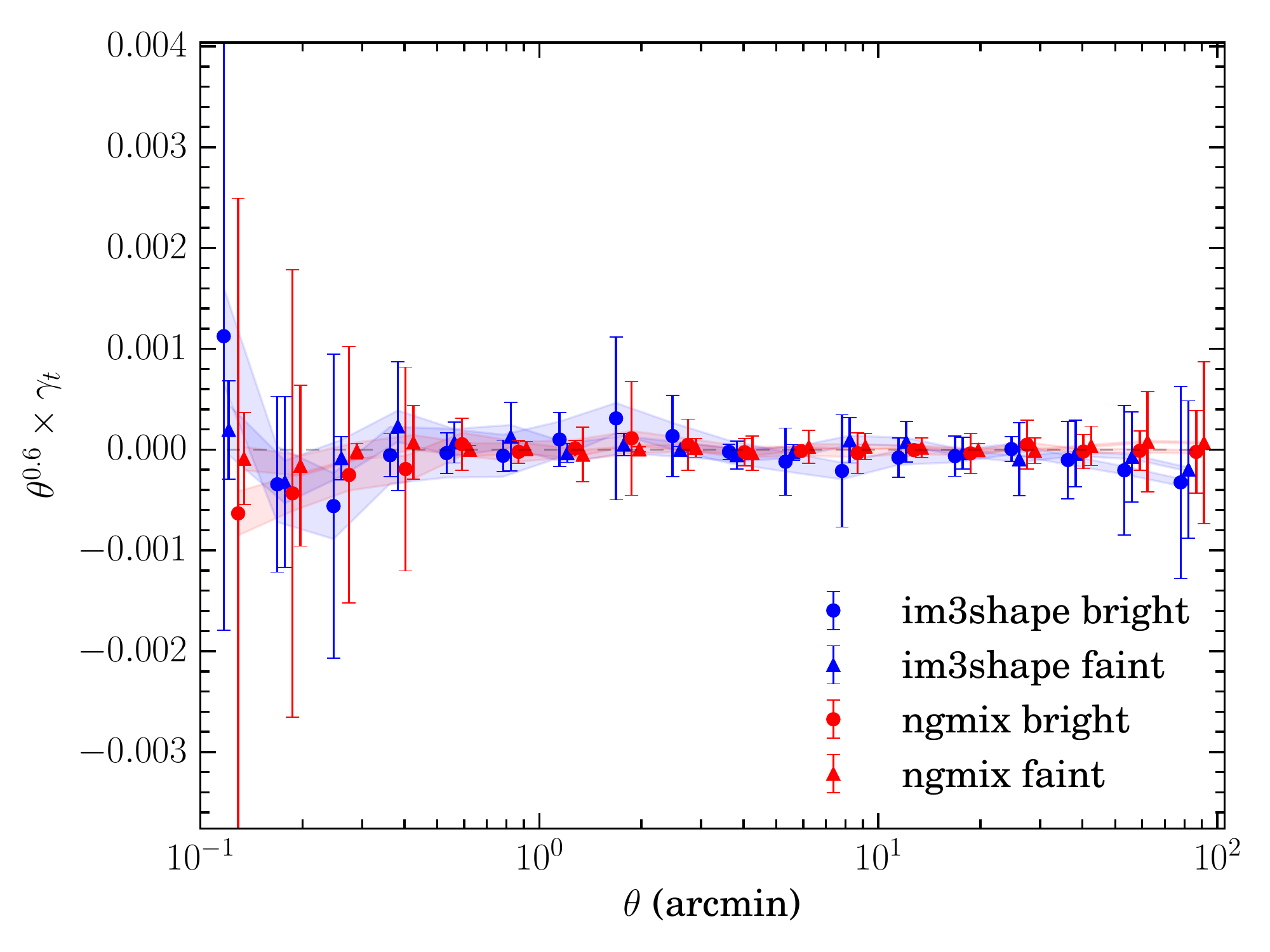}
\caption{Tangential shear around stars for \imshape\ (blue) and \ngmix\ (red). 
Stars are split into to two bins of $i$-band magnitude, where ``bright'' means $14 < m_i < 18.3$ and ``faint'' 
means
$18.3 < m_i < 22$.  The faint sample includes stars used for PSF modeling; bright stars are 
excluded to avoid the brighter-fatter effect (cf.~\S\ref{sec:psf:cuts}).
Shaded regions represent 1$\sigma$ shape noise
uncertainty, while error bars are from jackknifing the stars. 
\label{fig:tests:tanshearstars}
}
\end{figure}

If the PSF correction is incomplete, there may also be a residual signal seen in the
mean tangential shear around stars, which could potentially
contaminate galaxy-galaxy lensing studies.  To test for this, we measured
the tangential shear around the positions of stars in
both \imshape\ and \ngmix\ for ``bright'' ($14 < m_i < 18.3$) and ``faint'' ($18.3
< m_i < 22$) stellar populations. In all cases we found the signal, shown in \fig{fig:tests:tanshearstars},
to be consistent with
zero. The shape noise uncertainty is shown as the shaded regions.  The error bars are
jackknife uncertainty estimates.

The test using the faint stars primarily checks for effects related to PSF interpolation and PSF modeling.
The bright stars are not themselves used to constrain the PSF model (cf.~\fig{fig:psf:bfe}), so these stars
instead check for problems related to deblending and sky estimation errors in the outskirts of 
bright stellar haloes.  We see no evidence of any systematic errors around either set of stars.

\subsection{Galaxy Property Tests}
\label{sec:tests:gal}
\assign{Mike}
\contrib{Vinu, Erin, Daniel}

There are many properties of the galaxy images that should be independent of
the shear, but which in practice can be correlated with systematic errors in
the shear measurement.  For example, some of the properties we tested during
the course of our analysis were: the size of the postage stamp, the number of
neighbors being masked, the fraction of the stamp area being masked, the
estimated bulge-to-\disk\ ratio, the galaxy's signal-to-noise, and the galaxy
size.
These were all extremely helpful diagnostic tools during the analysis, but here
we only present the final two, which initially showed evidence of systematic errors and
took the most effort to resolve.

Using \snr\ or the galaxy size for selections is quite natural, since estimating the 
shear for
small, faint galaxies is more challenging than for large,
bright galaxies.  However, measurements of these quantities can be correlated
with the galaxy ellipticity, and thus an applied shear.  Binning the data for
the null test by these properties can thus induce selection effects and produce
a net mean spurious ellipticity.  This was already discussed in
\S\ref{sec:shear:snr} with respect to \snr.  We need to
do something similar to construct a proper null test for the galaxy size.

\subsubsection{Galaxy signal-to-noise}
\label{sec:tests:snr}

\begin{figure}
\includegraphics[width=\columnwidth]{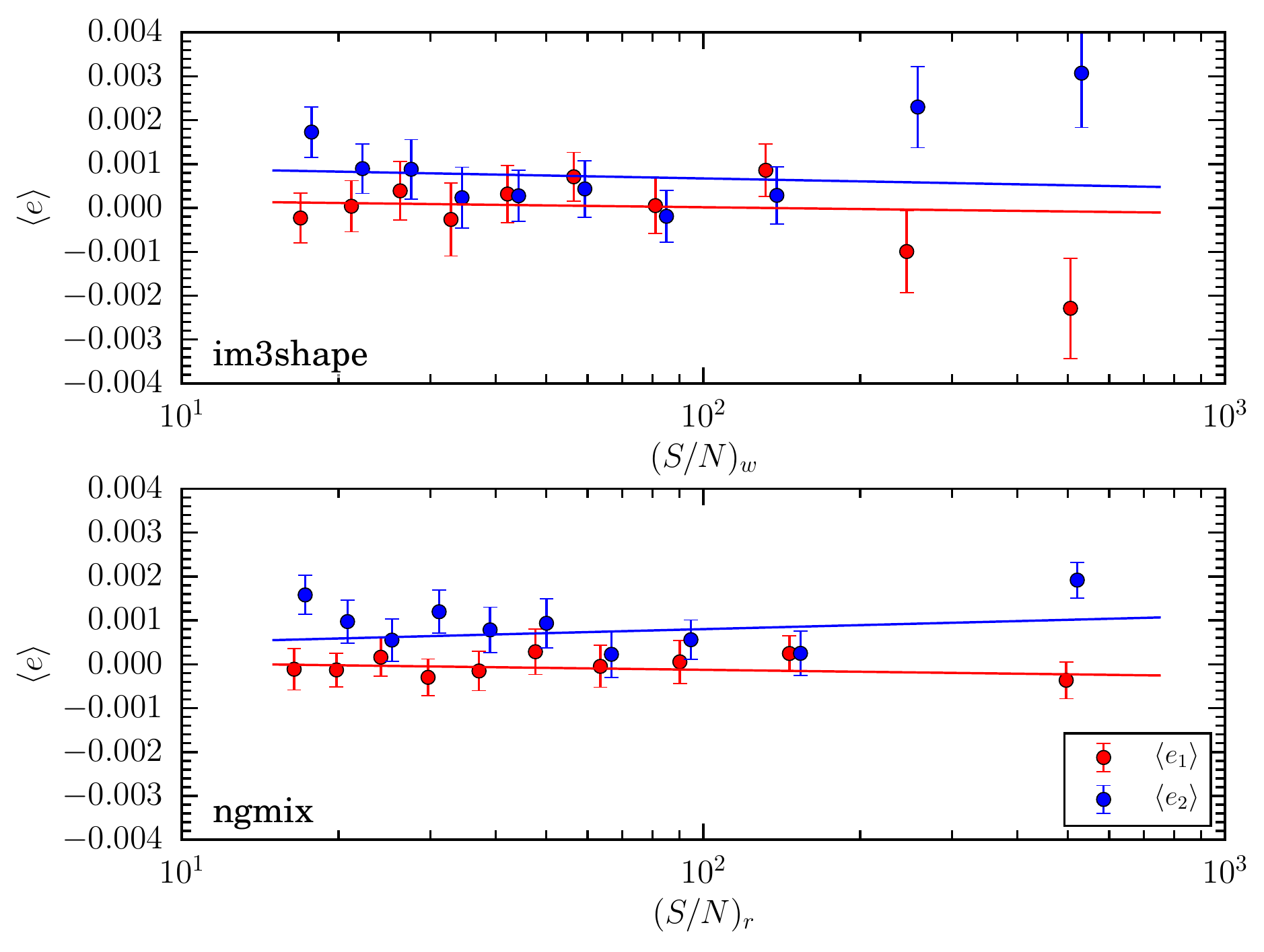}
\caption{The mean galaxy shear as a function of the signal-to-noise for
\imshape\ (top) and \ngmix\ (bottom).  
For \imshape\ we test against \snrw\ (cf.~\eqn{eq:snrw}), which is one
of the parameters used for the calibration, so it includes corrections for selection bias.  
For \ngmix, we test against \snrr\ (cf.~\eqn{eq:snrr}), 
which does not induce any significant selection bias from the binning.
\label{fig:tests:snr}
}
\end{figure}

The null test for checking that the galaxy shapes are independent of \snr\
requires different measures of \snr\ for each catalogue.  As described in
\S\ref{sec:im3shape:noisebias}, for \imshape\ we calibrated the bias in the
shear measurements from simulations as a function of \snrw\ and \rgp.  As such,
the selection bias that is induced by binning on \snrw\ 
(cf.~\S\ref{sec:shear:snr}) is accounted for as part of the calibration.  Thus, the
appropriate null test on the data is to check that the mean shear is
independent of \snrw, as shown in the top panel of \fig{fig:tests:snr}.  There
is no apparent bias in the mean shear down to $\snrw = 15$.

We did not apply any external calibration to \ngmix, so the null test for it requires a
\snr\ measure that does not induce selection biases from the binning.  For
\ngmix, we used \snrr\ (cf.~\eqn{eq:snrr}), which did not induce any selection
biases when we tested it on simulated data.  In the bottom panel of
\fig{fig:tests:snr} we show that the mean estimated shear for \ngmix\ is
independent of this ``roundified'' signal-to-noise measure down to $\snrr =
15$.

Previous versions of the \ngmix\ catalogue had shown a very significant bias
in the lowest \snr\ bin in this plot before we realized that the bias was being
induced by our galaxy selection from the cut on \snrw.  \imshape\ calibrates
this kind of selection bias, but when that calibration is faulty, it too could
show a bias in the lowest \snr\ bin.  In \fig{fig:tests:snr} we show that the final 
catalogues do not have any such bias.  The points are consistent with the mean
value (which, as we mentioned, is not necessarily expected to be zero) and
show maximal deviations less than our required $c_\mathrm{rms} < \crequirement$
(\eqn{eq:req:c}).

\subsubsection{Galaxy size}
\label{sec:tests:size}

\begin{figure}
\includegraphics[width=\columnwidth]{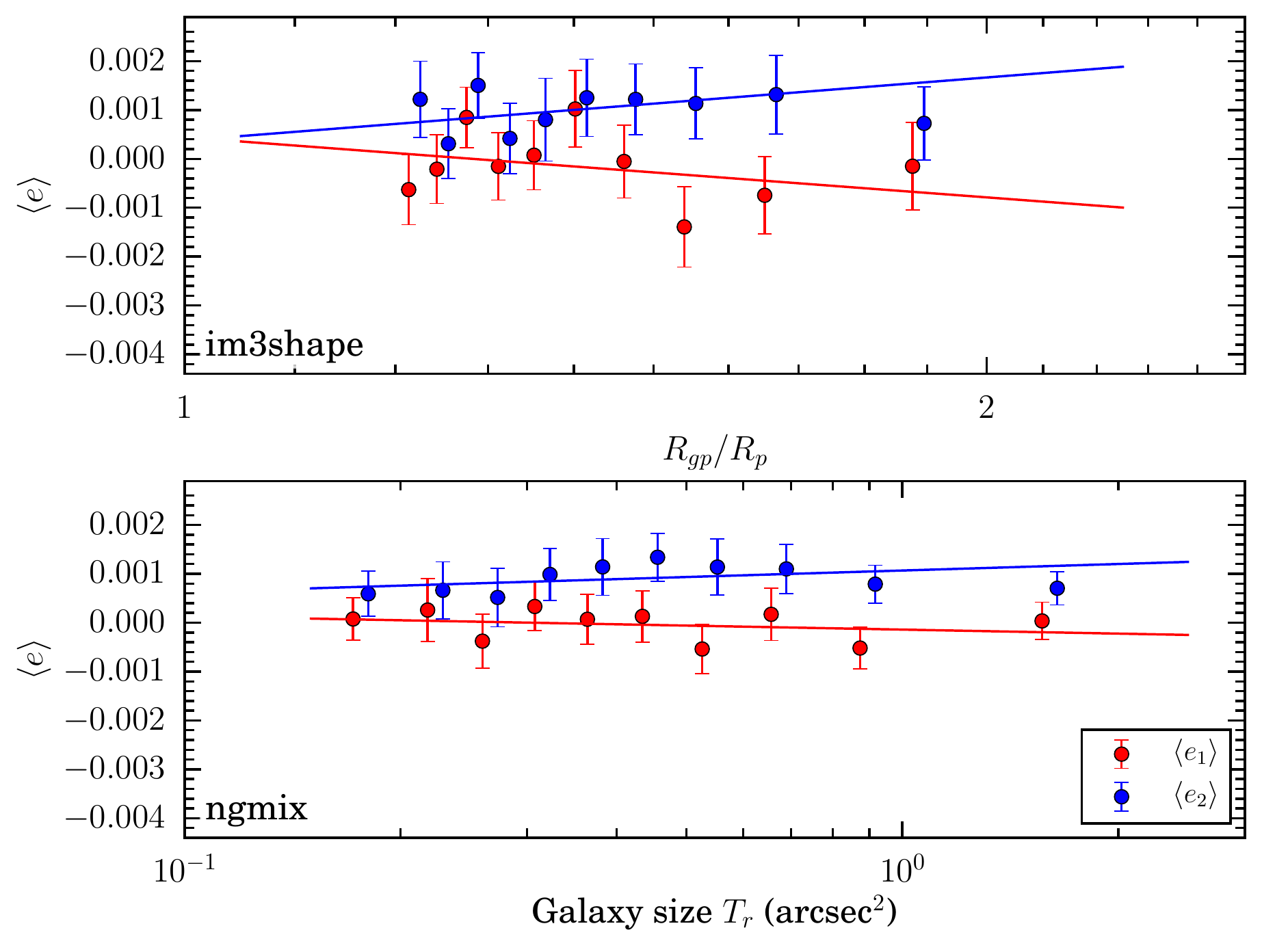}
\caption{The mean galaxy shear as a function of the galaxy size for
\imshape\ (top) and \ngmix\ (bottom).  
For \imshape\ we test against \rgp, which is one of the parameters
used for the calibration, so it includes corrections for selection bias.  
For \ngmix, we test against a size
measure, $T_r$ (cf.~\eqn{eq:tr}), that does not induce and significant selection bias from the binning. 
\label{fig:tests:galsize}
}
\end{figure}

Similar considerations apply to the null tests for galaxy size.  Since \imshape\ corrects for 
selection bias using the measured \rgp, this is the appropriate quantity to use for the 
null test regarding galaxy size.
In the top panel of \fig{fig:tests:galsize} we show the mean estimated shear binned by \rgp.
The \imshape\ measurements show no evidence of any dependence of the
shear estimates on the size of the galaxy.  

For \ngmix, we need to use a size measure that is independent of the shape of the
galaxy.  The internal parameter that \ngmix\ uses for the size of the galaxy in its model
is $T = I_{xx} + I_{yy}$, the standard second moment measure of the size of a galaxy;
however, this quantity changes with applied shear.  If a round galaxy is sheared by 
an area-preserving\footnote{By area-preserving, we mean that the determinant of the 
distortion matrix is unity: $A = \frac{1}{\sqrt{1-|g|^2}}
\left( \begin{array}{cc}
1-g_1 & -g_2 \\
-g_2 & 1+g_1 \end{array} \right)$.
} shear $g$, then the measured size will be
\begin{equation}
T^{(g)} = T^{(g=0)} \left( \frac{1+|g|^2}{1-|g|^2} \right).
\end{equation}

For non-round galaxies, an applied shear tends to make the estimated size $T$ larger
when the shear is aligned with 
the galaxy shape and smaller when it is anti-aligned.  This can lead to an
apparent bias in the measured shapes with respect to the measured value of $T$.  
If the mean
PSF shape were precisely round, this bias should average out over an ensemble of galaxies;
however, our PSFs have a preferred direction, which breaks the symmetry and leads
to an apparent bias in the mean shape with respect to $T$.

In parallel to our definition of \snrr\ as the signal-to-noise that the 
galaxy \emph{would have had if it were round}, we similarly construct an estimate of
the size that the galaxy would have had if it were round:
\begin{equation}
T_r \equiv T \left( \frac{1-|e|^2}{1+|e|^2} \right),
\label{eq:tr}
\end{equation}
where $e$ is the estimated shape of the galaxy.  Binning the shears by this quantity
should thus not induce any selection bias from the binning itself.
In the lower panel of \fig{fig:tests:galsize}
we show the results of this test for \ngmix.  There is no apparent dependence of the mean
shape on this ``roundified'' measure of the size of the galaxy.

In both cases the slopes are consistent with zero and
show maximal deviations well below our required $c_\mathrm{rms} < \crequirement$
(\eqn{eq:req:c}).

\subsection{B-mode Statistics}
\label{sec:tests:bmode}
\assign{Matt}
\contrib{Tim}

The deflection field induced by gravitational lensing has the special property that it is 
essentially curl-free.  Since this is also true of electric fields, the shear field is generally
referred to as being an ``E-mode'' field.  The corresponding divergence-free ``B-mode''
field can be considered as corresponding to an imaginary convergence \citep{schneider02}.

In fact gravity can induce a slight B-mode from source clustering \citep{schneider02},
multiple deflections \citep{krause10}, and intrinsic alignments \citep{crittenden01}.
But in practice all of these effects are well below the level at which we could measure 
them, which means that any significant measured B-mode is almost certainly a sign
of uncorrected systematic errors in the shears.

We calculated B-mode
statistics of the shear field by computing linear combinations of the binned
shear two-point correlation function values that are insensitive to any E-mode
signals, modulo a very small amount of computable E- to B-mode leakage. See
\citet{becker2013} for details. In this application, we chose linear
combinations that approximate band-powers in Fourier space as described in
\citet{becker2014}. Finally, we used the mock catalogues described in 
\citet{Becker15} to compute the shape noise and sample variance uncertainty
for the statistics. These mock catalogues include the survey mask and match
the shape noise and source photometric redshift distribution 
for each of the two shear catalogues.
We used 126 mock catalogues in total.

\begin{figure}
\includegraphics[width=\columnwidth]{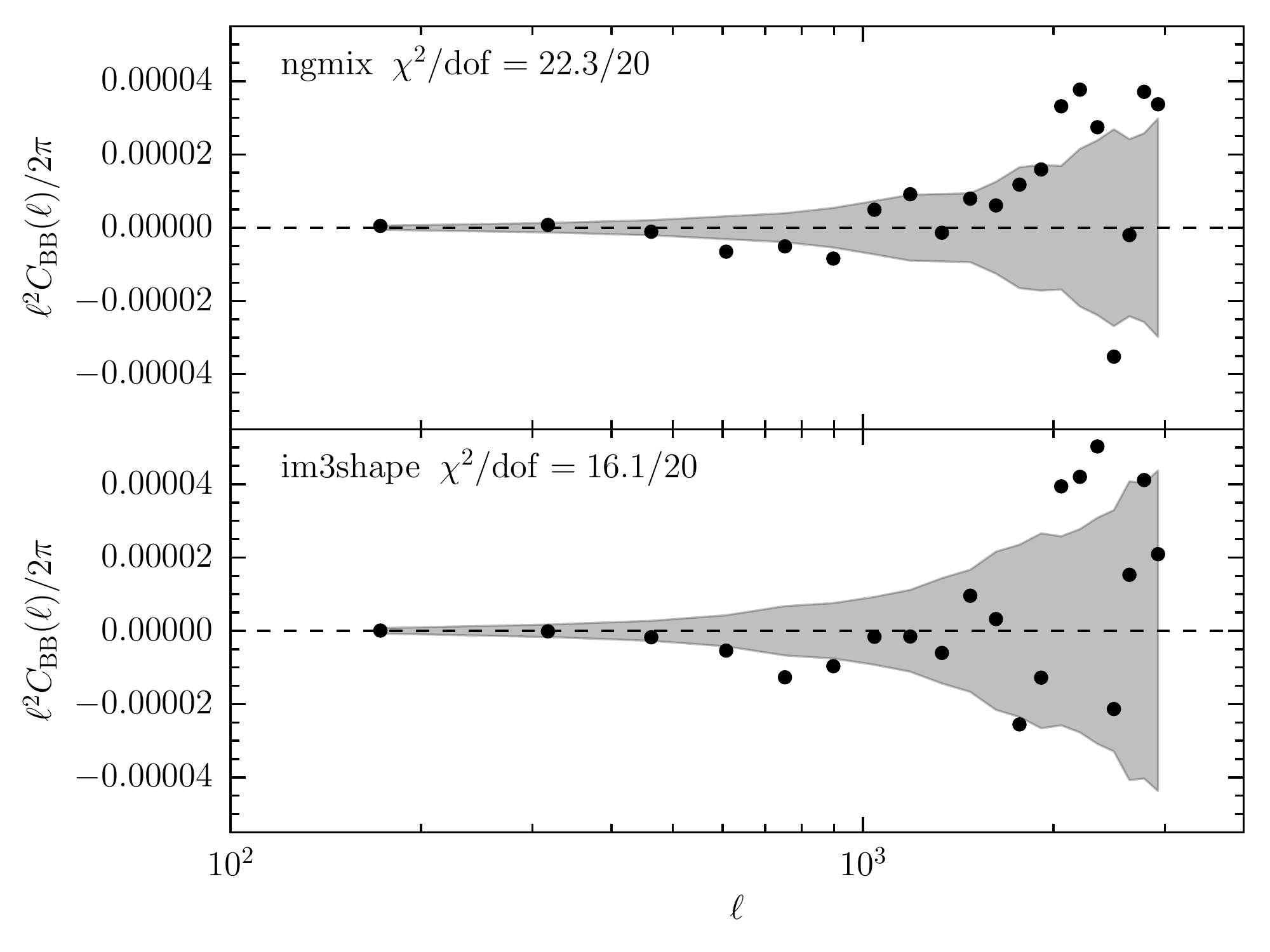}
\caption{The measured B-mode for \ngmix\ (top) and \imshape\ (bottom). Each band power measurement
is plotted at its central location in $\ell$. 
The grey band shows the uncertainty on the measurement due to both sample variance and shape noise. 
Adjacent points are highly correlated and the indicated $\chi^{2}$ accounts for the correlations.
\label{fig:tests:bmode}
}
\end{figure}

In \fig{fig:tests:bmode} we show the measured B-mode for each catalogue using the most conservative selection described below. Each 
band-power measurement is plotted at its central location in $\ell$. Adjacent points are highly correlated and the $\chi^{2}$ given in the figure 
accounts for the correlation. We find a $\chi^{2}/\mathrm{d.o.f.}$ of 22.3/20 for \ngmix\ and 16.1/20 for \imshape\ indicating no
significant B-mode contamination in the shear field. 

\subsection{Calibration Tests}
\label{sec:tests:sims}
\assign{Tomek}
\contrib{Mike, Erin}

It is difficult to test the overall shear calibration using the data alone.
However, we can use the \greatdes\ simulation described in \S\ref{sec:sims:greatdes}
to test the performance of the two shear algorithms on relatively realistic images
with known applied shear.

Since \imshape\ uses this simulation to calibrate the shear measurements
(cf.~\S\ref{sec:im3shape:noisebias}), the overall corrected shears should be
accurate, almost by construction.  The calibration was done without weighting,
but for this test we used the same weights that we recommend for the data (cf.
\S\ref{sec:im3shape:weights}).  The mean shear is thus not mathematically
guaranteed to be exactly zero.  We detect a net bias after applying the
calibrations, but it less than our requirements for DES SV data: $m_1=0.0008
\pm 0.0015$ and $m_2=-0.0068 \pm 0.0015$.

For \ngmix, the overall calibration error is a more relevant test.
The priors used for \greatdes\ were the same as used for the DES SV data, which
is expected to be appropriate given the general agreement between the galaxy properties
in the simulation and the data (cf.~\S\ref{sec:sims:greatdes}).
We found the overall calibration error for \ngmix\ to be
$m_1=-0.030 \pm 0.0015$ and $m_2=-0.035 \pm 0.0015$.
This does not quite meet our requirement of $|m| < 0.03$ from \eqn{eq:req:m}.

Considering that many science applications will use tomography to investigate the
evolution of the shear signal with redshift, it is interesting to look at
the calibration of both shear codes as a function of redshift.
To test the redshift dependence of the bias, we used the known
photometric redshifts of the galaxies from the COSMOS data, from which we drew
the galaxy images used in the \greatdes\ simulation.  With this test, we can
also learn if the tomographic selection process itself leads to any significant
selection biases.

\begin{figure}
\includegraphics[width=\columnwidth]{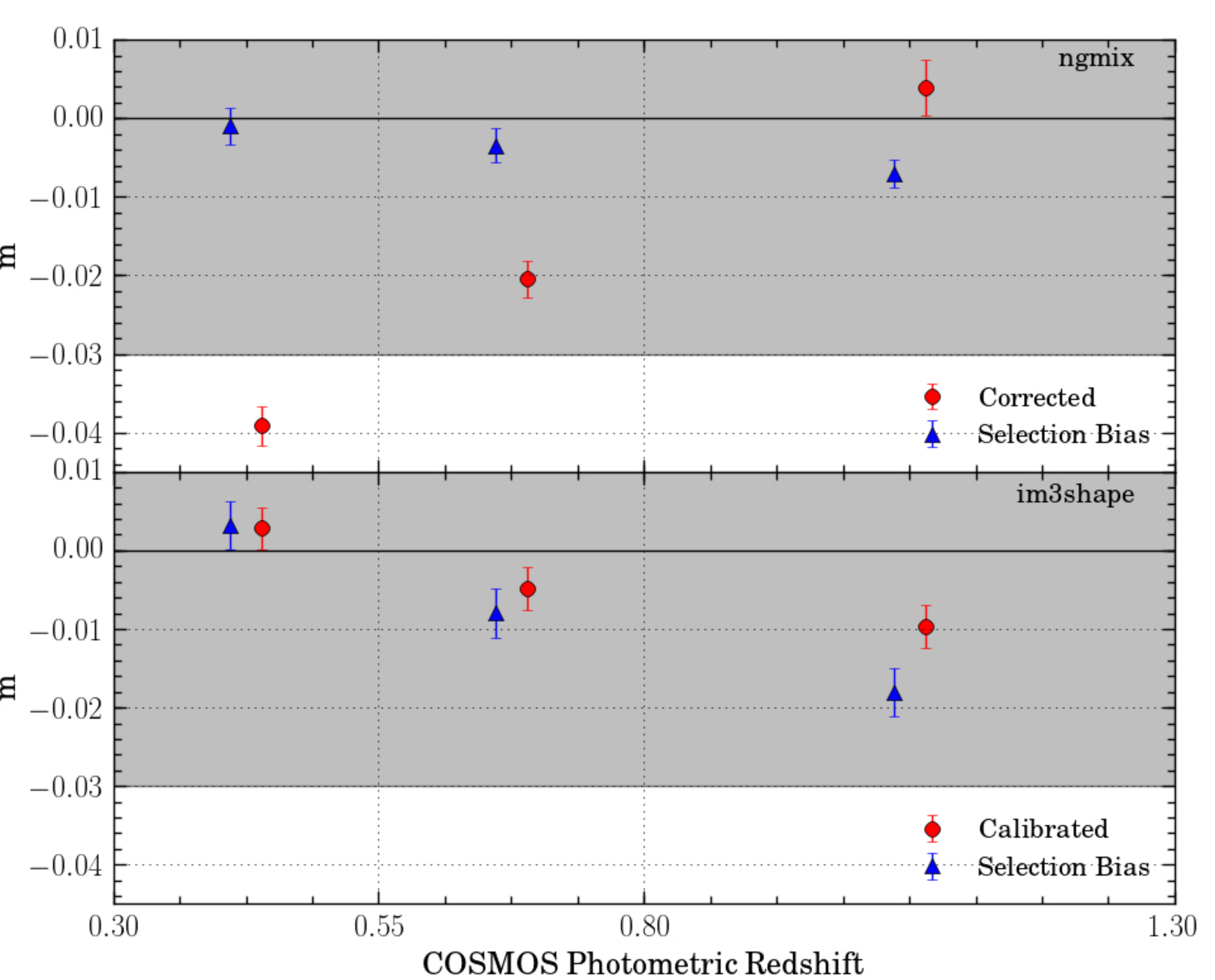}
\caption{Multiplicative shear bias in three bins of photometric redshift for \ngmix\ (top) and \imshape\ (bottom), as calculated
using the \greatdes\ simulation. In both cases, the same selection and weights were used as for the real data. The red
circles denote the average bias in each bin after correcting for the sensitivity (\ngmix) or the calibration (\imshape). 
The blue triangles show an estimate of selection bias, calculated using the 
known true ellipticities. The grey band in both panels marks the $\pm$3\% requirement for SV data. 
\label{fig:tests:calibration}
}
\end{figure}

In \fig{fig:tests:calibration} we show the results of performing this test for \ngmix\ (top) and
\imshape\ (bottom), taking galaxies in different ranges of photometric redshift, using
the same redshift bins that will be used for the cosmology constraints \citep{SVCosmology}.
Note that the redshift information was not used in the calibration process for \imshape,
so the variation with redshift is a non-trivial test of the correction. 
Prior to calibration, we find a significant bias in each of the three
redshift bins, $m=(-0.039,-0.058,-0.072)$. 
After calibration (red circles, \fig{fig:tests:calibration}), the net multiplicative bias for \imshape\ 
is reduced to a level well within the requirements.
This indicates that the derived corrections are robust to galaxy selections based on redshift.

We also tested the performance of \imshape's PSF leakage calibration as a
function of redshift (not shown).
We found the overall leakage before calibration was $\alpha= (0.070,  0.112, 0.102)$ 
for the three redshift bins.
After calibration, we found $\alpha = (0.001, 0.021, -0.005)$, 
which demonstrates good performance of the leakage calibration as well.

We found the multiplicative bias for \ngmix\ to be outside of the requirement band for the lowest redshift bin, 
and then rose to acceptable levels in the two higher bins.
We believe this is because the proportion of bulge galaxies is highest at low redshift,
and the \ngmix\ exponential model has significant model bias for these galaxies.
As the proportion of bulges decreases at higher redshift, the mean model bias decreases,
and the calibration is within our requirements.

To test the hypothesis that we are measuring a model bias for the exponential
\disk\ model, we implemented a more flexible model and applied it to this
simulation.  This model is a simple two-component bulge and \disk\ model, where
the bulge fraction is determined not by a simultaneous fit with other
parameters but by an initial comparison of two separate bulge and \disk\ fits to
the galaxy image.  A similar model used in the Sloan Digital Sky Survey for
galaxy fluxes was known as the ``composite'' model \citep{sdssdr2}.
For the composite model we found biases $\sim 1$\%, suggesting that the larger bias evident
for the exponential model is principally model bias.  Unfortunately, we were unable
to apply this new composite model to the DES data in time to be used for this paper.

The blue triangles in \fig{fig:tests:calibration} represent the estimated selection bias in each bin
induced by our various selection criteria.
To calculate these values, we applied the known shear to the COSMOS shape estimates \citep{Kannawadi2014} of the galaxies used for the 
simulation, and then applied the same selection criteria
we used for each of the two algorithms.
We found the selection bias from the \imshape\ cuts was at most $2\%$ in the
highest redshift bin, which is largely corrected by the calibration scheme.
We found the selection bias for \ngmix\ was less than $1\%$ for all redshift bins.

\subsection{Cross-catalogue Comparisons}
\label{sec:tests:cross}
\assign{Mike}
\contrib{Andres, Erin, Joe, Niall,Simon}

Another powerful test is to compare the two independent shear catalogues,
\imshape\ and \ngmix.  We used two very different strategies when generating
these catalogues.   For \imshape\ we used simulations to determine the shear
calibration, and applied corrections to the shear measurements on real data.
For \ngmix\ we expected relatively little noise bias, but the sensitivity of
the shear estimator was calculated from the data itself and applied to the
shear measurements, a process that was worth testing in detail.  And we did expect
some model bias for \ngmix.  Furthermore, the PSF was treated differently by the
two methods: for \imshape\ we used the reconstructed PSF image directly, and for
\ngmix\ we fit models to the PSF.

A direct galaxy-by-galaxy test is not appropriate for a cross-catalogue
comparison, since there is not a 
unique unbiased shear estimate for a single galaxy.  Rather, we wished
to test that both methods produced consistent shear statistics for an ensemble
of galaxies \citep[cf.][]{Velander11}.  Two potential shear statistics that can be used are a
galaxy-galaxy lensing signal and the two-point shear correlation function.  We
tested if the results were consistent when using the same ensemble of
galaxies with the same weighting.  

Disagreement between the catalogues would be proof that at least one catalogue is biased,
but we would not be able to determine which one, nor the magnitude of this bias.  
Agreement between the two
catalogues is subjectively reassuring, but we wish to emphasize that agreement
does not prove that both catalogues are ``correct'' in the sense that they
can be used to generate unbiased shear estimates.

\subsubsection{Tangential shear ratio}
\label{sec:tests:tanratio}

Galaxy-galaxy lensing provides one of the cleanest tests of the relative
calibration of the two catalogues, because the azimuthal symmetry
inherent in the tangential shear signal largely cancels most sources of
additive systematic error.  Thus the ratio of two tangential shear signals
is primarily a measure of the relative multiplicative errors between the 
two catalogues.

For this test, we used the tangential shear signal around Luminous Red Galaxies
(LRGs) as determined by redMaGiC \citep[red sequence Matched-filter Galaxies
Catalogue;][]{redmagic} from the same DES \spte\ data.  For this purpose, we
did not require sources to be behind the lenses.  Rather, we took the full LRG
catalogue as the lenses, and for the sources, we used all galaxies that were
well-measured by both \ngmix\ and \imshape.  Regardless of the redshifts of the
LRGs and the source galaxies, we expected the signal to be the same for both
catalogues in the absence of a multiplicative bias.

The observed signal $\langle e_{t,i}(\theta)\rangle$ for each method 
$i\in$ \{\imshape, \ngmix \} can be written as: 
\begin{equation}
\langle e_{t,i}(\theta)\rangle = 
(1+m_i) \langle \gamma_t(\theta)\rangle + \langle\eta_i(\theta)\rangle,
\end{equation}
where $\langle \gamma_t \rangle$ is the true underlying signal, $\langle\eta_i\rangle$
is a noise term including both intrinsic shape noise and measurement noise, and
$m_i$ is a possible calibration error for each method.
We mostly drop the argument $\theta$ in the following for brevity. 
For the same ensemble of galaxies, the two catalogues have identical 
values of $\langle \gamma_t \rangle$ and a similar shape noise contribution to $\langle\eta_i\rangle$ 
(though not 
identical, since the two methods use different bands). The contribution to $\langle\eta_i\rangle$ from 
shape measurement noise, however, is expected to be somewhat different.

\begin{figure}
\includegraphics[width=\columnwidth]{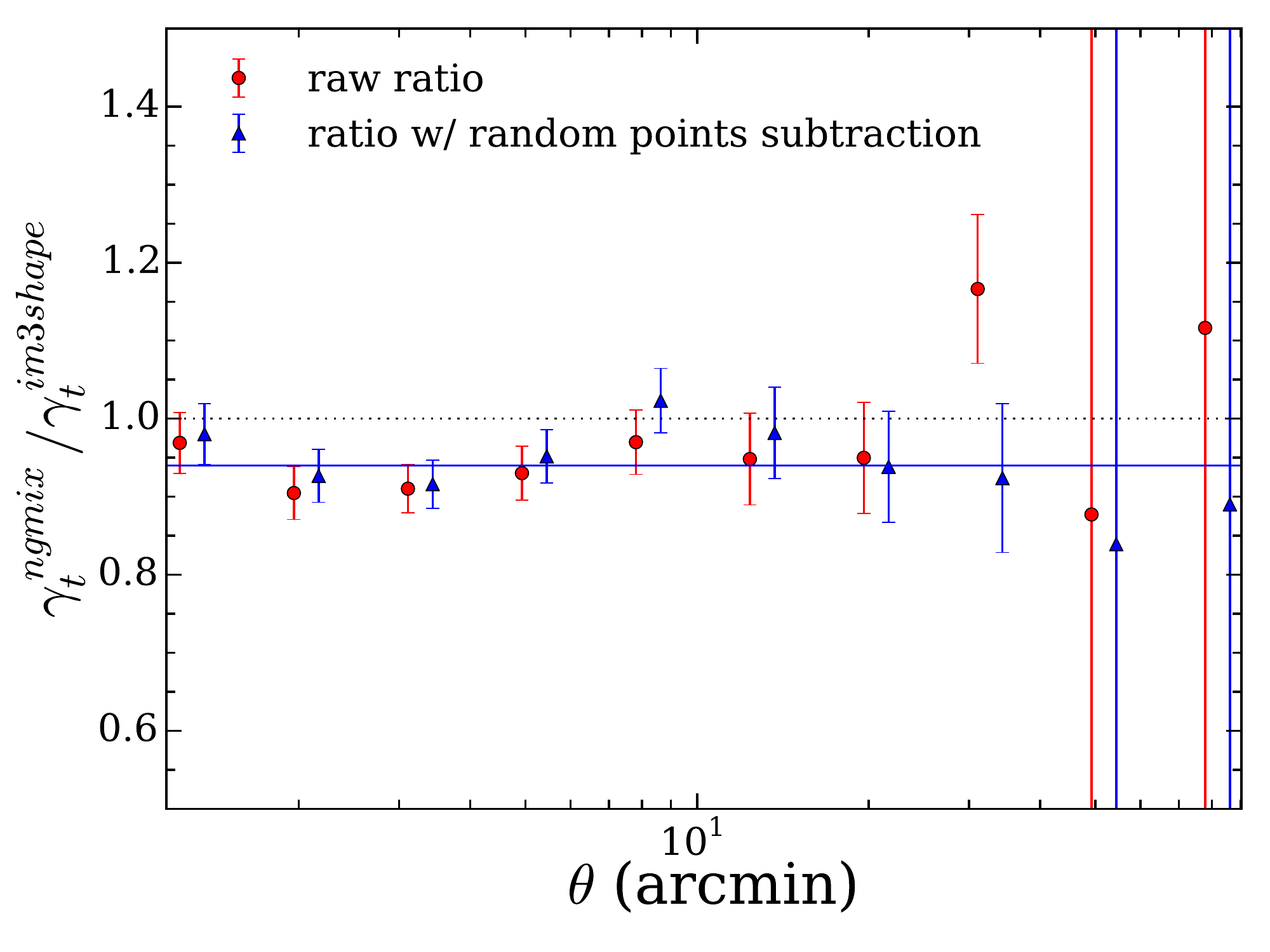}
\caption{The ratios of tangential shear measurements around LRG galaxies
from shears measured by \ngmix\ to those measured by \imshape.
The red circles show the direct ratio and the triangles correspond to the ratio after subtraction of the tangential
shear around random points. The weighted mean ratio in the scale range $1 - 20$ arcminutes is $0.954 \pm 0.018$. 
The blue line shows a prediction of the ratio ($0.94$) based on the \greatdes\ simulation, which accounts for a 
selection bias induced by the intersection of the two shape catalogues. This result is in good agreement with the data points. 
\label{fig:tests:ratio}
}
\end{figure}

The red points in \fig{fig:tests:ratio} represent the ratio of measured tangential shear 
using the two shear catalogues.  
The weighted mean of the ratio over the range from 1 to 20 arcminutes (the typical scales of interest
for weak lensing) is $0.932 \pm 0.018$. We would naively expect this to be an estimate of 
$(1+m_{\ngmix})/(1+m_{\imshape}) \approx 1+m_{\ngmix}-m_{\imshape}$.
However, three corrections are required before any conclusions can be drawn from this result about potential differences
in the relative calibration.

First, additive systematic errors only cancel if the sources are distributed
uniformly around the lenses.  This is approximately true, but masking can break
the symmetry, especially at large scales.  One solution is to subtract off the
measured tangential shear around random points, drawn from the same region and with
the same masking as the LRGs.  No signal is expected around such points, but any 
additive bias will affect both measurements equally.  Thus the
difference is a cleaner estimate of the true tangential shear than the uncorrected signal. 
The blue points in \fig{fig:tests:ratio} represent the signal after this subtraction, and have a 
mean ratio of $0.954 \pm 0.018$,

Second, the ratio of two noisy quantities with the same mean does not in general have an expectation value equal to 1.
If the denominator is a random variable, $X$, with a symmetric probability distribution
(e.g.~$X \sim \mathcal{N}(\bar{X}, \sigma_X)$), the ratio will be approximately $1 + \sigma_X^2/\bar X^2$.  
To account for this bias, we created simulated realizations of the ratio, and compared the measured signal to the mean and 
variance of these. We generated a ratio realization in the following way:
\begin{enumerate} 
\item Fit a polynomial, $\log(\langle e_t\rangle(\theta)) \!=\! p(\log(\theta))$ to the measured \ngmix\ 
signal, and take this to be the true signal, $\hat\gamma_t(\theta)$.
\item For each source in the ensemble, rotate both the \ngmix\ and \imshape\ shear by
the same random angle. 
\item Re-measure the two tangential shear signals, which now give estimates of the noise,
$\langle \eta^r(\theta)\rangle$, as the true signal is removed by the random rotations.
\item Compute the realization ratio as
\begin{equation}
(\hat\gamma_t + \langle\eta_{\ngmix}^r\rangle)/(\hat\gamma_t + \langle\eta_{\imshape}^r\rangle).
\end{equation}
\end{enumerate}
We found the mean of these realizations to be consistent with a ratio of 1 on all scales, which
means that the \snr of the tangential shear is high enough that we can
neglect the noise term in the denominator.

Finally, we found that the act of matching the two catalogues caused a selection bias in 
the \ngmix\ catalogue, for two reasons.  First, the \imshape\ algorithm failed more often
for objects with low \sersic\ index ($n<1$).  And second, the cuts we made on the \imshape\ measurements
of \snrw\ and \rgp\ altered the mix of galaxy properties in the matched catalogue.
These two selection effects, when applied to the \ngmix\ catalogue imparted a net bias
on the \ngmix\ shear estimates in the matched catalogue that was not present in the full \ngmix\ catalogue.

\begin{figure}
\includegraphics[width=\columnwidth]{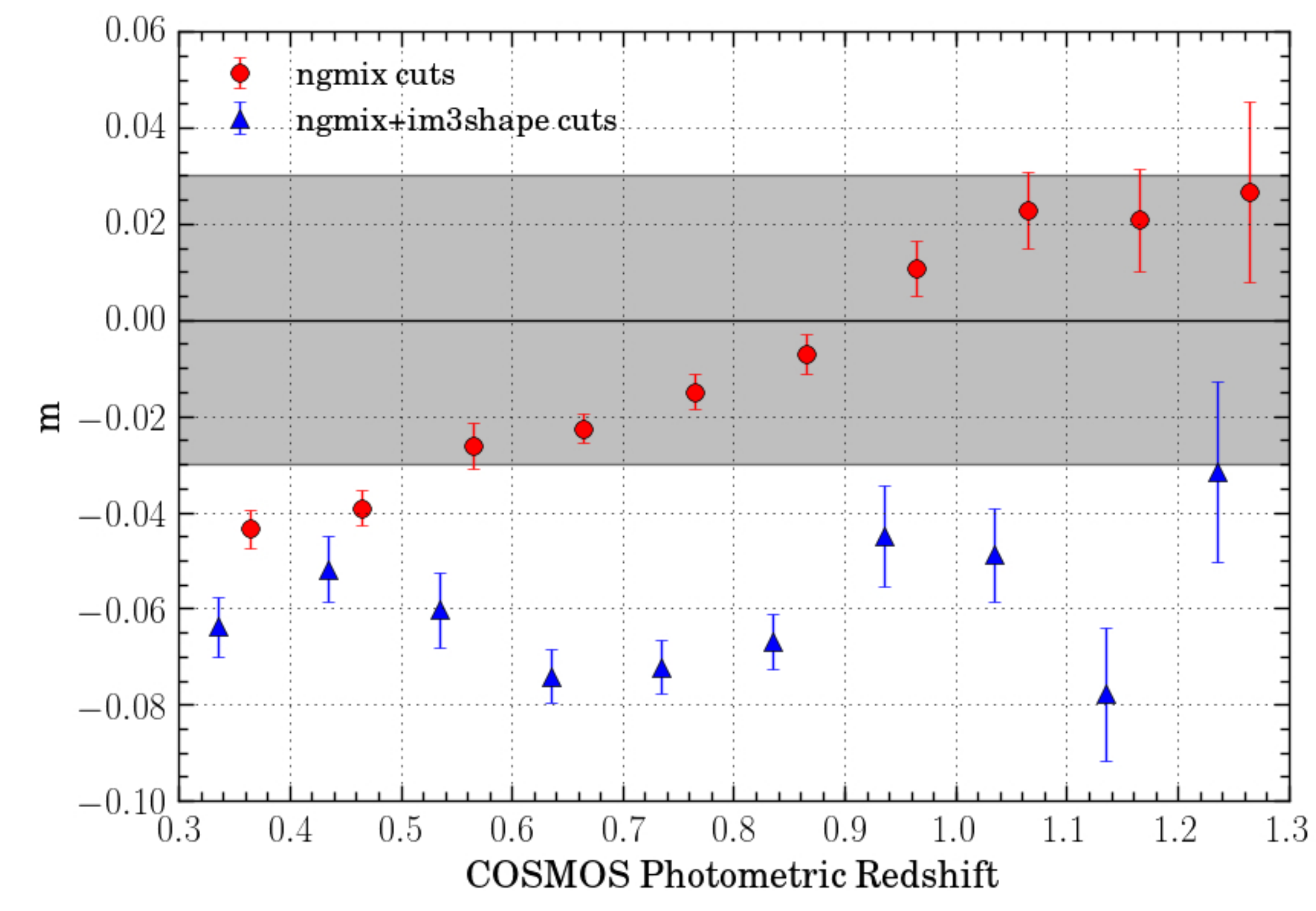}
\caption{Multiplicative bias for \ngmix\ shear measurements on
\greatdes\ simulated data as a function of redshift.
The red circles show the bias calculated using all galaxies that pass the \ngmix\ selection criteria
(as in the upper panel of in \fig{fig:tests:calibration}). 
The blue triangles show the bias when also including the recommended \imshape\ selection,
as we do to obtain the matched catalogue used for \fig{fig:tests:ratio}.
As in \fig{fig:tests:calibration}, the grey band represents the $\pm$3\% requirement for the SV data.    
\label{fig:tests:selection_bias}
}
\end{figure}

We quantified the level of this selection bias by performing the same procedure
on the \greatdes\ simulation.  We compared the mean bias for \ngmix\ using the
canonical \ngmix\ selection criteria to the bias after applying the \imshape\
selection, as a function of redshift.  The result is shown in
\fig{fig:tests:selection_bias}. The matching induced a mean selection bias of
about $-3\%$. Furthermore, we found that this bias increased with redshift.
Weighting the bias according to the lens redshift distribution and the lensing
efficiency of the source galaxies used in the tangential shear ratio test (and
assuming that the lenses do not evolve with redshift), we found a net selection
bias of $-6\%$ for \ngmix\ in the matched catalogue relative to whatever bias
might be present in the full \ngmix\ catalogue\footnote{We tested for a similar
    selection bias in the \imshape\ catalogue due to imposition of the \ngmix\
    cuts. The impact of the matching was found to be negligible, in part
    because the \ngmix\ catalogue is deeper, so its cuts have very little
impact on the \imshape\ selection.}. 

The mean ratio of $0.954 \pm 0.018$ is thus consistent with the prediction from 
\greatdes\ of $-6\%$ selection bias (which would 
produce a ratio of $0.94$). This bias induced by the combination of \imshape\ and \ngmix\ selection criteria
in the matched shape catalogues is shown by the blue
line in \fig{fig:tests:ratio}. 
Our finding is therefore consistent with no relative multiplicative bias between
the two catalogues.

We cannot of course prove that neither catalogue is affected by a significant multiplicative bias based on this
test.  They could both be biased by the same amount in either direction.  Furthermore, there are 
significant uncertainties in the calculation of the predicted selection bias described above
that may be at the $\sim3\%$ level.

\subsubsection{Differential shear correlations}
\label{sec:tests:dede}

The two-point shear correlation function is much more sensitive to additive
shear errors than the tangential shear, as mentioned above; it would be
difficult to disentangle multiplicative and additive errors in
a ratio test. Even in the absence of additive errors, the ratio of shear correlation
functions is much noisier than the ratio of tangential shears, making it a
less stringent test of calibration.

For these reasons, we instead use the two point function of the
\emph{difference} in the shear estimates from \ngmix\ and \imshape\ to compare
the shear catalogues:
\begin{align}
\xi_{+,\Delta e}(\theta) &= \langle (e_{\ngmix}(\mathbf{x})-e_{\imshape}(\mathbf{x}))^* \nonumber \\
&\qquad (e_{\ngmix}(\mathbf{x} + \boldsymbol\theta)-e_{\imshape}(\mathbf{x} + \boldsymbol\theta)) \rangle.
\label{eq:xi_de}
\end{align}

Consider the following model for the additive systematic errors in each catalogue (labelled $i$ here):
\begin{equation} \label{eq:simplebiasmodel}
e_i = (1+m_i) \gamma + \eta_i + a_i c_{\rm common} + c_i,
\end{equation}
where $m_i$ is the calibration error, $\eta_i$ is the noise in the estimate, $c_{\rm common}$
includes any additive systematic errors present in both catalogues, possibly multiplied by different
coefficients $a_i$, and $c_i$ is the additive error particular to each catalogue.

By construction, the additive bias terms in \eqn{eq:simplebiasmodel} are
independent.  If we further make the assumption that the systematic errors are
uncorrelated with the applied shear and the noise, and that $m$ and $c$ are uncorrelated, we find
that 
\begin{align} \label{eq:diffbiasprop}
\xi_{+,\Delta e}(\theta) &= (\Delta m)^2 \xi_+(\theta) \nonumber \\
&\quad + (\Delta a)^2 \langle c_{\rm common}^* c_{\rm common} \rangle(\theta) \nonumber\\
&\quad + \langle c_\ngmix^* c_\ngmix \rangle(\theta) \nonumber \\
&\quad + \langle c_\imshape^* c_\imshape\ \rangle(\theta).
\end{align}
This test is sensitive to the spatial correlations of the systematic errors in either
catalogue, but particularly to additive errors, rather than multiplicative.
The $(\Delta m)^2$ factor for the multiplicative term typically makes this term
insignificant.

There is one subtlety in the construction of this test.  As we found in \S\ref{sec:tests:tanratio},
the act of matching the two catalogues can induce selection biases that are not present in either catalogue
separately when using its own individual selection criteria.  
In this case, the 
salient selection effects are a spurious PSF leakage $\alpha$ and an overall mean $\langle c \rangle$
that can be induced by the match.  

The estimated value of $\alpha$ for \ngmix\ changed by less than $0.1\%$
in the matched catalogue relative to the full \ngmix\ catalogue.
But for \imshape, the matching changed $\alpha$ by $-1.5\%$.  Therefore, to make this a fair test
of the additive systematic errors, we
added back $0.015 \times \epsf$ to the \imshape\ galaxy shapes to account for this
selection effect.\footnote{We also subtracted the corresponding value for \ngmix, although
it makes no discernible difference.}

Even after correcting for the above effect, we found that the mean shear
changed by $(3.9 + 2.2 i) \times 10^{-4}$ for \ngmix\ and by $(2.0 - 3.0 i)
\times 10^{-4}$ for \imshape.  We interpreted these changes as due to selection
biases from the matching itself, leading to a spurious overall $\langle c
\rangle$ for each catalogue.  We thus subtracted these values as well from the
shape estimates in each catalogue.

\begin{figure}
\includegraphics[width=\columnwidth]{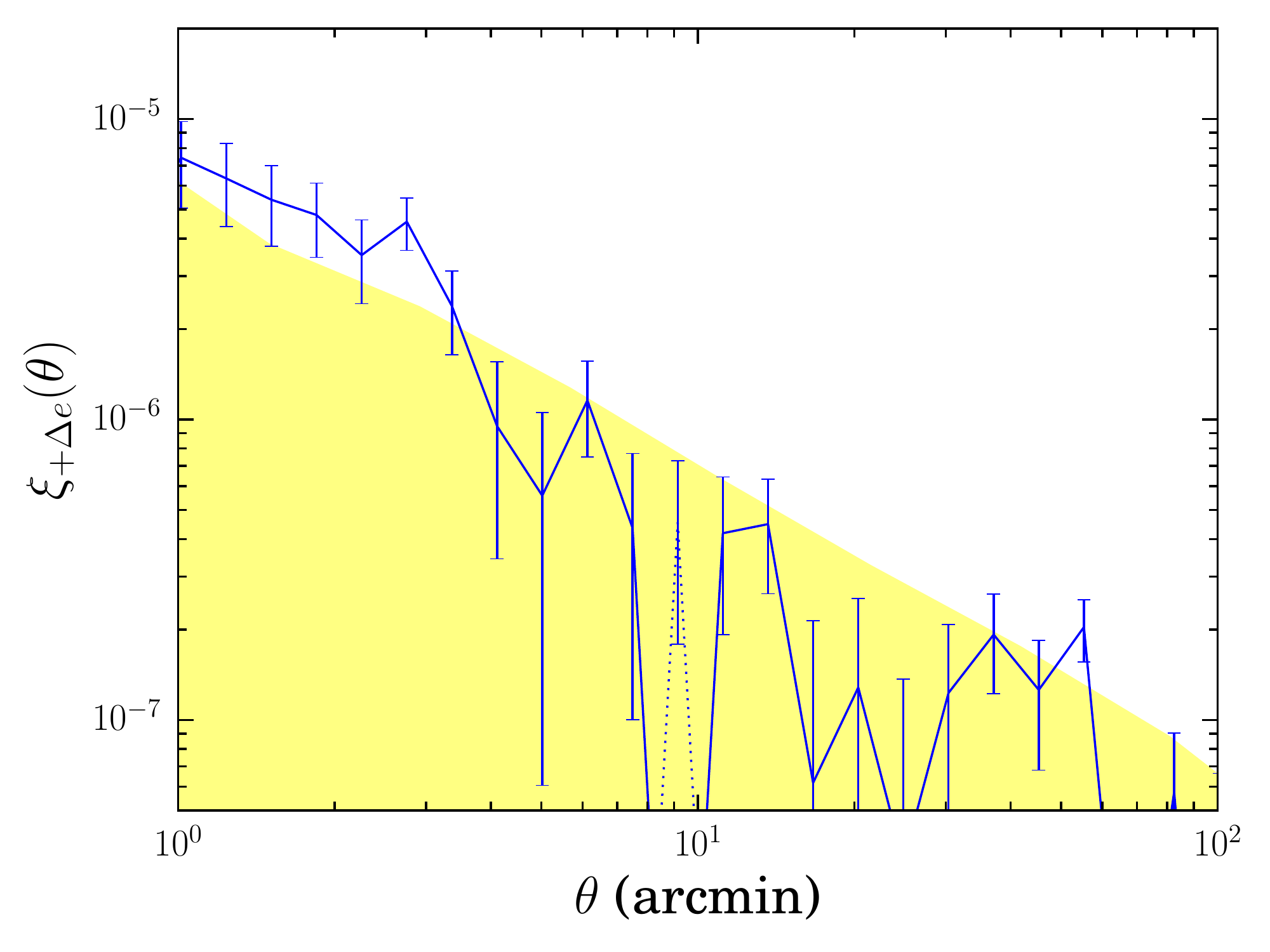}
\caption{The shear auto-correlation function of the difference in shear estimates of 
\ngmix\ and \imshape.  This test shows the level of additive systematic errors that may
still be present in one catalogue that is not present in the other.  The yellow band is 
the requirement, $\dximax_+$ from \fig{fig:req:dxi}.
\label{fig:tests:decorr}
}
\end{figure}

% NOTE: This too is placed earlier than its reference to get it onto the same page where it is referenced.
\begin{table*}
\begin{tabularx}{1\textwidth}{ l >{\raggedright\arraybackslash}X >{\raggedright\arraybackslash}X}
\hline
% I think there's a better way to do this centering, but I forget now how to do it.
\multicolumn{1}{c}{\large\textrm{Test}} &     \multicolumn{2}{l}{\large \hspace{8em}\textrm{Upper Limit on Systematic Error}} \\
& \imshape & \ngmix \\
\hline
\textbf{PSF Model Tests} &&\\
\S\ref{sec:psf:cuts} Mean PSF size error & $|m| < 0.005$ & $|m| < 0.01$  \\
\S\ref{sec:psf:rho} PSF model diagnostics $\rho_{1,3,4}$
 & $\xi_+^{cc}(1\arcmin) <$  $ 2\times10^{-6}$\textsuperscript{$\dagger$}  &  $\xi_+^{cc}(1\arcmin) < 5\times10^{-6}$ \\
 & $\xi_+^{cc}(30\arcmin) <$  $ 7\times10^{-8}$ &  $\xi_+^{cc}(30\arcmin) < 9\times10^{-8}$ \\
\S\ref{sec:psf:rho} PSF model diagnostics $\rho_{2,5}$ 
 & $\xi_+^{cc}(1\arcmin) <$  $2\times10^{-7}$ &  $\xi_+^{cc}(1\arcmin) < 8\times10^{-8}$ \\
 & $\xi_+^{cc}(30\arcmin) <$  $1.5\times10^{-7}$ &  $\xi_+^{cc}(30\arcmin) < 1.4\times10^{-7}$ \\
\hline
\textbf{Spatial Tests} && \\
\S\ref{sec:tests:fieldofview} Position in the field of view 
& No evidence of systematic errors & No evidence of systematic errors \\
\S\ref{sec:tests:ccd} Position on CCD
 & $\xi_+^{cc}(1\arcmin) < 6\times10^{-7}$  &  $\xi_+^{cc}(1\arcmin) < 6\times10^{-7}$ \\
\S\ref{sec:tests:fieldcenters} Tangential shear around field centres
 & $\xi_+^{cc}(1\arcmin) < 4\times10^{-8}$  &  $\xi_+^{cc}(1\arcmin) < 4\times10^{-8}$ \\
\hline
\textbf{PSF Tests} && \\
\S\ref{sec:tests:alpha} PSF leakage & $|\alpha| < 0.04$ & $|\alpha| < 0.01$ \\
\S\ref{sec:tests:alpha} Dependence on PSF size
& No evidence of systematic errors & No evidence of systematic errors \\
\S\ref{sec:tests:alpha2} Star-galaxy cross correlation & $|\alpha| < 0.03$ & $|\alpha| < 0.015$ \\
\S\ref{sec:tests:ggstars} Tangential shear around stars
& No evidence of systematic errors & No evidence of systematic errors \\
\hline
\textbf{Galaxy Property Tests} &&\\
\S\ref{sec:tests:snr} Galaxy \snr
& No evidence of systematic errors & No evidence of systematic errors \\
% This next line should have worked, but it is inserting an extra newline for some reason...
%& \phantom{Marginal:} at high \snr &\\ 
\S\ref{sec:tests:size} Galaxy size 
& No evidence of systematic errors & No evidence of systematic errors \\
\hline
\textbf{B-mode Statistics} && \\
\S\ref{sec:tests:bmode} $\ell^2C_\mathrm{BB}(\ell)/2\pi$ & No evidence of B-mode & No evidence of B-mode \\
\hline
\textbf{Calibration Tests} &&\\
\S\ref{sec:tests:sims} Redshift dependence in \greatdes\ & $|m| < 0.02$ & $|m| < 0.04$ \\
\hline
\textbf{Cross-catalogue Comparison} &&\\
\S\ref{sec:tests:tanratio} Tangential shear ratio 
& \multicolumn{2}{l}{\hspace{12em} $|\Delta m| \lesssim 0.04$} \\
\S\ref{sec:tests:dede} Differential shear correlations 
& \multicolumn{2}{l}{\hspace{12em} $|\xi_+^{cc}(1\arcmin)| < 9\times10^{-6}$} \\
& \multicolumn{2}{l}{\hspace{12em} $|\xi_+^{cc}(30\arcmin)| <$  $2\times10^{-7}$}\\
\hline
\end{tabularx}
\begin{flushleft}
\textsuperscript{$\dagger$}
Since \imshape\ use only the $r$-band images, the values quoted here are based on the $\rho$ statistics measured for the
$r$-band-only PSFs.
These curves are a bit higher than what is shown in \fig{fig:psf:rho}, which uses $r,i,z$ bands.  
\end{flushleft}
\caption{Summary of the results of our suite of null tests (including tests in \S\ref{sec:psf:rho}).
For reference, our nominal requirements from \S\ref{sec:req} are $|\alpha| < 0.03$, $|m| < 0.03$, 
$\xi_+^{cc}(1\arcmin) < 7\times10^{-6}$, and $\xi_+^{cc}(30\arcmin) < 2.5\times10^{-7}$.
\label{tab:tests:summary}
}
\end{table*}

In \fig{fig:tests:decorr} we show the resulting correlation function (\eqn{eq:xi_de}) after subtracting 
these selection biases.
For the weights, we used $w = \sqrt{w_\ngmix \times w_\imshape}$.
The yellow band represents
our requirement for additive systematic errors from \eqn{eq:req:xic}.
We see that, at scales less than 3 arcminutes,
we do not quite meet the requirements. 
Either one or both catalogues apparently have non-negligible additive
systematic errors at these scales.
We recommend that science applications sensitive to additive
systematic errors check carefully how these small-scale systematic errors may
affect their science results.

\subsection{Summary of Systematics Tests}
\label{sec:tests:summary}

We now attempt to synthesize the results of our large suite of null tests.  With this many tests, 
even if all the tests pass individually, it would not necessarily imply that the total systematic error is below our 
requirements.  In this section we attempt to quantify an upper limit on the
level of systematic error that may be in the shear catalogues, given all of the information we have available.

In \tab{tab:tests:summary} we provide a summary of the results from the previous sections (including the tests in \S\ref{sec:psf}).
For each, we have converted the result of the test into the impact that the result could have on 4 possible values.
For PSF leakage, we give the
maximum allowed value of $\alpha$.  
For other kinds of additive systematic errors, we give the maximum value of
$\xi_+^{cc}(\theta)$ at $\theta = 1$ arcminute and (when relevant) $30$ arcminutes.  
And for multiplicative errors, we give the maximum $|m|$ that is consistent with the test.  
Some tests do not lend themselves to a quantitative
upper limit.  Fortunately, in each of these cases, there is no evidence from the test that there is any systematic
error.

There are two tests that give constraints on the PSF leakage coefficient $\alpha$.  In all cases, the tests are
completely consistent with $\alpha = 0$.  However, given the uncertainties in each case, we think it is 
appropriate to take the upper limit from the star-galaxy correlation function estimate, since it is
the more precise estimate in both cases.  This gives us limits
of $\alpha < 0.03$ for \imshape\ and $\alpha < 0.015$ for \ngmix.  We can multiply this by $\xi_+^{pp}$ to 
give a limit on the maximum additive systematic error we may have at $1\arcmin$ and $30\arcmin$
due to PSF leakage.

For the other additive systematic errors, we can add them together linearly.  $\xi$ acts like a variance,
so systematic uncertainties add linearly, not in quadrature.  However, the differential shear correlation test
is different from the others.  It includes many of the additive systematic errors tested by other tests, and 
in particular would almost certainly incorporate any systematic error due to PSF leakage, as the
mechanism for any such leakage would be different for the two algorithms.  Thus, it actually places
a tighter limit on the potential systematic error from PSF leakage at $30\arcmin$
than the direct estimate of $\alpha$.

The differential shear correlation does not however include all of the additive errors from the PSF model tests.
The two codes use the 
same PSF model for the $r$-band exposures, although \ngmix\ also uses $i$ and $z$-bands.  
We conservatively assume that the PSF modeling systematic errors act as $c_\mathrm{common}$ terms in the 
nomenclature of \S\ref{sec:tests:dede} and 
add them to the estimate from the differential shear correlation
to get our final estimate on the possible additive systematic error in each catalogue:
\begin{align}
\imshape \quad&
\begin{array}{l}
\left |\xi_+^\mathrm{sys}(1\arcmin) \right | < 1.1\times10^{-5}\\
\left |\xi_+^\mathrm{sys}(30\arcmin) \right | < 4\times10^{-7}
\end{array}
\\
\ngmix \quad&
\begin{array}{l}
\left |\xi_+^\mathrm{sys}(1\arcmin) \right | < 1.4\times10^{-5}\\
\left |\xi_+^\mathrm{sys}(30\arcmin) \right | < 4\times10^{-7}.
\end{array}
\end{align}
Note that we are not claiming that either catalogue has systematic errors as large as this.  
Rather, we are claiming at $\sim 1\sigma$ level of confidence that the additive systematic
errors in the two catalogues are smaller than this.

The limits on the multiplicative systematic errors come from two sources.  We have estimated the 
bias on simulated data, and we have measured the relative bias of the two catalogues with respect
to each other.  With the exception of the lowest redshift bin for \ngmix, where we found a bias of
$m \simeq -0.04$, all of the tests are consistent
with $|m| < 0.02$ for both catalogues.  

Investigation of the low redshift result for \ngmix\ indicates
that it is largely due to that bin having more bulge galaxies than the higher-redshift bins, leading to increased
model bias there.  However, \fig{fig:greatdes:histograms} shows that the distribution of bulges in \greatdes\
may not match the data very well, in particular as a function of \snr, which is correlated with redshift.
This makes us uncertain how applicable the $m\!=\!-0.04$ result is to the SV data.

Furthermore, while the tangential shear ratio test showed that the two catalogues were consistent
to within $|\Delta m|\!<\!0.02$, this was only after correcting for a matching-induced selection effect of
$\Delta m\!\simeq\!0.06$.  This correction involves a number of assumptions, so we are not confident
that it is more precise than about $\pm 0.03$.  

For these reasons, we
feel that an appropriate upper limit on $m$ for both catalogues is
\begin{equation}
|m| < 0.05.
\end{equation}
We recommend science applications that are sensitive to multiplicative bias marginalize over a
Bayesian prior on $m$ centred at $0$ with a standard deviation of $0.05$.

\section{Shear Catalogues}
\label{sec:cats}

\edit{The final shear catalogues \edit{are} publicly available on the DESDM Releases web
page\footnote{\url{http://des.ncsa.illinois.edu/releases/sva1}}.  See that
web page for complete documentation about how to access these catalogues, as well as the
other DES SV catalogues that are available.}

In this section we describe the final galaxy selection,
how to correctly apply the calibrations and sensitivities
to ensembles of galaxies, and what final number density we achieve.
\app{app:cats} has further details about the content and structure of the catalogues.

\subsection{Final Galaxy Selection}
\label{sec:cats:flags}

The starting point for our galaxy catalogues was described in \S\ref{sec:gold} and \S\ref{sec:modest}.
The former described how we selected regions of the survey where we trust the images,
and the latter described our initial galaxy selection function.
We now describe further cuts
informed by the suite of null tests in \S\ref{sec:tests}, such that the 
final shear catalogues were found to pass our tests.

We removed individual objects according to the following criteria:
\begin{itemize}
\item \sex\ flags = 1 or 2.  Objects with higher \sex\ flags were
    already been removed from the input catalogues, since they are clearly problematic
    for a shape measurement code.
    But these two flags indicate that the object is likely to be blended, and
    thus the shape measurement was likely to be corrupted.

\item ``Crazy colours''\footnote{``Crazy colours'' mean any of the following:
    {\footnotesize ${g-r} < -1,~ {g-r} > 4,~ {i-z} < -1,$ or ${i-z} > 4$}}.  Individual
    objects with questionable colours are probably contaminated by cosmic rays
    or other defects, so their shapes are also likely to be inaccurate.

\item Very low surface brightness.  We found a class of spurious objects with very large
    sizes, but relatively low flux that were usually associated with various image
    artefacts.  We excluded objects with $i +
    3.5\log(f_i / T) < 28$, where $f_i$ is the $i$-band flux, and $T=I_{xx} +
    I_{yy}$ is the (deconvolved) object size estimated by \ngmix.  This cut is
    efficient at bright fluxes, but less so at faint fluxes, removing 
    a significant number of real galaxies.

\item Tiny size.   If the \ngmix\ estimate of the object size is very small,
    then the object is probably a star.  Specifically, we removed objects with
    $T + \sigma_T < 0.02$ square arcseconds.
\end{itemize}

From the resulting set of ``good galaxies'', we then made a further selection
based on both \snr\ and
the size of the galaxy relative to the PSF size, such that the resulting
ensembles of shear estimates passed the null tests.

As we have already mentioned in \S\ref{sec:im3shape:noisebias},
the \imshape\ selection needs to be made using \snrw\ and \rgp, since these are the
parameters used for the shear calibration.  \ngmix\ does not do any calibration,
so its selection is made using \snrr\ and $T_r/\Tpsf$ (cf.~\eqnb{eq:snrr}{eq:tr}) to
avoid inducing a selection bias.

The selection that we find passes the suite of null tests is the following:
\begin{align*}
\imshape: \quad&
\begin{array}{l}
\snrw > 15 \\
\rgp > 1.2
\end{array}
\\
\ngmix: \quad&
\begin{array}{l}
\snrr > 15 \\
T_r / \Tpsf > 0.15.
\end{array}
\end{align*}
We used this selection for all of the test results shown in \S\ref{sec:tests}.

\begin{figure}
\includegraphics[width=\columnwidth]
{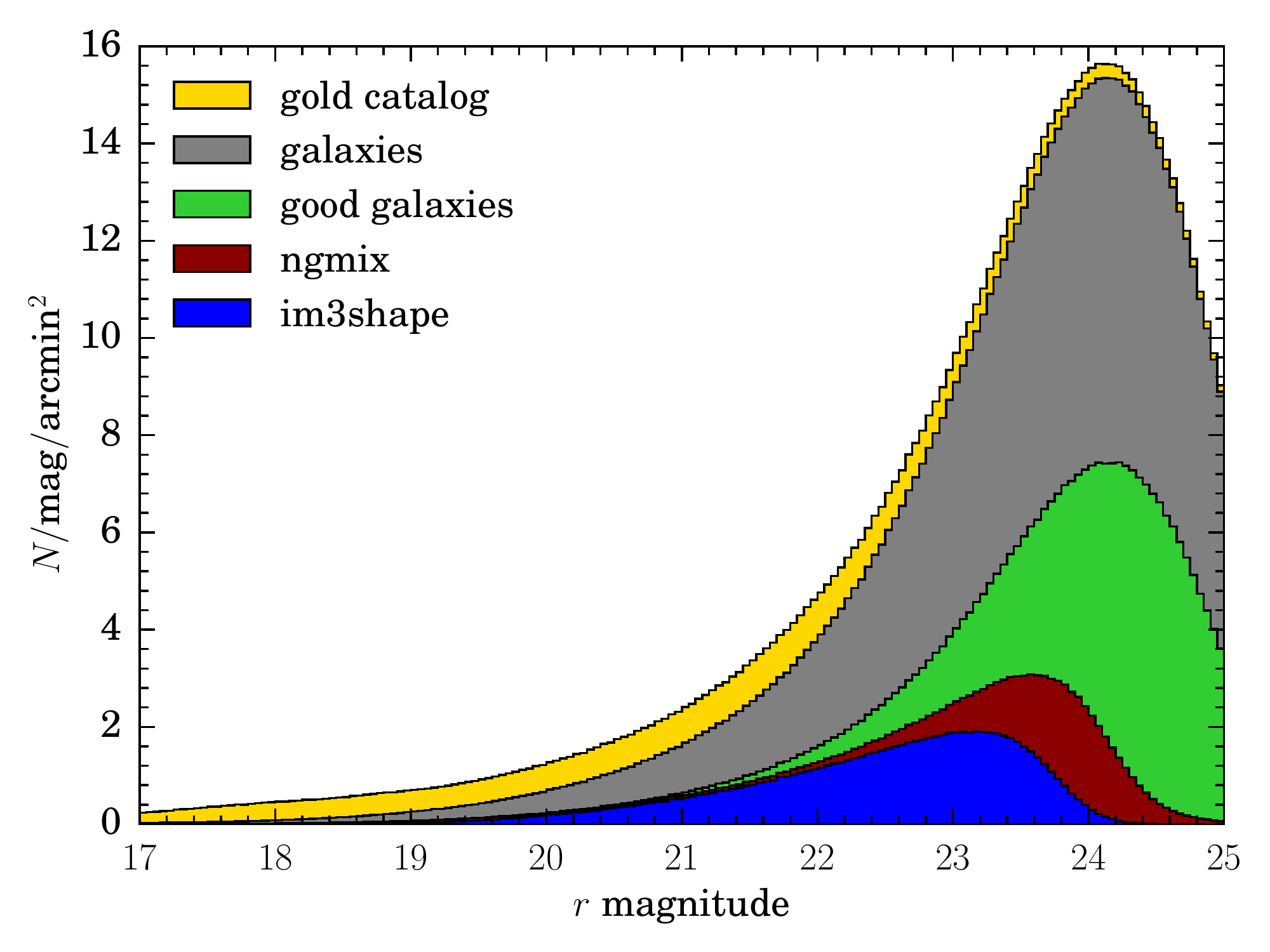}
\caption{A histogram of the $r$-band magnitude distribution showing the application of the
various selection criteria from the initial ``Gold'' Catalogue to the two final shear catalogues.
The dark red and blue show the galaxies with sufficiently accurate shape for 
\ngmix\ and \imshape\ respectively.
\label{fig:cats:maghist}
}
\end{figure}

In \fig{fig:cats:maghist} we show the effect that successively applying each round of selections has on the 
distribution of $r$-band magnitudes, starting with the original Gold Catalogue,
selecting possible galaxies, removing problematic galaxies, and then applying
the \snr\ and size criteria for the two shear catalogues.

\subsection{Applying the Calibration/Sensitivity}
\label{sec:cats:calib}

For both \imshape\ and \ngmix, the raw galaxy shape values given in the catalogue are
intrinsically biased estimators of the shear.  
In the case of \imshape, simulation-based calibration is used (cf.~\S\ref{sec:im3shape:noisebias}). 
For \ngmix, the expectation value of the ellipticity was estimated from the
posterior likelihood surface with a centrally-concentrated prior applied, which reduces
the sensitivity of the estimator to an applied shear. An estimate of
this sensitivity was calculated and given in the catalogue (cf.~\S\ref{sec:ngmix:shear}).  

In both cases, the correction factor is a noisy estimate of the true correction.
It is therefore not advisable to correct each galaxy's shape by the corresponding
correction factor directly as this will introduce a bias.  
Rather the mean shear of an ensemble of galaxies should be corrected
by the mean of the correction factors:
\begin{equation}
\langle\gamma\rangle = \frac{\sum (e_i - c_i)}{\sum s_i},
\end{equation}
where $c_i$ is the additive correction for \imshape\ ($c_i \equiv 0$ for \ngmix) and
$s_i$ is the multiplicative correction $1+m$ for \imshape\ or the estimated sensitivity for \ngmix.

The corrections in both cases are accurate in the limit of large numbers of galaxies.  
In practice, the ensemble should contain at least hundreds to thousands of galaxies to avoid 
dividing by noisy estimates of the mean sensitivity or shear bias correction.

In addition, each catalogue comes with a recommended weight $w_i$ to use for making these
ensemble averages:
\begin{equation}
\langle\gamma\rangle = \frac{\sum w_i (e_i - c_i)}{\sum w_i s_i}.
\label{eq:meangamma}
\end{equation}

For statistics such as tangential shear, you would apply the correction separately in each
bin where you are computing a mean shear.  This will apply the appropriate correction 
to the subset of the galaxies that fall into each bin.

The correction method is slightly more complicated for two-point correlation
functions, since each product involves two correction factors.  In this case,
the proper estimate is
\begin{equation}
\langle\gamma^a \gamma^b \rangle =  \frac
{\sum w_i^a w_j^b (e_i^a - c_i^a) (e_j^b - c_j^b)}
{\sum w_i^a w_j^b s_i^a s_j^b}.
\label{eq:meangammagamma}
\end{equation}
The denominator is just the two-point function of the scalar numbers $s^a$ and $s^b$.
The ratio is then taken for each bin in $\theta$.

\subsection{Effective Number Density}
\label{sec:cats:neff}

The effective number density of a weak lensing survey is defined implicitly in terms of the 
expected variance of either component of the estimated mean shear over its solid angle $\Omega$ \citep{Chang13}:
\begin{equation}
\mathrm{var}(\langle\gamma_{1,2}\rangle) \equiv  \frac{\SN^2}{\Omega n_\mathrm{eff}},
\label{eq:neffdef}
\end{equation}
where $\SN$ is the shape noise per component.

Applying all of the selections defined in \S\ref{sec:cats:flags} to our shape catalogues
results in \ngalimshape\ million galaxies for \imshape\ and \ngalngmix\ million galaxies for \ngmix.
The total useable area of SPTE is $\Omega = $ \sptearea\ square degrees (cf.~\S\ref{sec:gold}),
which leads to direct number
densities of 4.2 and 6.9 galaxies per square arcminute respectively.

To turn these numbers into proper effective number densities, we first need to calculate the 
shape noise $\SN^2$.  
\begin{align}
\SN^2 &= \frac{\sum w_i^2 \left(|e_i|^2 - 2\sigma_{e,i}^2\right)}{2 \sum w_i^2 s_i^2},
\label{eq:sndef}
\end{align}
where $2\sigma_e^2$ is the trace of the covariance matrix of $e_1,e_2$\footnote{\imshape\
does not produce a useful estimate of the covariance matrix, so we instead estimate 
$\sigma_e^2$ from the weights, which are designed to be an estimate of 
$1/(\SN^2 + \sigma_e^2)$ (cf.~\S\ref{sec:im3shape:weights}).
}, and
the $2$ in the denominator is to match the standard convention of quoting shape
noise \emph{per component}.  As described above, $s_i$ is the calibration factor or sensitivity
correction for the two catalogues.  For the \imshape\ catalogue, this number comes to 
$\SN = 0.233$, and for \ngmix, $\SN = 0.243$.

The variance of each component of the mean shear over the entire survey area can be calculated from \eqn{eq:meangamma}:
\begin{equation}
\mathrm{var}(\langle\gamma_{1,2}\rangle) 
= \frac{\sum w_i^2 \left(s_i^2 \SN^2 + \sigma_{e,i}^2\right)}{\left(\sum w_i s_i\right)^2},
\end{equation}
which, using \eqnb{eq:neffdef}{eq:sndef}, leads to
\begin{align}
n_{\rm eff} &= \frac{1}{\Omega} 
\frac{\SN^2 \left(\sum w_i s_i\right)^2}{\sum w_i^2 \left(s_i^2 \SN^2 + \sigma_{e,i}^2\right)}.
\\
&= \frac{1}{\Omega}
\frac{\left(\sum w_i s_i\right)^2}{\sum w_i^2 s_i^2} 
\left(1-\frac{2 \sum w_i^2 \sigma_{e,i}^2}{\sum w_i^2 |e_i|^2}\right)
%\left(\frac{\sum w_i^2 \left(|e_i|^2 - 2\sigma_{e,i}^2\right)}{\sum w_i^2 |e_i|^2}\right)
\end{align}
For \imshape, we find $n_{\rm eff}$ = \neffimshape\ galaxies per square arcminute, and
for \ngmix, $n_{\rm eff}$ = \neffngmix\ galaxies per square arcminute.

Note that other authors use different definitions of $n_{\rm eff}$ than this.
For instance, \citet{Heymans12} uses the definition
\begin{equation}
n_\mathrm{eff} = \frac{1}{\Omega} \frac{\left(\sum w_i \right)^2}{\sum w_i^2}.
\end{equation}
Using this definition we obtain $n_{\rm eff} = $ 4.1 and 6.8 for \imshape\ and \ngmix,
respectively.  With this definition however, the appropriate numerator in the 
ratio $\sigma_\epsilon^2/n_{\rm eff}$ is not the intrinsic shape noise $\SN^2$, but rather the
total shear noise including measurement noise.  For our data, the values
to use would be $\sigma_\epsilon = 0.245$ for \imshape\ and 
$\sigma_\epsilon = 0.265$ for \ngmix.

These number densities are quite a bit below the 10 galaxies per square arcminute
that was predicted for DES \citep{DESWhitePaper}.  This is in part because of
our decision to cut both catalogues at $\snr > 15$ rather than $10$ as we had
originally hoped to be able to do.  This  removed about 0.5 million galaxies
from the \imshape\ catalogue and 1.0 million from the \ngmix\ catalogue.
Moving the \imshape\ size cut down to $\rgp > 1.15$ as well
would add another 0.8 million galaxies.  We hope that algorithm improvements
to both catalogues will make these looser selection criteria possible in future DES analyses.

Furthermore, the average depth of the SV survey was not the full $\sim$\nomepochs\ exposures
we expect for DES after five years.  Instead, the mean is approximately $7$ exposures
averaged across the \spte\ area.  If we reach an average of $10$ exposures,
This will lead to a 20\% increase in the mean \snr\ and
a corresponding increase in the number of usable galaxies.

In addition, the predicted value was based on an expected median seeing of 0.9\arcsec,
while the median seeing during science verification was slightly above 1.0\arcsec.
We are closer to achieving our goal of 0.9\arcsec\ in the main survey observations \citep{Diehl14}, so this
will help to increase $n_\mathrm{eff}$.

Another reason for the low number count
is the rejection of objects with neighbors.  The \sex\ flags related to blended objects
removed almost 1 million galaxies from the catalogues.
We are currently working on an algorithm to
model the profiles of neighboring objects so their light profiles can be
effectively removed from the image and not contaminate the shapes of nearby
objects, thus allowing us to keep more of these objects in the catalogue.

Another obvious improvement will be to use multi-band fitting in \imshape, 
which would increase the \snr\ of each galaxy by using more pixels of information.  
This is already implemented, but it was not complete in time
to be run and tested on these data.   It will be used in the next DES analysis.

Finally, the detection of image artefacts in the data management pipeline has been improved
from the version used for the SV data.  
The removal of low surface brightness objects, which was designed to remove a large proportion
of these artefacts,
removed 1.5 million objects.  Presumably many of these are real galaxies rather than
image artefacts, so if we can omit this step, we will keep more galaxies in the catalogue.

With all of these improvements to both the data and the algorithms, we are optimistic
that we will be able 
achieve our forecasted $n_\mathrm{eff} \!=\! 10$ galaxies per square arcminute in the 5-year DES analysis.

\section{Summary and Discussion}
\label{sec:conclusions}
\assign{Mike}

We present here two shear catalogues for the \spte\ region observed as part of the DES science verification
time.  Both catalogues, \ngmix\ and \imshape, have passed a comprehensive suite of null tests that show that they are
accurate enough to be used for weak lensing science with these data.
The catalogues have 4.2 and 6.9 galaxies per square arcminute (for \imshape\ and  \ngmix\ respectively), 
which corresponds to \ngalimshape\ and \ngalngmix\ million galaxies over the \sptearea\ square degree footprint.
These correspond to effective number densities of \neffimshape\ and \neffngmix\ galaxies per square
arcminute, respectively (cf.~\S\ref{sec:cats:neff}).

For creation of both shear catalogues we used the original \SE\ pixel data to jointly constrain the galaxy models, 
thereby avoiding issues of correlated noise and complex PSF interpolation that occur
when using stacked images.
This is a relatively new technique in weak lensing, having only previously been employed on 
real data by \citet{Heymans12} and \citet{Kuijken15}.
%However, given its significant advantages,

%we believe it will become the standard algorithm employed by most future surveys.

%\ESS{You are probably right, but I'm not sure we should get into the business of speculating}
% MJ: Fair point.

In addition to passing null tests on the data individually,
the two catalogues are consistent with each other, both in terms of 
possible additive systematic errors and the overall calibration (i.e. multiplicative systematic errors).  
This is a non-trivial
result, considering that the calibration strategies of the two catalogues are completely different;
\imshape\ calibrates the shear bias from simulations, and \ngmix\ uses a Bayesian algorithm
that is relatively insensitive to noise bias, but does require a prior on the ellipticity distribution.  
This is the first significant weak lensing analysis to present two accurate and independent shear catalogues,
and thus the first to be able to show this kind of consistency.

In \S\ref{sec:tests:summary} we estimated upper limits on the level of additive
systematic errors that may be present in the two catalogues at $1\arcmin$ and
$30\arcmin$.  We recommend a Bayesian prior of $|m| < 0.05$ for the
systematic uncertainty on the calibration for both catalogues.  \edit{Since
    \ngmix\ is the deeper catalogue (due to using multi-band data rather than
    just $r$ band), users desiring the most precise measurement may adopt
    \ngmix\ as their canonical shear catalogue.  However, we strongly recommend
also comparing with results using the \imshape\ catalogue, while carefully
taking into account any selection effects; any discrepancy 
indicates a systematic error in one or both catalogues.}

While the catalogues were seen to be sufficiently accurate for SV weak lensing science, they 
do not yet pass the tests at the level that will be needed for the full 5-year DES data.  There is still
a significant amount of work required to improve the algorithms to meet those requirements.

One area that needs improvement is our PSF modeling (cf.~\figb{fig:psf:bfe}{fig:psf:rho}).
Fortunately, there has
been a significant amount of work in recent years on improved PSF modeling and interpolation algorithms
\citep[e.g.][]{Chang12, Li12, Kitching13, Gentile13}.
We have also been working on an algorithm to model the PSF using the actual optical
aberrations measured from the wavefront sensors in the corners of the DECam field of view \citep{Roodman14}. 
We will investigate
whether incorporating this information can lead to more accurate PSF interpolation.

We also expect a significant improvement in the astrometric solution in the next round of analysis.
It will include a more accurate functional form
for the telescope distortion and also take into account effects like edge distortions and tree rings
that are present in the data (cf.~\fig{fig:tests:colnum}).  
We expect this to reduce some of the spurious features seen 
in \figb{fig:psf:fov}{fig:tests:fieldcenter}.

We have recently implemented an algorithmic correction to the brighter-fatter relation discussed in \S\ref{sec:psf:cuts}
\citep{Gruen15}.  This will allow us to use brighter stars for constraining the PSF
than we were able to use in this analysis, which is expected to lead to better estimated PSFs.

We are working on an improved algorithm for handling neighbors by subtracting off an estimate of their
light profile rather than merely masking contaminated pixels.  While not a perfect subtraction, 
we expect this will let us use more pixels for constraining the galaxy models, which will lead
to fewer galaxies being removed from the final catalogues.  Contamination by neighbors was one of the 
more significant cuts that led to the drop in number density for the ``good galaxies'' seen in \fig{fig:cats:maghist}.

There are also two new shear algorithms being developed for DES.  One is based on the 
Bayesian Fourier domain (BFD) algorithm of \citet{BernsteinArmstrong2014}.  The other is based on the
MetaCalibration strategy, a preliminary version of which was tested
in the GREAT3 challenge \citep{great3results}.  Considering how useful we found it
to have two catalogues, we are looking forward to the prospect of additional catalogues
to compare in various ways.

We also plan to begin implementing corrections for the chromatic effects of the PSF described by
\citet{Myers15}.  According to their estimates of the effects of PSF chromaticity, it is not expected
to be a significant problem for
the current analysis, but we will need to correct for these effects in the 5-year data analysis.

In addition to these planned algorithmic improvements, the data itself will be somewhat better
in the main survey.  Part of the reason for taking the science verification data was to find problems with the
camera and telescope hardware.  As such, quite a few hardware improvements were made during
this time, as well as some in the following year \citep[cf.][]{Diehl14}.
The image quality for the main survey is thus significantly better than the already quite good
image quality in the SV data.

We therefore believe that we will be able to significantly improve the quality of the shear catalogues
in future DES analyses.  We must keep the level of systematic errors below the improved statistical uncertainty
for these data.  The full 5-year DES data will cover about 30 times more area, so our requirements
for the systematic errors will drop by roughly a factor of 5.
By implementing the improvements discussed here, we hope to keep pace with the requirements.

\section*{Acknowledgements}

We are grateful for the extraordinary contributions of our CTIO colleagues and the DECam 
Construction, Commissioning and Science Verification teams in achieving the excellent 
instrument and telescope conditions that have made this work possible. The success of this 
project also relies critically on the expertise and dedication of the DES Data Management group.

We thank the many DES-internal reviewers whose suggestions have vastly improved this paper during the
collaboration-wide review process.
The DES publication number for this article is DES-2015-0059. 
The Fermilab preprint number is FERMILAB-PUB-15-309-AE.

We also thank the anonymous referee, whose careful reading helped to significantly
improve the paper.

Jarvis has been supported on this project by NSF grants AST-0812790 and AST-1138729.
Jarvis, Bernstein, Clampitt, and Jain are partially supported by DoE grant DE-SC0007901.
Sheldon is supported by DoE grant DE-AC02-98CH10886.
Zuntz, Kacprzak, Bridle, and Troxel acknowledge support from the European Research Council in the form of a Starting Grant with number 240672.
Das was funded by DoE Grant DE-SC0007859.
Gruen was supported by SFB-Transregio 33 `The Dark Universe' by the Deutsche Forschungsgemeinschaft (DFG) and the DFG cluster of excellence `Origin and Structure of the Universe'.
Gangkofner acknowledges the support by the DFG Cluster of Excellence `Origin and Structure of the Universe'.
Melchior was supported by DoE grant DE-FG02-91ER40690.
Plazas was supported by DoE grant DE-AC02-98CH10886 and by JPL, run by Caltech under a contract for NASA. 
Lima is partially supported by FAPESP and CNPq.

Funding for the DES Projects has been provided by the U.S. Department of Energy, the U.S. National Science 
Foundation, the Ministry of Science and Education of Spain, the Science and Technology Facilities Council of 
the United Kingdom, the Higher Education Funding Council for England, the National Center for Supercomputing 
Applications at the University of Illinois at Urbana-Champaign, the Kavli Institute of Cosmological Physics 
at the University of Chicago, the Center for Cosmology and Astro-Particle Physics at the Ohio State University,
the Mitchell Institute for Fundamental Physics and Astronomy at Texas A\&M University, Financiadora de 
Estudos e Projetos, Funda{\c c}{\~a}o Carlos Chagas Filho de Amparo {\`a} Pesquisa do Estado do Rio de 
Janeiro, Conselho Nacional de Desenvolvimento Cient{\'i}fico e Tecnol{\'o}gico and the Minist{\'e}rio da 
Ci{\^e}ncia e Tecnologia, the Deutsche Forschungsgemeinschaft and the Collaborating Institutions in the 
Dark Energy Survey. 

The DES data management system is supported by the National Science Foundation under Grant Number 
AST-1138766. The DES participants from Spanish institutions are partially supported by MINECO under 
grants AYA2012-39559, ESP2013-48274, FPA2013-47986, and Centro de Excelencia Severo Ochoa 
SEV-2012-0234, some of which include ERDF funds from the European Union.

The Collaborating Institutions are Argonne National Laboratory, the University of California at Santa Cruz, 
the University of Cambridge, Centro de Investigaciones Energeticas, Medioambientales y Tecnologicas-Madrid, 
the University of Chicago, University College London, the DES-Brazil Consortium, the Eidgen{\"o}ssische 
Technische Hochschule (ETH) Z{\"u}rich, Fermi National Accelerator Laboratory,
the University of Edinburgh, 
the University of Illinois at Urbana-Champaign, the Institut de Ci\`encies de l'Espai (IEEC/CSIC), 
the Institut de F\'{\i}sica d'Altes Energies, Lawrence Berkeley National Laboratory, the Ludwig-Maximilians 
Universit{\"a}t and the associated Excellence Cluster Universe, the University of Michigan, the National Optical 
Astronomy Observatory, the University of Nottingham, The Ohio State University, the University of Pennsylvania, 
the University of Portsmouth, SLAC National Accelerator Laboratory, Stanford University, the University of 
Sussex, and Texas A\&M University.

\bibliography{bibliography}

\appendix

\section{\medsfull}
\label{app:medsappendix}

Each \meds\ file corresponds to a single \coadd\ image.
For each one, we gather the list of all \SE\ images that were used to construct the \coadd\ image.
Then for each object in the corresponding \coadd\ detection catalogue, we identify the location of the object
in all \SE\ images where that object appears using each image's WCS transformation to convert
between the coordinate systems.  We then identify a region
around each object in each \SE\ image and save it as a postage stamp in
the \meds\ file.  A postage stamp from the \coadd\ image is also stored in the file as the first
entry for each object.

The size of the cutout is determined from the basic \sex\ measurements \frad,
\aworld\ and \bworld\ as follows:
\begin{align} \label{eq:boxsize}
    s             &= 2 \times 5 \times \sigma \times (1 + \epsilon) \\
    \sigma        &= \mathrm{FWHM}/fac \\
    \mathrm{FWHM} &= 2 \times \code{FLUX\_RADIUS} \\
    \epsilon      &= 1 - \code{B\_WORLD}/\code{A\_WORLD},
\end{align}
where $fac \sim 
2.35$ is the conversion between FWHM and $\sigma$.
The \frad\ is a robustly measured quantity, being the radius of the circular
aperture enclosing half the estimated total flux of the object.  We find that
\aworld\ and \bworld, while not suitable for a lensing shear analysis, are
measured well enough for the purpose of estimating the eccentricity $\epsilon$.

We take the maximum of the size $s$ from all \SE\ measurements as the fiducial cutout size.
To facilitate fast FFT calculations on the cutouts, we round the fiducial
cutout sizes upward to either a power of two or 3 times a power of two.

In addition to the image cutouts, we also store the \sex\ weight map and
segmentation map.  The cutouts are background subtracted using the background
maps output by \sex.  The weight maps are modified to be zero anywhere that a flag
is set in the \sex\ maskplane, which includes defects such as bad pixels.
The different image types are stored in
separate extensions of the file, along with a plethora of metadata. 

All images, including the coadd image, are placed on the same photometric
system such that the magnitude zero point is 30.0.  The weight maps are also
scaled appropriately.

Because the full set of data to be stored in the \meds\ file
does not fit into memory simultaneously, we use the ability of
CFITSIO\footnote{\url{http://heasarc.gsfc.nasa.gov/fitsio/fitsio.html}} to
write chunks of images directly to disk without keeping the full image in main
memory.  

The code for creating \meds\ files, including the WCS transformation
library, is hosted publicly as part of a larger package
\code{deswl\_shapelets}\footnote{\url{https://github.com/rmjarvis/deswl\_shapelets}}.
The code that generates the input object list, including cutout sizes, is part
of the \code{meds} software library.

The \meds\ data, including all of the images of each object observed in a single \coadd\ tile,
along with appropriate catalogue information, are stored in a single \fits\ file.

To simplify access to the data in the \meds\ filess,
we provide an Application Programmer's Interface (API) library,
\code{meds}, which is available for download\footnote{\url{https://github.com/esheldon/meds}}
and is free software.  A full API is
provided for the Python language. A smaller subset of the
full functionality is available as a library for the C programming language.
\edit{For complete documentation of the meds file structure and the API for reading these
files,} we direct the reader to the \code{meds}
repository URL.

\section{Details of the Shear Catalogues}
\label{app:cats}

\subsection{\imshape\ Flags}
\label{app:im3shapeflags}

\imshape\ reports two kinds of flags. \tab{tab:im3shape:errorflag} lists ``error'' flags, and 
\tab{tab:im3shape:infoflag} lists ``info'' flags.  For the most conservative treatment, users
should select galaxies where both are zero.  However, using \code{INFO\_FLAG} $> 0$ may be appropriate
in some cases. 

\begin{table}
\begin{tabulary}{1\columnwidth}{ c r J }
\hline
Value & \multicolumn{1}{c}{Decimal}  & \multicolumn{1}{c}{Meaning} \\
\hline
        \edit{ $2^{0}$}  &           1  &  \imshape\ failed completely. \\
        \edit{ $2^{1}$}  &           2  &  Minimizer failed to converge. \\
        \edit{ $2^{2}$}  &           4  &  Tiny ellipticty $|e|<10^{-4}$: \imshape\ fit failed. \\
        \edit{ $2^{3}$}  &           8  &  $e_1$ or $e_2$ outside $(-1,1)$. \\
        \edit{ $2^{4}$}  &          16  &  Radius $> 20$ arcseconds. \\
        \edit{ $2^{5}$}  &          32  &  $\rgp > 6$ - huge galaxy. \\
        \edit{ $2^{6}$}  &          64  &  Negative or nan $\rgp$. \\
        \edit{ $2^{7}$}  &         128  &  $\snrw<1$. \\
        \edit{ $2^{8}$}  &         256  &  $\chi^2$ per effective pixel $> 3$. \\
        \edit{ $2^{9}$}  &         512  &  Normed residuals $< -20$ somewhere. \\
        \edit{$2^{10}$}  &        1024  &  Normed residuals $> 20$ somewhere. \\
        \edit{$2^{11}$}  &        2048  &  $\delta u$ more than 10 arcseconds from nominal. \\
        \edit{$2^{12}$}  &        4096  &  $\delta v$ more than 10 arcseconds from nominal. \\
        \edit{$2^{13}$}  &        8192  &  Failed to measure the FWHM of PSF or galaxy. \\
        \edit{$2^{14}$}  &       16384  &  $r$-band \sex\ flag has \code{0x4} or above. \\
        \edit{$2^{30}$}  &  \edit{1073741824}  &  \edit{No attempt at a fit was made due to cuts prior to running im3shape.}\\
  \hline
\end{tabulary}
\caption{Error flags in the \imshape\ catalogue. Objects with non-zero \code{ERROR\_FLAG} should be removed 
from any science analysis.
\label{tab:im3shape:errorflag}
}
\end{table}

\begin{table}
\begin{tabulary}{1\columnwidth}{ c r J }
\hline
Value & \multicolumn{1}{c}{Decimal}  & \multicolumn{1}{c}{Meaning} \\
\hline
         \edit{$2^{0}$}     &        1  &  $r$-band \sex\ flagged with \code{0x1}, indicating bright neigbours. \\
         \edit{$2^{1}$}     &        2  &  $r$-band \sex\ flagged with \code{0x2}, indicating blending. \\
         \edit{$2^{2}$}     &        4  &  Mask fraction $> 0.5$. \\
         \edit{$2^{3}$}     &        8  &  Model image $< -0.01$ somewhere. \\
         \edit{$2^{4}$}     &       16  &  $\rgp < 1.15$. \\
         \edit{$2^{5}$}     &       32  &  Radius $> 5$ arcseconds. \\
         \edit{$2^{6}$}     &       64  &  Centroid more than 0.6 arcseconds from nominal. \\
         \edit{$2^{7}$}     &      128  &  $\chi^2$ per effective pixel $> 1.25$. \\
         \edit{$2^{8}$}     &      256  &  $\rgp > 3.5$ (very large galaxy). \\
         \edit{$2^{9}$}     &      512  &  Normed residuals $< -2$ somewhere. \\
        \edit{$2^{10}$}     &     1024  &  Normed residuals $> 2$ somewhere. \\
        \edit{$2^{11}$}     &     2048  &  Declination below limit where we have good photometry. \\
        \edit{$2^{12}$}     &     4096  &  \edit{$\snrw > 10^5$}. \\
        \edit{$2^{13}$}     &     8192  &  Radius $> 10$ arcseconds. \\
        \edit{$2^{14}$}     &    16384  &  $\snrw < 10$. \\
        \edit{$2^{15}$}     &    32768  &  Model image $< -0.05$ somewhere. \\
        \edit{$2^{16}$}     &    65536  &  $\chi^2$ per effective pixel $< 0.8$. \\
        \edit{$2^{17}$}     &   131072  &  More than 70\% of fitted flux is in masked region. \\
        \edit{$2^{18}$}     &   262144  &  Model completely negative. \\
        \edit{$2^{19}$}     &   524288  &  $\chi^2$ per effective pixel $> 2$. \\
        \edit{$2^{20}$}     &  1048576  &  \edit{PSF FWHM $> 10$ arcsec}. \\
        \edit{$2^{21}$}     &  2097152  &  Negative PSF FWHM. \\
        \edit{$2^{22}$}     &  4194304  &  \edit{$\rgp < 1.1$}. \\
        \edit{$2^{23}$}     &  8388608  &  Centroid more than one arcsecond from nominal. \\
        \edit{$2^{24}$}     & 16777216  &  Mask fraction $> 0.75$. \\
        \edit{$2^{25}$}     & 33554432  &  One or more error flags is set. \\
  \hline
\end{tabulary}
\caption{Info flags in the \imshape\ catalogue. Objects with non-zero \code{INFO\_FLAG} may be acceptable
depending on the scientific application.
\label{tab:im3shape:infoflag}
}
\end{table}

\subsection{\ngmix\ Flags}
\label{app:ngmixflags}

The \ngmix\ catalogue has an error flag that indicate when some kind of error occurred during the
fitting procedure.
Users should only use galaxies with \code{ERROR\_FLAG == 0}. 
The meanings of the various possible error flag values are given in \tab{tab:ngmix:flags}.

\begin{table}
\begin{tabulary}{1\columnwidth}{ c r J }
\hline
Value & \multicolumn{1}{c}{Decimal}  & \multicolumn{1}{c}{Meaning} \\
\hline
\edit{$2^{0}$}   &           1  &  There were no cutouts for this object. \\
\edit{$2^{1}$}   &           2  &  PSF fitting failed for all epochs. \\
\edit{$2^{2}$}   &           4  &  Not used. \\
\edit{$2^{3}$}   &           8  &  Galaxy fitting failed. \\
\edit{$2^{4}$}   &          16  &  Box size was larger than 2048. \\
\edit{$2^{5}$}   &          32  &  Not used. \\
\edit{$2^{6}$}   &          64  &  The \snrw\ of the PSF flux was lower than 4 in all bands. \\
\edit{$2^{7}$}   &         128  &  Utter failure of the fitting. For this release, the flag was set when no valid guess for the fitters could be generated. \\
\edit{$2^{8}$}   &         \edit{256}  &  \edit{Not used.} \\
\edit{$2^{9}$}   &         \edit{512}  &  \edit{No attempt was made for round measures because of prior failure.} \\
\edit{$2^{10}$}  &        \edit{1024}  &  \edit{Round model could not be evaluated within the allowed region of parameter space.}\\
\edit{$2^{11}$}  &        \edit{2048}  &  \edit{Fitting failed when trying to estimate \snr\ of $T_r$.}\\
\edit{$2^{30}$}  &  1073741824  &  No attempt of a fit was made due to other flags. \\
  \hline
\end{tabulary}
\caption{Error flag values in the \ngmix\ catalogue.  
Objects with non-zero \code{ERROR\_FLAG} should be removed from any science analysis.
\label{tab:ngmix:flags}
}
\end{table}

\subsection{\edit{File Structure}}
\label{app:catfiles}

\edit{There are three files 
available on the DESDM SVA1 Release web
page\footnote{\url{http://des.ncsa.illinois.edu/releases/sva1}}
containing the final DES SV shear catalogues:}
\begin{itemize}
\item
\edit{\code{sva1\_gold\_r1.1\_im3shape.fits.gz}} is the \imshape\ catalogue.
The relevant columns in this catalogue are listed in \tab{tab:cats:im3shape}.

\item
\edit{\code{sva1\_gold\_r1.0\_ngmix.fits.gz}} is the \ngmix\ catalogue.
The relevant columns in this catalogue are listed in \tab{tab:cats:ngmix} 

\item
\edit{\code{sva1\_gold\_r1.0\_wlinfo.fits}} has flags that can be used to select
a set of galaxies with good shear estimates.  
It also has columns with information from the main \coadd\ catalogue,
such as RA and Declination, for convenience in using these catalogues without having to join them to
the main DES object catalogue.  Photometric redshift information is based on the 
SkyNet algorithm \citep{Sanchez14, SkyNet, Bonnett15}.
\edit{The columns in this catalogue are listed in \tab{tab:cats:info}.}

\end{itemize}

\edit{In addition to these three weak lensing files, the SVA1 release includes files
for the Gold Catalogue, limiting magnitude maps, the good-region footprint,
photo-z catalogues (including both point estimates and full PDFs) \citep{Bonnett15}, 
redMaPPer cluster catalogues \citep{Rykoff16}, and redMaGiC galaxy catalogues \citep{redmagic}.
See the release documentation page for more details about these other files.}

Most users will want to select objects where \code{SVA1\_FLAG == 0}.  This selects
the objects that we are confident are actually galaxies, and not either stars
or some kind of spurious artefact in the data.  See \tab{tab:cats:sva1} for the 
meaning of non-zero values of this flag.
In addition we have two additional columns that indicate which galaxies fail the 
\imshape\ and \ngmix\ selection criteria.  The \code{IM3SHAPE\_FLAG} column is zero if

{\footnotesize
\begin{verbatim}
   (ERROR_FLAG==0) & (INFO_FLAG==0) & 
   (SNR_W>15) & (MEAN_RGPP_RP>1.2)
\end{verbatim}
}
The \code{NGMIX\_FLAG} column is zero if
{\footnotesize
\begin{verbatim}
   (ERROR_FLAG==0) & (0.4<ARATE<0.6) &
   (SENS_1>0.0) & (SENS_2>0.0) &
   (SNR_R>15) & (T_R/MEAN_PSF_T>0.15)
\end{verbatim}
}
In each case, these select the galaxies which have been found to pass all of the
null tests in \S\ref{sec:tests}.
Users can thus select galaxies with good shear estimates by simply selecting
\code{IM3SHAPE\_FLAG==0} or \code{NGMIX\_FLAG==0} as appropriate.

\begin{table}
\begin{tabulary}{1\columnwidth}{ c J }
\hline
Column & \multicolumn{1}{c}{Meaning} \\
\hline
\code{COADD\_OBJECTS\_ID} & A unique id number of the object. \\
\code{E\_1} & The raw $e_1$ shape estimate. \\
\code{E\_2} & The raw $e_2$ shape estimate. \\
\code{NBC\_M} & The multiplicative bias correction. \\
\code{NBC\_C}$i$ & ($i\!\!\,\in$ \{\code{1},\code{2}\}) 
The additive bias corrections. \\
\code{W} & The recommended weight. \\
\code{ERROR\_FLAG} & The error flag (cf.~\tab{tab:im3shape:errorflag} in \app{app:im3shapeflags}). \\
\code{INFO\_FLAG} & The info flag (cf.~\tab{tab:im3shape:infoflag} in \app{app:im3shapeflags}). \\
\code{SNR\_W} & The estimated \snrw. \\
\code{SNR\_R} & The estimated \snrr. \\
\edit{\code{MEAN\_RGPP\_RP}} & \edit{The mean value of $\rgp$ among the different observations of the galaxy.}\\
\hline
\end{tabulary}
\caption{The most relevant columns in the \imshape\ catalogue.
\edit{Additional columns included in the catalogues are described in the documentation on the
release web page.}
\label{tab:cats:im3shape}
}
\end{table}

\begin{table}
\begin{tabulary}{1\columnwidth}{ c J }
\hline
Column & \multicolumn{1}{c}{Meaning} \\
\hline
\code{COADD\_OBJECTS\_ID} & A unique id number of the object. \\
\code{E\_1} & The raw $e_1$ shape estimate. \\
\code{E\_2} & The raw $e_2$ shape estimate. \\
\code{SENS\_AVG} & \edit{The average of the two sensitivity estimates.} \\
\code{W} & The recommended weight. \\
\code{E\_COV\_}$i$\code{\_}$j$ & ($i,j\!\!\,\in$ \{\code{1},\code{2}\})  
The covariance matrix of the shape estimate. \\
\code{ERROR\_FLAG} & The error flag (cf. \tab{tab:ngmix:flags} in \app{app:ngmixflags}). \\
\code{SNR\_W} & The estimated \snrw. \\
\code{SNR\_R} & The estimated \snrr. \\
\edit{\code{T}} & \edit{An estimate of the size, $T$, in arcsec$^2$.}\\
\edit{\code{T\_R}} & \edit{An estimate of the size, $T$, that the object would have had if it were round.}\\
\edit{\code{MEAN\_PSF\_T}} & \edit{The mean measured size, $T$, of the PSF for the different exposures that went into the shear estimates for this galaxy.}\\
\edit{\code{ARATE}} & \edit{Acceptance rate of the MCMC chain.}\\
\hline
\end{tabulary}
\caption{The most relevant columns in the \ngmix\ catalogue.
\edit{Additional columns included in the catalogues are described in the documentation on the
release web page.}
\label{tab:cats:ngmix}
}
\end{table}

\begin{table}
\begin{tabulary}{1\columnwidth}{ c J }
\hline
Column & \multicolumn{1}{c}{Meaning} \\
\hline
\code{COADD\_OBJECTS\_ID} & A unique id number of the object \\
\code{RA} & The right ascension of the object in degrees \\
\code{DEC} & The declination of the object in degrees \\
\code{MAG\_AUTO\_G} & The $g$-band magnitude \\
\code{MAG\_AUTO\_R} & The $r$-band magnitude \\
\code{MAG\_AUTO\_I} & The $i$-band magnitude \\
\code{MAG\_AUTO\_Z} & The $z$-band magnitude \\
\code{PHOTOZ\_BIN} & The cosmological photometric redshift bin (0,1,2)
\edit{as described in \citet{SVCosmology}} \\
\code{MEAN\_PHOTOZ} & A point estimate of the photometric redshift
\edit{from the SkyNet photo-z catalogue \citep{Bonnett15}}  \\
\code{SVA1\_FLAG} & A flag indicating problematic galaxies (cf.~\tab{tab:cats:sva1}) \\
\code{IM3SHAPE\_FLAG} & A flag that is 0 if this object's shape in the \imshape\ catalogue is good to use; 1 if not. \\
\code{NGMIX\_FLAG} & A flag that is 0 if this object's shape in the \ngmix\ catalogue is good to use; 1 if not. \\
\hline
\end{tabulary}
\caption{The columns in the info catalogue.
\label{tab:cats:info}
}
\end{table}

\begin{table}
\begin{tabulary}{1\columnwidth}{ c r J }
\hline
Value & \multicolumn{1}{c}{Decimal}  & \multicolumn{1}{c}{Meaning} \\
\hline
\edit{$2^0$}    &    1      &  $i$-band \sex\ flag has bit \code{0x1} set, indicating bright neighbors. \\
\edit{$2^1$}    &    2      &  $i$-band \sex\ flag has bit \code{0x2} set, indicating blending. \\
\edit{$2^2$}    &    4      &  Modest Classification calls this object a star (\code{bright\_test} or \code{locus\_test} from \S\ref{sec:modest}). \\
\edit{$2^3$}    &    8      &  Modest Classification calls this object junk (\code{faint\_psf\_test} from \S\ref{sec:modest}). \\
\edit{$2^4$}    &   16     &  In region with high density of objects with ``crazy colours''
\edit{(i.e. any of the following: ${g-r} < -1$, ${g-r} > 4$, ${i-z} < -1$, or ${i-z} > 4$)}. \\
\edit{$2^5$}    &   32     &  \edit{In region with a high density of objects} 
with large centroid shifts between bandpasses. \\
\edit{$2^6$}    &   64     &  \edit{Near a 2MASS star with $J_M < 12$.
The mask radius is flux dependent, up to 120 arcmin for the brightest stars.} \\
\edit{$2^7$}    &  128    &  Large offset in $g$ and $i$ band windowed positions. \\
\edit{$2^8$}    &  256    &  Object was not measured by \ngmix. \\
\edit{$2^9$}    &  512    &  Likely star according to \ngmix\ $T + \sigma_T < 0.02$ square arcseconds. \\
\edit{$2^{10}$}  & 1024   &  Very low surface brightness according to \ngmix\ measurements. \\
\edit{$2^{11}$}  & 2048   &  Object does not satisfy good measurement flags in \ngmix. \\
\edit{$2^{12}$} & \edit{4096} & \edit{Object does not have a valid magnitude in all $g$,$r$,$i$,$z$ bands. (That is, at least one of them is invalid.)} \\
\hline
\end{tabulary}
\caption{Values of the \code{SVA1\_FLAG} in the info catalogue
\label{tab:cats:sva1}
}
\end{table}

\bsp % typesetting comment
\label{lastpage}
\end{document}